\pdfoutput=1
\documentclass[preprint,12pt]{elsarticle}

\usepackage{soul}

\graphicspath{{./figures/}}

\usepackage{hyperref}
\usepackage{multirow}
\usepackage{bm}
\usepackage{amsmath,amssymb}
\usepackage{cleveref}
\usepackage{soul}
\usepackage{float}
\usepackage[labelformat=simple]{subcaption}
\DeclareCaptionLabelFormat{bold}{\textbf{#2} }

\usepackage{color}

\newcommand{\norm}[1]{\left\lVert#1\right\rVert}

\newcommand{\NNodes}{N_{\Degree}}

\def\equationautorefname{Eq.}

\def\equationautorefname~#1\null{%
  Eq.~(#1)\null
}

\usepackage{algpseudocode}
\usepackage{algorithm}

\algrenewcommand\algorithmicindent{0.7em}

\algnewcommand\algorithmicswitch{\textbf{switch}}
\algnewcommand\algorithmiccase{\textbf{case}}
\algdef{SE}[SWITCH]{Switch}{EndSwitch}[1]{\algorithmicswitch\ #1\ \algorithmicdo}{\algorithmicend\ \algorithmicswitch}%
\algdef{SE}[CASE]{Case}{EndCase}[1]{\algorithmiccase\ #1}{\algorithmicend\ \algorithmiccase}%
\algtext*{EndSwitch}%
\algtext*{EndCase}%
\algnewcommand{\LineComment}[1]{\State \(\triangleright\) #1}
\makeatletter
\renewcommand{\Function}[2]{%
  \csname ALG@cmd@\ALG@L @Function\endcsname{#1}{#2}%
  \def\@currentfunction{#1}%
}
\newcommand{\funclabel}[1]{%
  \@bsphack
  \protected@write\@auxout{}{%
    \string\newlabel{#1}{{\@currentfunction}{\thepage}}%
  }%
  \@esphack
}
\makeatother

\newcommand{\Degree}{q}

\newcommand{\figsize}{0.40}

\renewcommand{\d}[1]{\ensuremath{\operatorname{d}\!{#1}}}

\newcommand{\NameSurfaceSubdivision}{SurfaceSubdivision}

\newcommand{\NameRecastFeatures}{SmoothFeatures}
\newcommand{\CreateMesh}{SetMesh}
\newcommand{\MapOntoLimit}{MapOntoLimitManifold} 
\newcommand{\MapOntoLimitSurface}{MapOntoLimitSurface} 
\newcommand{\MapOntoLimitCurve}{MapOntoLimitCurve} 
\newcommand{\NameAlgorithmInterpolation}{GenerateHOSurfaceMesh}

\newcommand{\Tri}{T}

\newcommand{\Seg}{S}

\newcommand{\MasterOneD}{\hat{\Seg}}
\newcommand{\MasterTwoD}{\hat{\Tri}}

\newcommand{\Model}{\Omega}
\newcommand{\ModelDegree}{\Model^{\Degree}}
\newcommand{\ModelLimit}{\Model^{\infty}}

\newcommand{\FeatSurface}{\mathcal{T}}
\newcommand{\FeatCurve}{\mathcal{S}}
\newcommand{\FeatPoint}{\mathcal{P}}

\newcommand{\IndexSurface}{s}
\newcommand{\IndexCurve}{c}
\newcommand{\IndexPoint}{p}

\newcommand{\Facet}{f}
\newcommand{\Edge}{e}
\newcommand{\Vertex}{v}

\newcommand{\LimitSurface}{\FeatSurface^{\infty}}
\newcommand{\LimitCurve}{\FeatCurve^{\infty}}

\newcommand{\SurfaceDegree}{\FeatSurface^{\Degree}}
\newcommand{\CurveDegree}{\FeatCurve^{\Degree}}

\newcommand{\PhiCurveElem}{\varphi_{\IndexCurve,\Edge}}

\newcommand{\PhiSurfaceElem}{\varphi_{\IndexSurface,\Facet}}

\newcommand{\MeshNodePhysicalLetter}{x}
\newcommand{\MeshNodePhysical}{\bm{\MeshNodePhysicalLetter}}

\newcommand{\Mesh}{\mathcal{M}}

\newcommand{\NormalVector}{\bm{n}}
\newcommand{\NormalFunction}{\alpha}

\newcommand{\PhiLim}{\phi^{\infty}}
\newcommand{\PhiDeg}{\phi^{\Degree}}

\newcommand{\PhiEdge}[1]{\phi_{f_{#1}}}

\newcommand{\SubIndex}[2]{f_{#1},#2}

\newcommand{\xIsoEdge}[2]{\bm{x}_{\SubIndex{#1}{#2}}}
\newcommand{\NSF}[2]{N_{\SubIndex{#1}{#2}}}
\newcommand{\PointPhysical}{\bm{x}}
\newcommand{\PointReferenceCartesianLetter}{\xi}
\newcommand{\PointReferenceCartesian}{\bm{\PointReferenceCartesianLetter}}
\newcommand{\PointReferenceBarycentricLetter}{\lambda}
\newcommand{\PointReferenceBarycentric}{\bm{\PointReferenceBarycentricLetter}}
\newcommand{\PRBProj}[2]{\PointReferenceBarycentric_{f_{#1}}^{#2}}
\newcommand{\PRCProj}[2]{\PointReferenceCartesian_{f_{#1}}^{#2}}
\newcommand{\PPProj}[2]{\PointPhysical_{f_{#1}}^{#2}}

\journal{Computer-Aided Design}

\begin{document}

\begin{frontmatter}

\title{\uppercase{Interpolation of subdivision features for curved geometry modeling}}
\author{Albert Jiménez-Ramos}
\ead{albert.jimenez@bsc.es}
\author{Abel Gargallo-Peiró}
\ead{abel.gargallo@bsc.es}
\author{Xevi Roca\corref{cor1}}
\ead{xevi.roca@bsc.es}
\cortext[cor1]{Corresponding author}
\address{Barcelona Supercomputing Center, Carrer Jordi Girona, 29, E-08304, Barcelona, Spain.}

\begin{abstract}
We present a nodal interpolation method to approximate a subdivision model. The main application is to model and represent curved geometry without gaps and preserving the required simulation intent. Accordingly, we devise the technique to maintain the necessary sharp features and smooth the indicated ones. This sharp-to-smooth modeling capability handles unstructured configurations of the simulation points, curves, and surfaces. The surfaces correspond to initial linear triangulations that determine the sharp point and curve features. The method automatically suggests a subset of sharp features to smooth which the user modifies to obtain a limit model preserving the initial points. This model reconstructs the curvature by subdivision of the initial mesh, with no need of an underlying curved geometry model. Finally, given a polynomial degree and a nodal distribution, the method generates a piece-wise polynomial representation interpolating the limit model. We show numerical evidence that this approximation, naturally aligned to the subdivision features, converges to the model geometrically with the polynomial degree for nodal distributions with sub-optimal Lebesgue constant. We also apply the method to prescribe the curved boundary of a high-order volume mesh. We conclude that our sharp-to-smooth modeling capability leads to curved geometry representations with enhanced preservation of the simulation intent.
\end{abstract}

%


\end{frontmatter}


\section{Introduction}

The capability to model and represent curved geometry preserving the simulation intent is critical for flow simulation with unstructured high-order methods. These high-order simulations require curved meshes that approximate a curved boundary representation \cite{PerssonPeraire,chaurasia2012coarse,johnen2013geometrical,tesi-abel,Gargallo-Peiro2015,gargallo2015optimization,ruiz2016generation,ruiz2016high,MOXEY2016130}. Ideally, this boundary representation should be composed of smooth and sharp features agreeing with the simulation intent \cite{thakur2009survey,white2005meshing,shapiro2011geometric,nolan2015defining}.

Flow simulation practitioners favor continuous normal vectors on smooth features where the intent is to obtain attached flow. In contrast, they only need model continuity on sharp features where the flow detaches. To illustrate both types of features, we can consider an aircraft model. We can find there smooth features such as the nose tip, leading edges, and wing surfaces; and sharp features such as trailing edges and points.

To model the previous features, it is standard to use CAD boundary representations based on trimmed NURBS. Unfortunately, these trim-based models might violate the simulation intent. This breach is so since they might present unintended gaps or discontinuities of the normal vector on irregular points and curves adjacent to trimmed surfaces. Nevertheless, together with virtual geometry \cite{sheffer2000virtual,sheffer2001model,foucault2008adaptation,foucault2013generalizing}, CAD boundary representations have shown to be crucial for generating an unstructured triangular mesh approximating the curved model boundary.

If the element size is fine but coarser than the model tolerance, we obtain a fair second-order approximation of the CAD boundary representation without gaps. However, since the triangles are planar, this approximation does not feature the normal vector continuity through any triangular edges, and thus, it is not adequate for flow simulation. Nevertheless, we can convert the triangular mesh to a gap-free curved geometry model \cite{On-the-Use-of-Loop-Subdivision-Surfaces-for-Surrogate-Geometry} that features normal vector continuity by using a subdivision scheme \cite{smooth-subdivision-surfaces-based-on-triangles}.

This subdivision-based conversion to a curved model is useful even when there is no underlying CAD model. The conversion only needs a model composed of triangulations, which boundaries determine the model points and polylines. Hence, this conversion is of practical interest in any application where the triangular mesh comes from legacy data or real samples, such as in onshore wind farm energy forecasting  \cite{gargallo:meshForABLandWindFarms,gargallo2018mesh}, transport of pollutants in urban areas \cite{gargallo2016representing}, and bio-engineering.

In these applications, the subdivision conversion provides a curved limit model. We can query this limit model by successive refinements \cite{On-the-Use-of-Loop-Subdivision-Surfaces-for-Surrogate-Geometry, smooth-subdivision-surfaces-based-on-triangles, LaneRiesenfeld1980}. However, this approach requires more refinement levels the closer a given query point is to an irregular point.

To avoid this unbalanced query, in our previous work \cite{jimenez2018incorporating,jimenez_ramos_albert_2020_3653357}, we proposed a method to interpolate with a continuous piece-wise quadratic or quartic mesh the limit model while exploiting the structure of iterative subdivision. Any posterior query to the interpolation model only uses the corresponding triangular-wise polynomial components, thus skipping the successive refinement step.

Although skipping posterior successive refinement, our previous approach only extends to interpolation degrees equal to powers of two on equispaced nodal distributions. Therefore, it does not allow using arbitrary polynomial degrees and nodal distributions. Recall that beyond degree four equispaced nodal distributions feature large Lebesgue constants that might hamper the corresponding interpolation accuracy.

The main contribution of this work is to propose a method to address the previous accuracy issues while still skipping successive refinement on posterior geometry queries. Specifically, we propose a method to interpolate the subdivision model with any degree and nodal distribution. The method evaluates the limit model parameterization \cite{Evaluation-of-Loop-Subdivision-Surfaces} once on each interpolation point to obtain the resulting nodal curved mesh model. Furthermore, we show that this nodal mesh is ready to prescribe the boundary of a curved high-order volume mesh.

We also propose an approximation of the distance between the interpolation and the limit model to check the geometric accuracy. To compute this distance, we only need to perform forward evaluations of the nodal parameterization. The distance allows studying how the nodal interpolation converges to the limit model in terms of the polynomial degree for different nodal distributions.

We finally propose an assisted sharp-to-smooth modeling capability. The resulting method automatically suggests a subset of sharp features to smooth, which the user modifies to obtain a limit model preserving the initial points.  It aims to reduce the amount of human labor required to describe the simulation intent of the model. We illustrate this assisted modeling capability to assign sharp and smooth features to the union of triangulations describing an aircraft model in a high-lift configuration.

The organization of the rest of the paper is as follows. First, in \autoref{sec:relatedWork}, we review the literature related to this work. Then, in \autoref{sec:problemStatementMethodology}, we present the problem statement and the methodology. Next, in \autoref{sec:curve_surface_subdivision}, we present some preliminary results on subdivision, and in \autoref{sec:TheLimitModel} we detail the subdivision limit model. In \autoref{sec:InterpolationArbitrary}, we detail the curved piece-wise polynomial surface mesh interpolation of the limit model. Following, in \autoref{sec:volume_ho}, we present our subdivision-based mesh curving approach, detailing the sharp-to-smooth modeling process. Then, in \autoref{sec:results}, we present several results to illustrate the capabilities and main features of the presented methods, and in \autoref{sec:discussion}, we discuss some aspects of the method as well as future contributions. Lastly, in \autoref{sec:concludingRemarks}, we present some concluding remarks of this work.

\section{Related Work}
\label{sec:relatedWork}

It is well documented that to obtain a geometric model suitable for simulation, we need to simplify the original model \cite{thakur2009survey,white2005meshing}.  To obtain a simplified model, one standard approach is to define a Virtual Topology on top of the original CAD model \cite{sheffer2000virtual,sheffer2001model}. A similar CAD simplification approach, emphasizing automation and posterior meshing needs, considers a Mesh Constraint Topology \cite{foucault2008adaptation,foucault2013generalizing}. These simplification frameworks have been proven to be successful in many practical applications. However, during the simplification process, these frameworks might present some geometric issues. These issues can be fixed by maintaining a firm link between the original CAD and the simulation model \cite{shapiro2011geometric}.The key idea is to store the simplification steps in a high-level framework connecting the design and simulation geometry, and thus, facilitating CAD de-featuring \cite{shapiro2011geometric,nolan2015defining}. Alternatively, we can automatically de-feature a faceted representation of the original CAD model \cite{quadros2012defeaturing}.
	
In mesh generation, Loop's subdivision surfaces have been already used to define surrogate geometry with a surface-wise continuous normal vector \cite{On-the-Use-of-Loop-Subdivision-Surfaces-for-Surrogate-Geometry}. This subdivision scheme has also been used to provide nodal quadratic, and quartic surface and volume meshes \cite{jimenez2018incorporating,jimenez_ramos_albert_2020_3653357}. Alternatively, the butterfly subdivision scheme \cite{dyn1990butterfly} has been used to relocate the boundary nodes when the geometry is unavailable \cite{yang2019open}. However, the limit model does not feature continuous normal vectors on irregular points, and the volume is not curved using hierarchical blending. Finally, there are volume subdivision methods to generate curved volume meshes, featuring parallel implementations, but they need a curved volume mesh to define the surrogate geometry \cite{GARGALLOPEIRO2017310}. 

There are alternatives to subdivision schemes for describing the surrogate geometry. The work presented in \cite{Jiao2012} proposes two curving methods based on weighted least squares approximations and piece-wise polynomial fittings to generate curved meshes of the target surfaces. However, the method only enforces the continuity of the model. Herein, the limit model ensures continuity in the sharp features and normal vector continuity on the smooth features.

The curved surface meshes provided by the method in \cite{Jiao2012} can also be used to limit the mesh volumes and, thus, to generate curved high-order meshes when the underlying curved geometry is not available. Furthermore, it is possible to remove sharp features by selecting one-by-one the mesh entities defining it \cite{ims2015meshcurve}. Besides the surface mesh curving method, the method herein has other differences: a hierarchical blending approach to curve the mesh volume, an all-in-one feature selection to perform sharp-to-smooth modeling, and a robust and tunable assisted smooth feature detector.

\section{Problem Statement and Methodology}
\label{sec:problemStatementMethodology}

\subsection{Problem Statement}
\label{sec:problemStatement}

We consider the problem of converting a linear mesh model to a piece-wise polynomial curved mesh of degree $\Degree$ that preserves both the sharp and smooth features determined in the model. Our solution approximates a curved limit model based on a subdivision scheme of the linear model. The rest of this section describes the inputs and the output of the proposed method.

The input data is a linear tetrahedral mesh, a linear model $\Model^1$, and a list of features to be smoothed. A model of degree $\Degree$, $\Model^{\Degree}$, represents the geometry as the union of the feature points ${\FeatPoint}_{\IndexPoint}$, curves $\FeatCurve^{\Degree}_{\IndexCurve}$, and surfaces $\FeatSurface^{\Degree}_{\IndexSurface}$:
\[
\Model^{\Degree} = \bigcup_{\IndexPoint=1}^{n_{\IndexPoint}} {\FeatPoint}_{\IndexPoint} \cup \bigcup_{\IndexCurve=1}^{n_{\IndexCurve}} \FeatCurve^{\Degree}_{\IndexCurve} \cup \bigcup_{\IndexSurface=1}^{n_{\IndexSurface}} \FeatSurface^{\Degree}_{\IndexSurface} .
\]
A geometry feature is characterized by a set of entities of the boundary of the volume mesh with the same identifier. Then, a point feature describes a vertex to preserve, and it is characterized by the global identifier of the point to be preserved. A curve feature describes a smooth curve to preserve. A curve of $\Model^{\Degree}$ is described by the union of segments of polynomial degree $\Degree$ forming a polycurve. These segments correspond to boundary edges of the tetrahedral mesh with the same curve identifier. Finally, a surface feature describes a smooth surface to preserve. A surface of $\Model^{\Degree}$ is described by the union of facets of polynomial degree $\Degree$ forming a triangulation. These facets correspond to boundary triangles of the tetrahedral mesh with the same surface identifier.

Alternatively, we can obtain the model when only the surface features are described. That is, if only the boundary triangles are marked, we can retrieve the feature curves from the intersection of the boundary of two or more feature surfaces. Similarly, point features can be determined by the intersection of two or more feature curves.

In addition to the tetrahedral mesh, the linear model and the geometry features, we have an optional input determining the sharp features to smooth. Smoothing a geometry feature corresponds to removing it from the list of features to preserve and  merging adjacent regions to ensure $\mathcal{C}^1$-continuity of the model along the removed feature. That is, if we smooth a curve, we remove the curve feature and merge the two adjacent surfaces. While when we smooth a point, we remove the point feature and merge the curves incident to the vertex. Since each geometry feature is associated with a unique identifier, the list of features to be smoothed is a sub-sequence of these unique identifiers.

The output of our method is a piece-wise polynomial mesh of degree $\Degree$ with a boundary preserving the sharp features of the model and satisfying three properties. First, high-order element vertices interpolate the initial linear mesh nodes. Second, the nodes of the high-order edges that belong to a feature curve and are not adjacent to a feature point (\emph{inner curve edges}) interpolate a cubic $\mathcal{C}^2$-continuous curve. Third, the nodes of the high-order elements that belong to a feature surface and are not adjacent to a feature curve or point (\emph{inner surface elements}) interpolate a $\mathcal{C}^1$-continuous surface. These properties provide regularity guarantees in the output mesh that are discussed in \autoref{sec:SmoothnessHOSurfaceMesh}. The geometry features of the boundary of the generated tetrahedral mesh of polynomial degree $\Degree$ characterize the model of polynomial degree $\Degree$, $\Model^{\Degree}$.

\subsection{Method: Hierarchical Subdivision and Blending}

\begin{figure}[t]
\centering
\renewcommand{\figsize}{0.2}
\begin{tabular}{cccc}
\begin{subfigure}[t]{\figsize\textwidth}
\centering
\includegraphics[width=\textwidth]{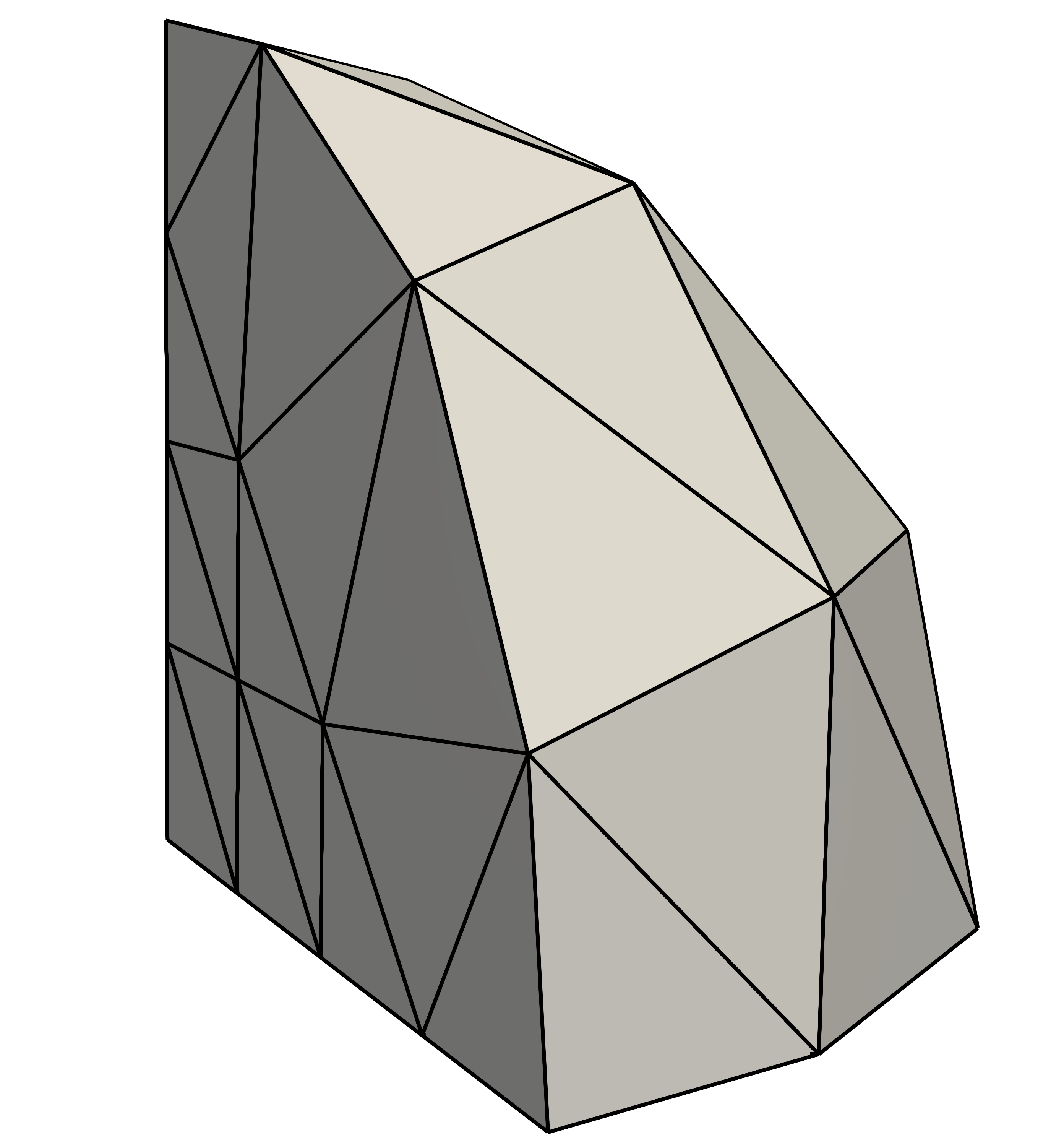}
\caption{}
\label{fig:sphere_initial}
\end{subfigure}
&
\begin{subfigure}[t]{\figsize\textwidth}
\centering
\includegraphics[width=\textwidth]{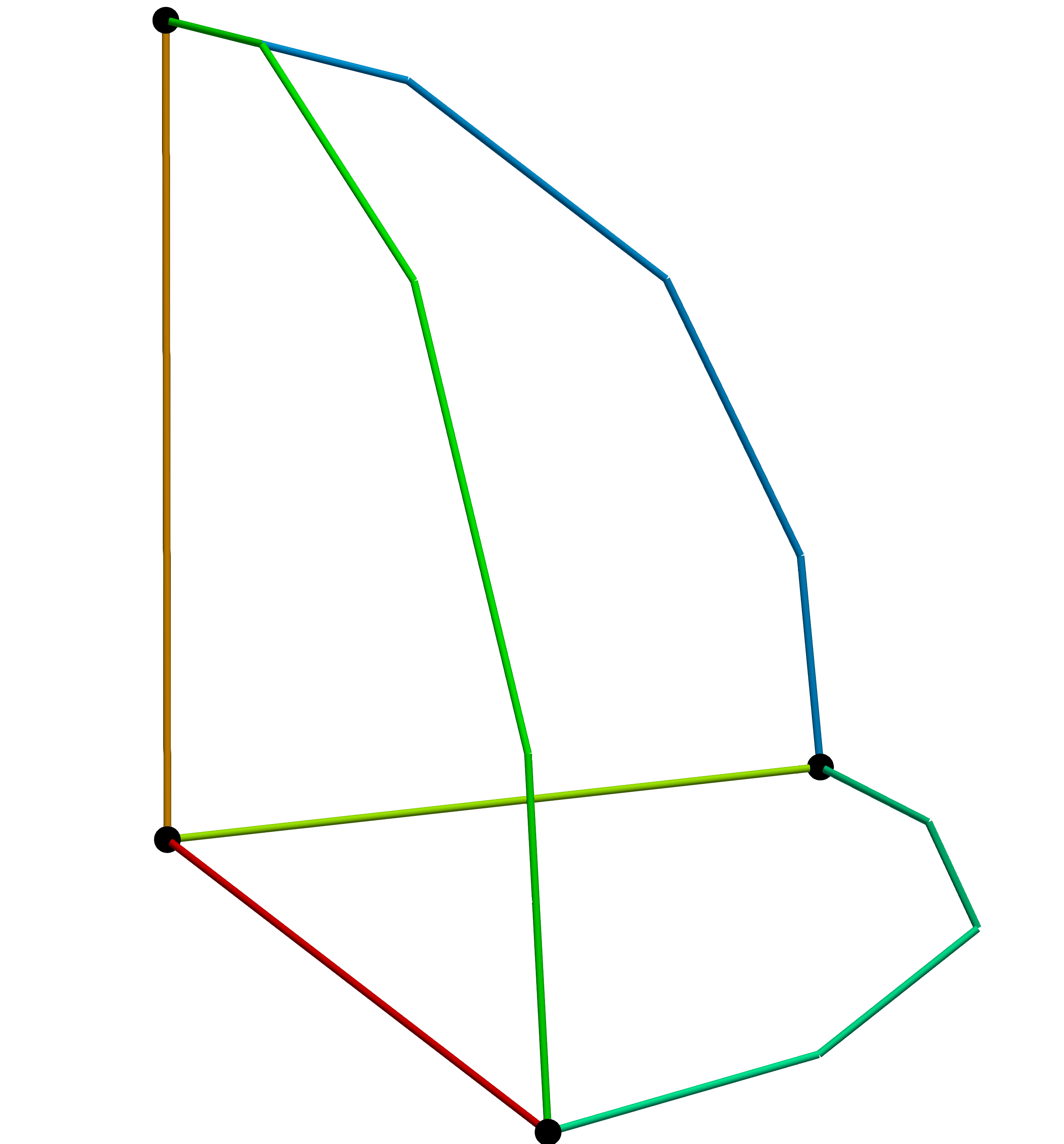}
\caption{}
\label{fig:sphere_curves}
\end{subfigure}
&
\begin{subfigure}[t]{\figsize\textwidth}
\centering
\includegraphics[width=\textwidth]{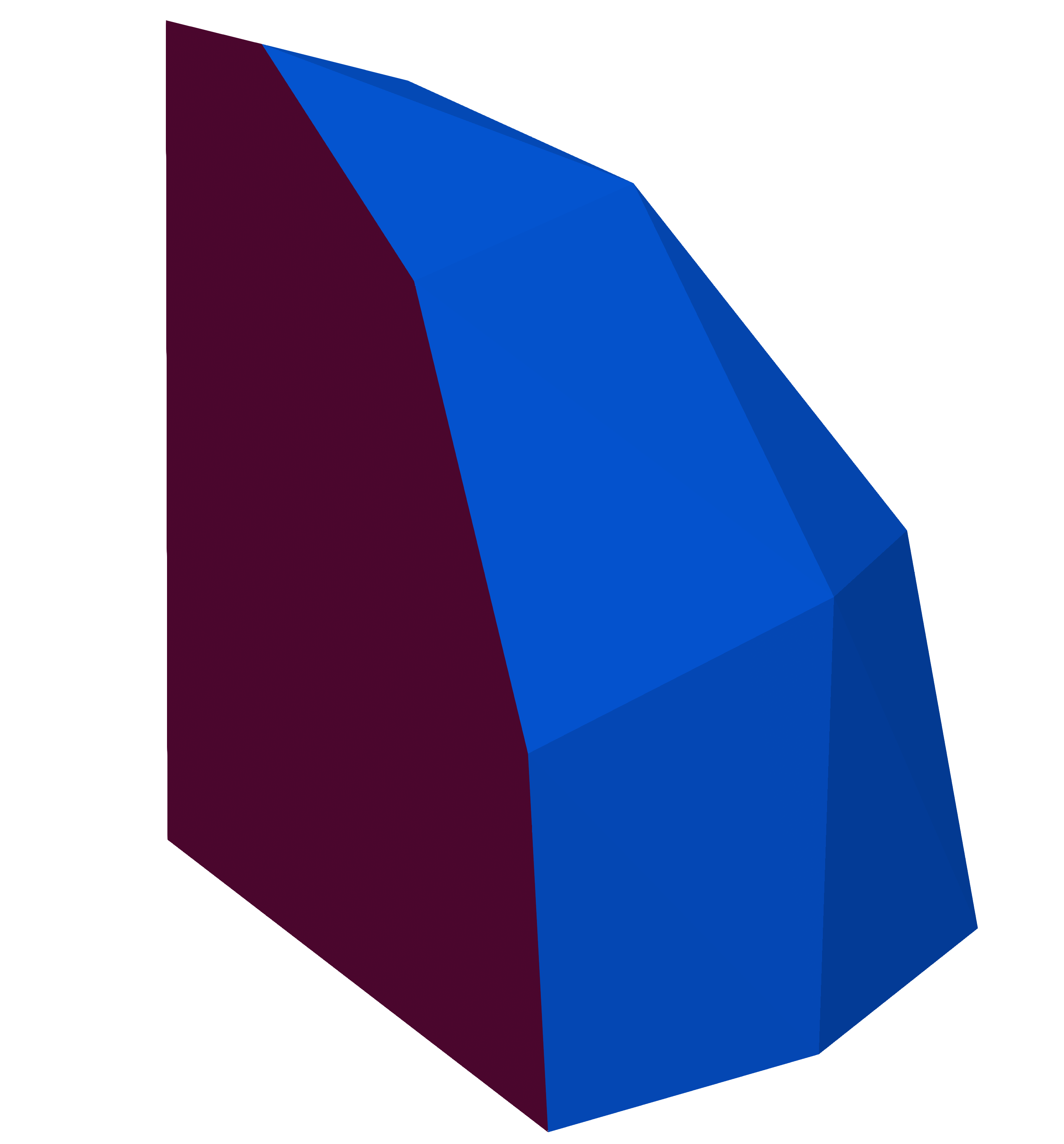}
\caption{}
\label{fig:sphere_surfaces}
\end{subfigure}
&
\begin{subfigure}[t]{\figsize\textwidth}
\centering
\includegraphics[width=\textwidth]{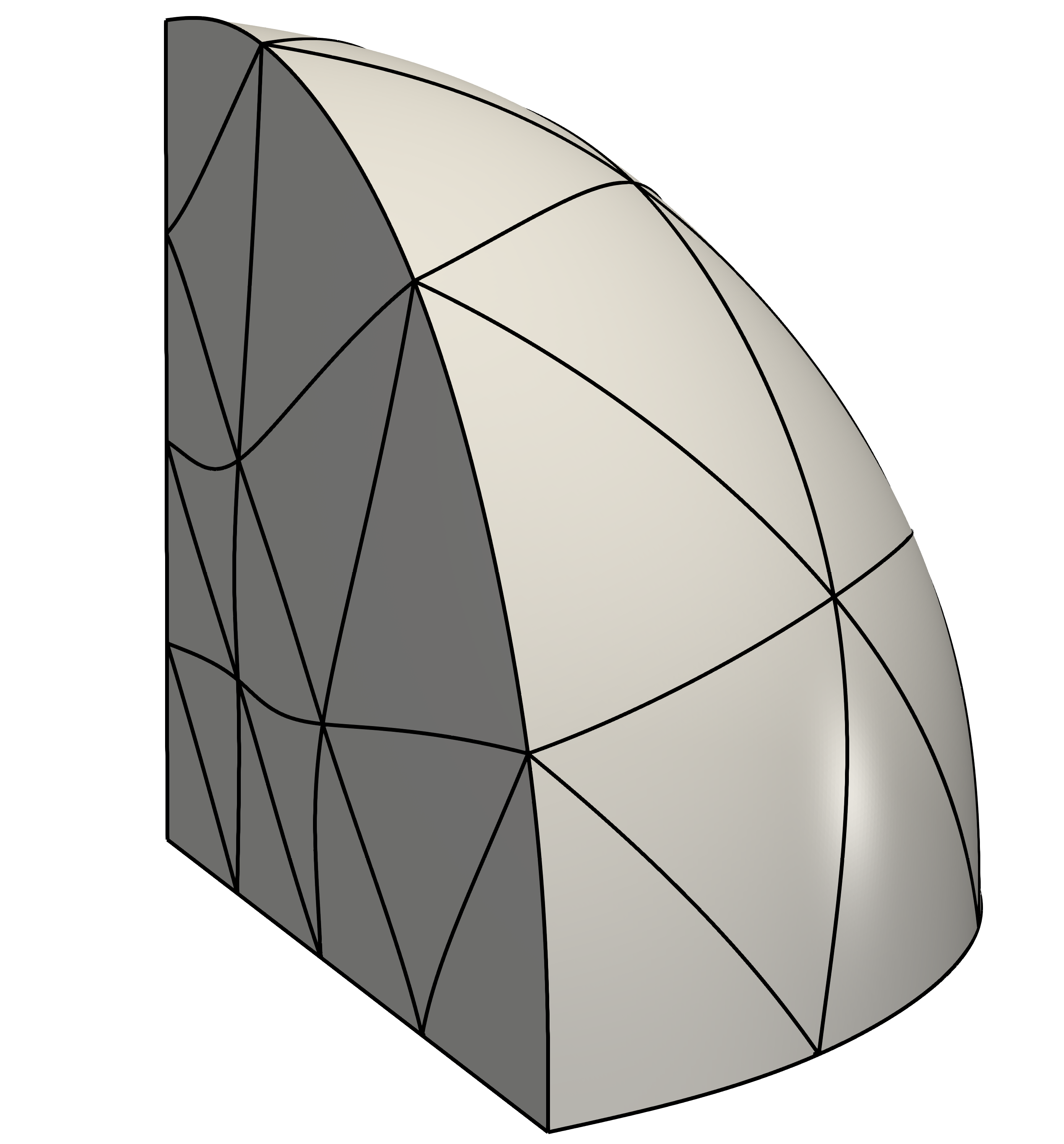}
\caption{}
\label{fig:sphere_p4}
\end{subfigure}
\end{tabular}
\caption{Method. Input: \subref{fig:sphere_initial} a linear tetrahedral mesh and the associated model with \subref{fig:sphere_curves} feature points, curves and \subref{fig:sphere_surfaces} surfaces. Output: \subref{fig:sphere_p4} a curved tetrahedral mesh of polynomial degree $\Degree$, $\Degree=4$ in this case.}
\label{fig:Methodology}
\end{figure}

The curved high-order mesh generation procedure proposed in this work is composed of five main steps:
\begin{enumerate}
\setcounter{enumi}{-1}
\item \textbf{Sharp-to-smooth modeling}. Some applications need the capability to perform sharp-to-smooth geometry modeling. Then, as a preprocess, we can smooth the non-desired geometry features accordingly to the list of features to remove provided as input to obtain a smoother surrogate geometry. We devise a technique to automatically suggest the features to smooth in \autoref{sec:AutomaticFeatureDetection}, and the smoothing process is detailed in \autoref{sec:curved_volume}.

\item \textbf{Approximate a surrogate boundary}. Given a linear tetrahedral mesh, \autoref{fig:sphere_initial}, we extract its boundary. The entities of the boundary mesh characterize the geometry features of the given linear model $\Model^1$, see \autoref{fig:sphere_curves} and \autoref{fig:sphere_surfaces}. The hierarchical subdivision of the geometry features determines a set of $\mathcal{C}^2$-continuous limit curves and $\mathcal{C}^1$-continuous limit surfaces that serves as surrogate geometry to generate a curved high-order triangular surface mesh, see \autoref{sec:TheLimitModel}. This curved surface mesh preserves the sharp features and smooth regions of the linear model $\Model^1$, and interpolates it at the nodes of the high-order mesh. See details in \autoref{sec:InterpolationArbitrary}.

\item \textbf{Substitute the boundary of the volume mesh}. We increase the polynomial degree of the volume mesh and replace the straight-sided boundary of the current high-order volume mesh by the high-order surface mesh obtained in the first step. See details in \autoref{sec:curved_volume}.

\item \textbf{Accommodate the curvature of the boundary}. We accommodate the curvature of the curved surface mesh to the boundary volume elements using an explicit hierarchical blending, see \autoref{fig:sphere_p4}. See details in \autoref{sec:blending}.

\item \textbf{Local smoothing and untangling}. If necessary, we optimize the low-quality elements locally following the approach detailed in \cite{Gargallo-Peiro2015,gargallo2015optimization}.
\end{enumerate}

\section{Preliminaries: Curve and Surface Mesh Subdivision}
\label{sec:curve_surface_subdivision}

In this section, we present the subdivision algorithms that determine the limit model. The boundary mesh is composed of points, curves, and surfaces. Then, the subdivision is performed hierarchically, that is, points remain fixed, curves are refined using a curve subdivision scheme and surfaces are subdivided using a surface subdivision scheme. 

In \autoref{sec:curve_subdivision}, we detail the curve subdivision algorithm we use in this work, and \autoref{sec:Loop_subdivision} recalls Loop's subdivision surface scheme. Although these methods do not preserve the position of the initial control points, a simple modification leads to a smooth limit manifold interpolating these initial points, see \ref{sec_app:interpolative}.

\subsection{Curve Subdivision Scheme}
\label{sec:curve_subdivision}

The curve subdivision scheme used in this work was originally presented in \cite{LaneRiesenfeld1980}. Successive applications of the algorithm generate finer polygons, all of them with the same limit curve determined by the initial mesh, referred to as control mesh. This curve is parametrized by a polynomial of degree three and it is $\mathcal{C}^2$-continuous. 

We remark that there exists an element-wise parameterization of the limit curve that allows mapping any polyline point onto the limit curve. Specifically, consider $\FeatCurve^1$ the $\IndexCurve$-th curve of the model $\Model^1$. Let $\MasterOneD$ be the master interval, $\Seg_{\Edge}$ the $\Edge$-th segment of $\FeatCurve^1$, $\PointReferenceBarycentric$ the position in barycentric coordinates of a point in $\MasterOneD$, and $\lbrace \PointPhysical_i^{\Edge} \rbrace_{i=1, \dotsc, 4}$ the set composed of the nodes of the segment $\Seg_{\Edge}$ and its neighbors elements. Then, if  $\Seg_{\Edge}$ is not adjacent to a feature vertex, we define the element-wise parameterization of the limit curve $\LimitCurve$ as
\begin{align}
\begin{split}
 \PhiCurveElem \colon \MasterOneD &\to \Seg^{\infty}_{\Edge} \subset \LimitCurve\\
 \PointReferenceBarycentric &\mapsto \PhiCurveElem \left( \PointReferenceBarycentric \right) \equiv \PhiCurveElem \left( \PointReferenceBarycentric ; \lbrace \PointPhysical_i^{\Edge} \rbrace_{i=1, \dotsc, 4} \right),
\end{split}
\label{eq:ParamCurves}
\end{align}
where $\Seg^{\infty}_{\Edge}$ is the section of the limit curve corresponding to $\Seg_{\Edge}$. The explicit expression of $ \PhiCurveElem$ can be found in \cite{On-the-Use-of-Loop-Subdivision-Surfaces-for-Surrogate-Geometry,de1978practical}. We remark that the parameterization features third-degree polynomial components in barycentric coordinates. The case in which segment $\Seg_{\Edge}$ is adjacent to a feature vertex is detailed in \autoref{sec:ParameterizationLimitManifold}.

\subsection{Surface Subdivision Scheme}
\label{sec:Loop_subdivision}

Loop's subdivision algorithm \cite{smooth-subdivision-surfaces-based-on-triangles} is used to subdivide the feature surfaces of the model. Successive applications of the algorithm generate finer triangular meshes, all of them with the same limit surface determined by the initial control mesh. This surface is $\mathcal{C}^1$-continuous, attaining $\mathcal{C}^2$-continuity around regular vertices. 

We start by introducing several definitions related to neighbor elements and nodes. The \emph{neighbor elements} of a node $\Vertex$ are the elements incident to $\Vertex$, and the \emph{neighbor nodes} of $\Vertex$ are the vertices of these elements. For the case of simplices, there exists an equivalent definition based on edges. The \emph{neighbor edges} of a node $\Vertex$ are the edges incident to $\Vertex$, and the \emph{neighbor nodes} of $\Vertex$ are the vertices of these edges. We say that a surface node is \emph{regular} if it has six neighbor nodes. Otherwise, we say the node is \emph{irregular} or \emph{extraordinary}. Around regular nodes, the limit surface is parametrized by a polynomial of degree four and is $\mathcal{C}^2$-continuous, while on irregular nodes it is of class $\mathcal{C}^1$ \cite{zorin2000method}. 

Similarly to the curve case, there also exists an element-wise parameterization of the limit surface for elements with at most one extraordinary vertex \cite{Evaluation-of-Loop-Subdivision-Surfaces} and not adjacent to a sharp feature. Specifically, consider $\FeatSurface^1$ the $\IndexSurface$-th surface of the model $\Model^1$.  Let $\MasterTwoD$ be the master triangle, $\Tri_{\Facet}$ the $\Facet$-th triangle of $\FeatSurface^1$, $\PointReferenceBarycentric$ the position in barycentric coordinates of a point in $\MasterTwoD$, and $\lbrace \PointPhysical_i^{\Facet} \rbrace_{i=1, \dotsc, N}$ the $N$-point set composed of the nodes of the triangle $\Tri_{\Facet}$ and its neighbors elements. Assume $\Tri_{\Facet}$ has at most one irregular vertex and is not adjacent to a sharp feature. Then, we define the element-wise parameterization of the limit surface $\LimitSurface$ as
\begin{align}
\begin{split}
 \PhiSurfaceElem \colon \MasterTwoD &\to \Tri^{\infty}_{\Facet} \subset \LimitSurface\\
 \PointReferenceBarycentric &\mapsto \PhiSurfaceElem \left( \PointReferenceBarycentric \right) \equiv \PhiSurfaceElem \left( \PointReferenceBarycentric ; \lbrace \PointPhysical_i^{\Facet} \rbrace_{i=1, \dotsc, N} \right).
\end{split}
\label{eq:ParamSurfaces}
\end{align}
The details of this parameterization can be found in \cite{Evaluation-of-Loop-Subdivision-Surfaces}. We remark that the parameterization features fourth-degree polynomial components in barycentric coordinates.

In order to compute the parameterization of an element featuring more than one extraordinary vertex, one subdivision step is performed. Since the new edge points obtained from Loop's algorithm are regular, one subdivision of the mesh ensures that every element contains at most one irregular vertex and, therefore, the parameterization can be evaluated for all the elements of the subdivided mesh. Moreover, this parameterization can be applied to elements not adjacent to a sharp feature. The case in which triangle $\Tri_{\Facet}$ is adjacent to a sharp feature is detailed in \autoref{sec:ParameterizationLimitManifold}.

\section{The Limit Model}
\label{sec:TheLimitModel}

The model describes the geometry to represent and is defined by the union of feature points, curves and surfaces. For each of the feature curves (surface) in the model, the curve (surface) subdivision scheme determines a limit curve (surface), see \autoref{sec:curve_surface_subdivision}. The union of feature points, limit curves and limit surfaces determines the limit model $\ModelLimit$,
\[
\ModelLimit = \bigcup_{\IndexPoint=1}^{n_{\IndexPoint}} \FeatPoint_{\IndexPoint} \cup \bigcup_{\IndexCurve=1}^{n_{\IndexCurve}} \LimitCurve_{\IndexCurve} \cup \bigcup_{\IndexSurface=1}^{n_{\IndexSurface}} \LimitSurface_{\IndexSurface} .
\]
This limit model preserves the sharp features of the linear model, inherits the smoothness properties of the limit curves and surfaces, and serves as surrogate geometry for the mesh curving problem. In this section, we detail the parameterization of the limit model, \autoref{sec:ParameterizationLimitManifold}, and study its smoothness, \autoref{sec:SmoothnessSurrogate}.

\subsection{Parameterization of the Limit Model}
\label{sec:ParameterizationLimitManifold}

The limit model can be queried using the element-wise parameterizations of the limit curves and limit surfaces. As noted in \autoref{sec:curve_surface_subdivision}, the evaluation of these parameterizations can only be computed in certain configurations of the mesh. Specifically, they can only be computed for inner elements and, in the case of surfaces, for triangles with at most one extraordinary vertex. However, in our models, there are segments (triangles) adjacent to feature points (feature points or curves), and triangles with more than one irregular vertex. Therefore, in order to compute the parameterization of an element in one of these two configurations the following procedures are applied. First, we successively subdivide a segment (triangle) adjacent to a feature point (feature point or curve) until the evaluation point belongs to an inner segment (triangle) of the subdivided polyline (triangular mesh). Second, if an inner triangle has more than one extraordinary vertex, we subdivide once the stencil needed for the evaluation of the parameterization of the limit surface. Since new edge points generated with Loop's scheme are regular, all the triangles of the subdivided patch feature at most one irregular vertex. In conclusion, the parameterization of the limit model can be computed for any element of the mesh.

\newcommand{\NameFunctionGetSegment}{GetContainerSegment}
\newcommand{\NameFunctionGetTriangle}{GetContainerTriangle}
\begin{algorithm}[t]
	\begin{algorithmic}[1]
		\renewcommand{\algorithmicrequire}{\textbf{Input:}}
		\renewcommand{\algorithmicensure}{\textbf{Output:}}
		\Require{Linear surface mesh $\Mesh$, Linear model $\Model^1$, Triangle $\Tri^1_{\Facet}$, Point $\PointReferenceBarycentric \in \MasterTwoD$}
		\Ensure{Point $\PointPhysical^{\infty}$}
		\Function{\MapOntoLimit}{}
		\State $\PointPhysical$ $\leftarrow$ \Call{ComputePhysicalCoordinate}{$\Tri^1_{\Facet}$, $\PointReferenceBarycentric$} \label{alg:ParameterizationLimitManifold_line:ComputePhysicalCoordinate}
		\Switch{Type of feature containing $\PointPhysical$}
		\Case{$\PointPhysical$ is feature point}
		\State $\PointPhysical^{\infty} = \PointPhysical$ \label{alg:ParameterizationLimitManifold_line:FeatureVertex}
		\EndCase
		\Case{$\PointPhysical$ belongs to the interior of $\IndexCurve$-th curve $\FeatCurve^1$ of $\Model^1$}
		\State $\PointPhysical^{\infty}$ $\leftarrow$ \Call{\MapOntoLimitCurve}{$\Mesh$, $\Model^1$, $\FeatCurve^1$, $\PointReferenceBarycentric $,  $\PointPhysical$}
		\EndCase
		\Case{$\PointPhysical$ belongs to the interior of $\IndexSurface$-th surface $\FeatSurface^1$ of $\Model^1$}
		\State $\PointPhysical^{\infty}$ $\leftarrow$ \Call{\MapOntoLimitSurface}{$\Mesh$, $\Model^1$, $\FeatSurface^1$, $\Tri^1_{\Facet}$, $\PointReferenceBarycentric $,  $\PointPhysical$}
		\EndCase
		\EndSwitch
		\State \Return $\PointPhysical^{\infty}$
		\EndFunction
	\end{algorithmic}
	\caption{Parameterization of the limit model.}
	\label{alg:ParameterizationLimitManifold}
\end{algorithm}
\begin{algorithm}[t]
	\begin{algorithmic}[1]
		\renewcommand{\algorithmicrequire}{\textbf{Input:}}
		\renewcommand{\algorithmicensure}{\textbf{Output:}}
		\Require{Linear surface mesh $\Mesh$, Linear model $\Model^1$, Curve $\FeatCurve^1$, Point $\PointReferenceBarycentric \in \MasterTwoD$, Point $\PointPhysical\in\FeatCurve^1$}
		\Ensure{Point $\PointPhysical^{\infty}$}
		\Function{\MapOntoLimitCurve}{}
		\State $\Seg_{\Edge}$ $\leftarrow$ \Call{\NameFunctionGetSegment}{$\FeatCurve^1$, $\PointPhysical$} \label{alg:ParameterizationLimitManifold_line:GetEdge}
		\State $\FeatCurve_{\Edge}$, $\Model_{\Edge}$ $\leftarrow$ \Call{SubCurve}{$\Mesh$, $\Model^1$, $\Seg_{\Edge}$} \label{alg:ParameterizationLimitManifold_line:SetSubMeshCurve}
		\While{$\Seg_{\Edge}$ is adjacent to feature point}
		\State $\FeatCurve^{\prime}$, $\Model^{\prime}$ $\leftarrow$ \Call{CurveSubdivision}{$\FeatCurve_{\Edge}$, $\Model_{\Edge}$} \label{alg:ParameterizationLimitManifold_line:CurveSubdivision}
		\State $\Seg_{\Edge}$ $\leftarrow$ \Call{\NameFunctionGetSegment}{$\FeatCurve^{\prime}$, $\PointPhysical$} \label{alg:ParameterizationLimitManifold_line:GetElementBelongingCurve}
		\State $\FeatCurve_{\Edge}$, $\Model_{\Edge}$ $\leftarrow$ $\FeatCurve^{\prime}$, $\Model^{\prime}$
		\EndWhile
		\State $\PointReferenceBarycentric_{\Edge}$ $\leftarrow$ \Call{ComputeBarycentricCoordinates}{$\Seg_{\Edge}$, $\PointPhysical$} \label{alg:ParameterizationLimitManifold_line:RecomputeBarCurve}
		\State $\lbrace \PointPhysical_i^{\Edge} \rbrace_{i=1, \dotsc, 4}$ $\leftarrow$ \Call{GetCurveStencil}{$\FeatCurve_{\Edge}$, $\Seg_{\Edge}$} \label{alg:ParameterizationLimitManifold_line:GetCurveStencil}
		\State $\PointPhysical^{\infty}$ $\leftarrow$ $\PhiCurveElem \left( \PointReferenceBarycentric_{\Edge} ; \lbrace \PointPhysical_i^{\Edge} \rbrace_{i=1, \dotsc, 4} \right)$ \label{alg:ParameterizationLimitManifold_line:SendToLimitCurve}
		\State \Return $\PointPhysical^{\infty}$
		\EndFunction
	\end{algorithmic}
	\caption{Parameterization of a limit curve of the limit model.}
	\label{alg:ParameterizationLimitManifold_curve}
\end{algorithm}

\begin{figure}[t]
	\centering
	\renewcommand{\figsize}{0.45}
	\begin{tabular}{cc}
		\begin{subfigure}[t]{\figsize\textwidth}
			\centering
			\includegraphics[width=\textwidth]{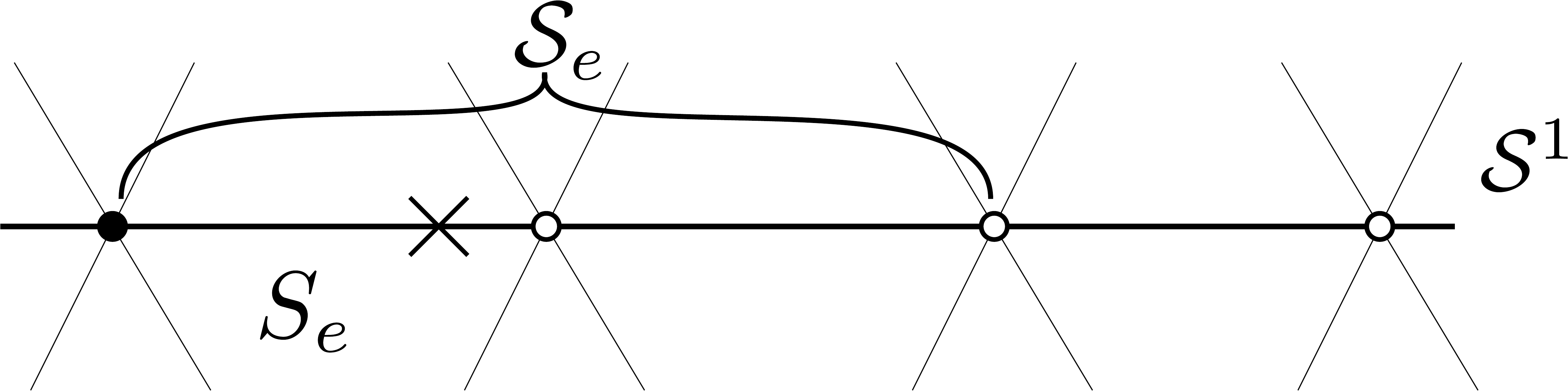}
			\caption{}
			\label{fig:PolylineAdjacentVertex}
		\end{subfigure}
		&
		\begin{subfigure}[t]{\figsize\textwidth}
			\centering
			\includegraphics[width=\textwidth]{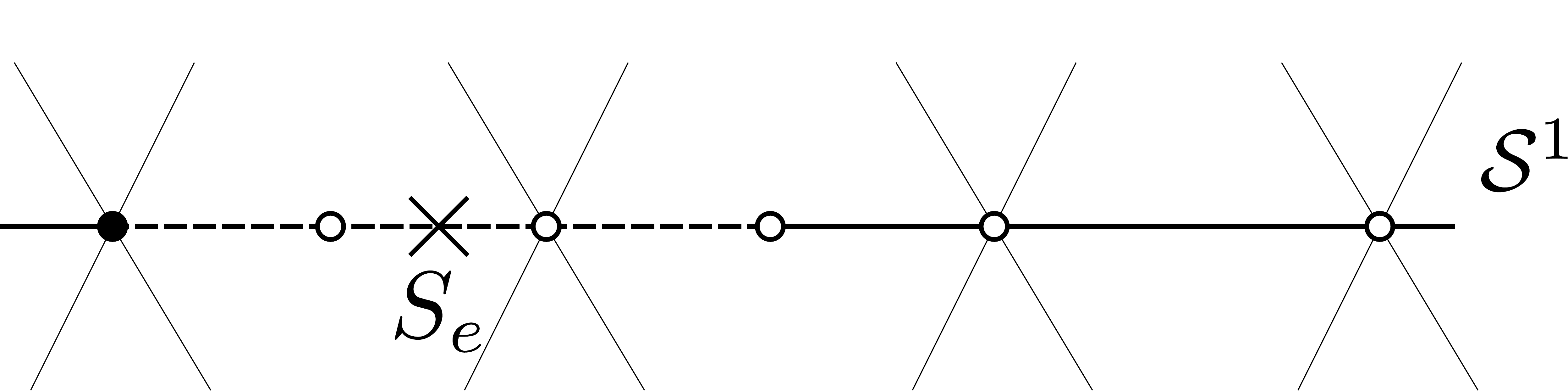}
			\caption{}
			\label{fig:PolylineAdjacentVertexSubdivided}
		\end{subfigure}
	\end{tabular}
	\caption{Subdivision to compute the element-wise parameterization of the limit curve on a point (cross) adjacent to a feature vertex (bold). Container edge in the:	\subref{fig:PolylineAdjacentVertex} original, and	\subref{fig:PolylineAdjacentVertexSubdivided} subdivided meshes. Dashed segments represent the curve stencil.}
	\label{fig:IrregularConfigurations_curve}
\end{figure}

\begin{algorithm}[t]
	\begin{algorithmic}[1]
		\renewcommand{\algorithmicrequire}{\textbf{Input:}}
		\renewcommand{\algorithmicensure}{\textbf{Output:}}
		\Require{Linear surface mesh $\Mesh$, Linear model $\Model^1$, Surface $\FeatSurface^1$, Triangle $\Tri^1_{\Facet}$, Point $\PointReferenceBarycentric \in \MasterTwoD$, Point $\PointPhysical\in\FeatSurface^1$}
		\Ensure{Point $\PointPhysical^{\infty}$}
		\Function{\MapOntoLimitSurface}{}
		\State $\Tri_{\Facet}$ $\leftarrow$ $\Tri^1_{\Facet}$
		\State $\FeatSurface_{\Facet}$, $\Model_{\Facet}$ $\leftarrow$ \Call{SubSurface}{$\Mesh$, $\Model^1$, $\Tri_{\Facet}$} \label{alg:ParameterizationLimitManifold_line:SetSubMeshSurface}
		\If{$\Tri_{\Facet}$ has more than one irregular vertex}
		\State $\FeatSurface^{\prime}$, $\Model^{\prime}$ $\leftarrow$ \Call{\NameSurfaceSubdivision}{$\FeatSurface_{\Facet}$, $\Model_{\Facet}$} \label{alg:ParameterizationLimitManifold_line:SurfaceSubdivisionIrregular}
		\State $\Tri_{\Facet}$ $\leftarrow$ \Call{\NameFunctionGetTriangle}{$\FeatSurface^{\prime}$, $\PointPhysical$} \label{alg:ParameterizationLimitManifold_line:ElementBelongingIrregular}
		\State $\FeatSurface_{\Facet}$,  $\Model_{\Facet}$ $\leftarrow$ $\FeatSurface^{\prime}$, $\Model^{\prime}$
		\EndIf
		\While{$\Tri_{\Facet}$ is adjacent to feature point or curve}
		\State $\FeatSurface^{\prime}$, $\Model^{\prime}$ $\leftarrow$ \Call{\NameSurfaceSubdivision}{$\FeatSurface_{\Facet}$, $\Model_{\Facet}$} \label{alg:ParameterizationLimitManifold_line:SurfaceSubdivisionAdjacent}
		\State $\Tri_{\Facet}$ $\leftarrow$ \Call{\NameFunctionGetTriangle}{$\FeatSurface^{\prime}$, $\PointPhysical$} \label{alg:ParameterizationLimitManifold_line:ElementBelongingAdjacent}
		\State $\FeatSurface_{\Facet}$,  $\Model_{\Facet}$ $\leftarrow$ $\FeatSurface^{\prime}$, $\Model^{\prime}$
		\EndWhile
		\State $\PointReferenceBarycentric_{\Facet}$ $\leftarrow$ \Call{ComputeBarycentricCoordinates}{$\Tri_{\Facet}$, $\PointPhysical$} \label{alg:ParameterizationLimitManifold_line:RecomputeBarSurface}
		\State $\lbrace \PointPhysical_i^{\Facet} \rbrace_{i=1, \dotsc, N}$ $\leftarrow$ \Call{GetSurfaceStencil}{$\FeatSurface_{\Facet}$, $\Tri_{\Facet}$} \label{alg:ParameterizationLimitManifold_line:GetSurfaceStencil}
		\State $\PointPhysical^{\infty}$ $\leftarrow$ $\PhiSurfaceElem \left( \PointReferenceBarycentric_{\Facet} ; \lbrace \PointPhysical_i^{\Facet} \rbrace_{i=1, \dotsc, N} \right)$ \label{alg:ParameterizationLimitManifold_line:SendToLimitSurface}
		\State \Return $\PointPhysical^{\infty}$
		\EndFunction
	\end{algorithmic}
	\caption{Parameterization of a limit surface of the limit model.}
	\label{alg:ParameterizationLimitManifold_surface}
\end{algorithm}

\begin{figure}[t]
	\centering
	\renewcommand{\figsize}{0.45}
	\begin{tabular}{cc}
		\begin{subfigure}[t]{\figsize\textwidth}
			\centering
			\includegraphics[width=\textwidth]{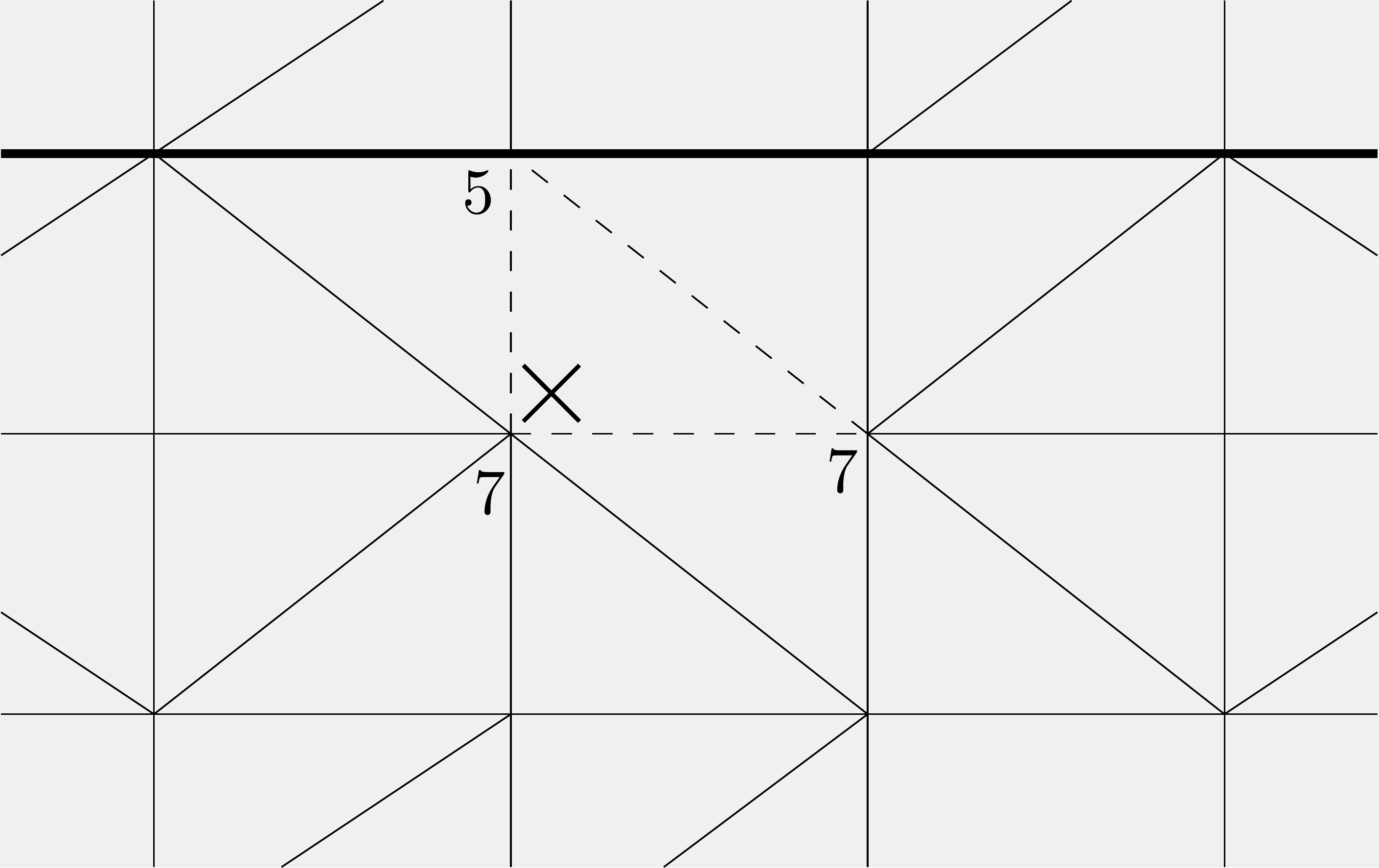}
			\caption{}
			\label{fig:TriangleAdjacentCurve}
		\end{subfigure}
		&
		\begin{subfigure}[t]{\figsize\textwidth}
			\centering
			\includegraphics[width=\textwidth]{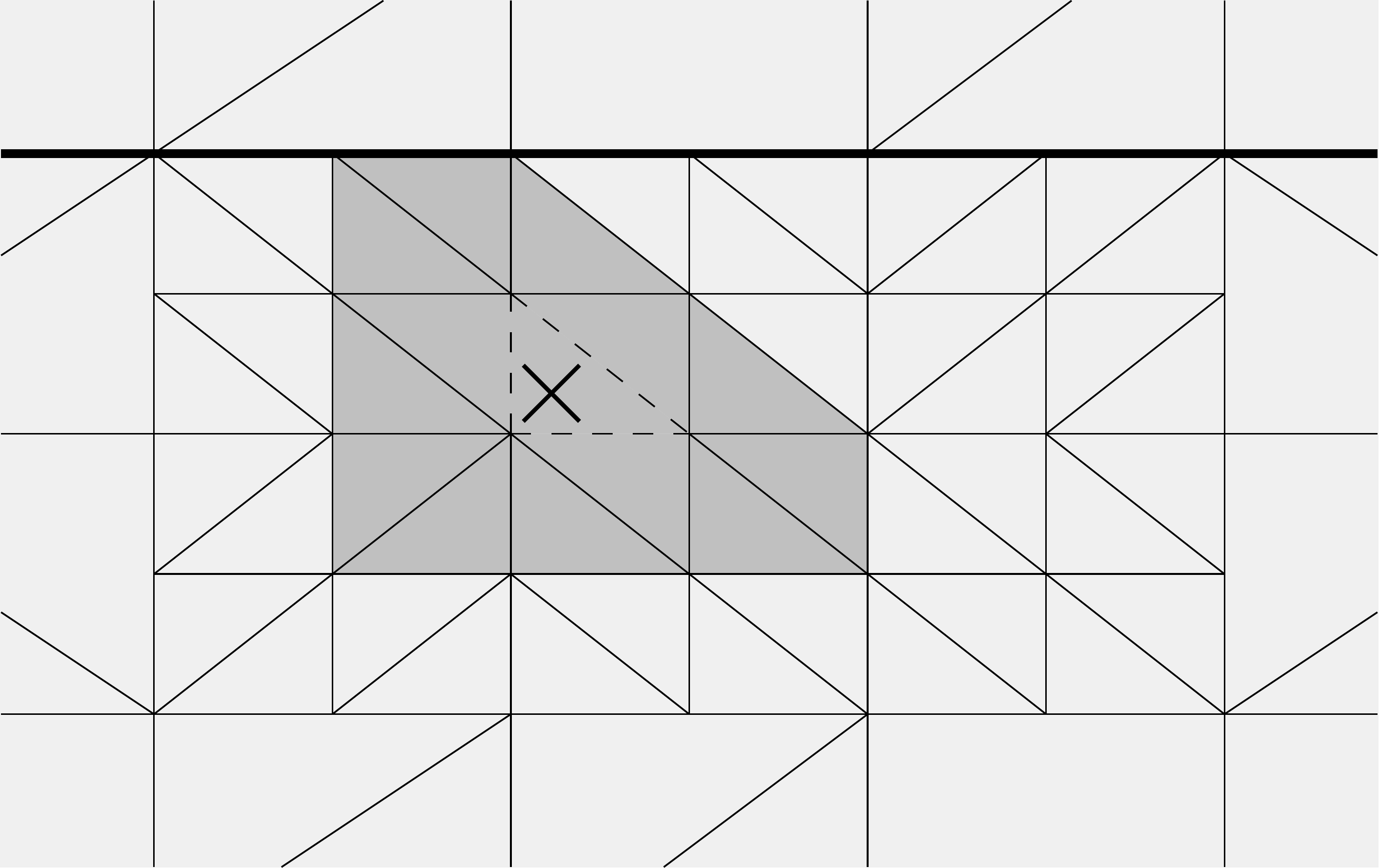}
			\caption{}
			\label{fig:TriangleAdjacentCurveSubdivided}
		\end{subfigure}
	\end{tabular}
	\caption{Subdivision to compute the element-wise parameterization of the limit surface on a point (cross) adjacent to a feature curve (bold). Container triangle (dashed edges) in the: \subref{fig:TriangleAdjacentCurve} original, and \subref{fig:TriangleAdjacentCurveSubdivided} subdivided meshes. Dark gray region represents the surface stencil.}
	\label{fig:IrregularConfigurations_surface}
\end{figure}

Following, we detail the evaluation of the limit model for all the possible configurations. Consider a linear surface mesh $\Mesh$, a triangle $\Tri^1_{\Facet}$ belonging to a surface $\FeatSurface^{1}$ of the linear model $\Model^1$, and denote by $\MasterTwoD$ the master triangle. Then, we define the element-wise parameterization of the limit model $\ModelLimit$ as
\begin{align}
\begin{split}
 \PhiLim_{\Facet} \colon \MasterTwoD &\to  \LimitSurface \subset \ModelLimit\\
 \PointReferenceBarycentric &\mapsto \PhiLim_{\Facet} \left( \PointReferenceBarycentric \right) \equiv \PhiLim \left( \Mesh, \Tri^{1}_{\Facet}, \PointReferenceBarycentric \right).
\end{split}
\label{eq:ParamLimitManifold}
\end{align}

The computation of function $\PhiLim$ is detailed in \autoref{alg:ParameterizationLimitManifold}. Given a linear surface mesh $\Mesh$, a linear model $\Model^1$, a triangle $\Tri^{1}_{\Facet}$ of $\Mesh$, and a point in the master triangle with barycentric coordinates $\PointReferenceBarycentric$, the function \texttt{\MapOntoLimit} maps $\PointReferenceBarycentric$ onto the limit model using the parameterization on triangle $\Tri^{1}_{\Facet}$. Firstly, in Line \ref{alg:ParameterizationLimitManifold_line:ComputePhysicalCoordinate}, we map $\PointReferenceBarycentric$ onto the linear triangle $\Tri^{1}_{\Facet}$, and obtain the point $\PointPhysical \in \Tri^{1}_{\Facet}$. Now, we distinguish three cases in terms of the type of feature the point $\PointPhysical$ belongs to. If $\PointPhysical$ belongs to a feature:
\begin{itemize}
\item point, the limit position does not change, $\PointPhysical^{\infty} = \PointPhysical$, Line \ref{alg:ParameterizationLimitManifold_line:FeatureVertex}. 

\item curve, say the $\IndexCurve$-th curve $\FeatCurve^1$, the parameterization of its limit curve  \texttt{\MapOntoLimitCurve} is computed, see \autoref{alg:ParameterizationLimitManifold_curve}. First, we get the segment $\Seg_{\Edge}$ of the curve $\FeatCurve^1$ that contains the point $\PointPhysical$, Line \ref{alg:ParameterizationLimitManifold_line:GetEdge}. Then, we define as $\FeatCurve_{\Edge}$ the polygonal mesh composed of the segment $\Seg_{\Edge}$ and the neighbor segments of the curve, and as $\Model_{\Edge}$ the associated linear model, Line \ref{alg:ParameterizationLimitManifold_line:SetSubMeshCurve}, see \autoref{fig:PolylineAdjacentVertex}. Then, while $\Seg_{\Edge}$ is adjacent to a feature point, in Line \ref{alg:ParameterizationLimitManifold_line:CurveSubdivision}, we subdivide $\FeatCurve_{\Edge}$ using the curve subdivision scheme, and compute the segment of the subdivided polygonal mesh to which $\PointPhysical$ belongs, Line \ref{alg:ParameterizationLimitManifold_line:GetElementBelongingCurve}. Note that if the original segment $\Seg_{\Edge}$ is not adjacent to a feature point, no subdivisions are performed. Now, since element $\Seg_{\Edge}$ is not adjacent to a feature point, see \autoref{fig:PolylineAdjacentVertexSubdivided}, in Line \ref{alg:ParameterizationLimitManifold_line:RecomputeBarCurve}, we compute the barycentric coordinates of the point $\PointPhysical$ with respect to $\Seg_{\Edge}$, $\PointReferenceBarycentric_{\Edge}$, and the stencil $\lbrace \PointPhysical_i^{\Edge} \rbrace_{i=1, \dotsc, 4}$ needed for the evaluation of the limit curve, Line \ref{alg:ParameterizationLimitManifold_line:GetCurveStencil}. Finally, in Line \ref{alg:ParameterizationLimitManifold_line:SendToLimitCurve}, we map $\PointReferenceBarycentric_{\Edge}$ onto its limit position, $\PointPhysical^{\infty} = \PhiCurveElem \left( \PointReferenceBarycentric_{\Edge} ; \lbrace \PointPhysical_i^{\Edge} \rbrace_{i=1, \dotsc, 4} \right)$.

\item surface, say the $\IndexSurface$-th surface $\FeatSurface^1$, the parameterization of its limit surface  \texttt{\MapOntoLimitSurface} is computed, see \autoref{alg:ParameterizationLimitManifold_surface}. First, we set $\Tri_{\Facet} \leftarrow \Tri^1_{\Facet}$. Then, we define as $\FeatSurface_{\Facet}$ the triangular mesh composed of the triangle $\Tri_{\Facet}$ and its neighbor triangles of the surface, and as $\Model_{\Facet}$ the associated linear model, Line \ref{alg:ParameterizationLimitManifold_line:SetSubMeshSurface}, see \autoref{fig:TriangleAdjacentCurve}. Next, we check if the triangle has more than one irregular vertex. If so, in Line \ref{alg:ParameterizationLimitManifold_line:SurfaceSubdivisionIrregular}, we subdivide the triangular mesh $\FeatSurface_{\Facet}$ to isolate the irregularity, and, in Line \ref{alg:ParameterizationLimitManifold_line:ElementBelongingIrregular}, compute the triangle $\Tri_{\Facet}$ of the subdivided mesh containing the point $\PointPhysical$, see \autoref{fig:TriangleAdjacentCurveSubdivided}. Then, analogously to the curve case, while element $\Tri_{\Facet}$ is adjacent to a sharp feature, we subdivide $\FeatSurface_{\Facet}$, Line \ref{alg:ParameterizationLimitManifold_line:SurfaceSubdivisionAdjacent}, and get the triangle of the subdivided mesh which $\PointPhysical$ belongs to, Line \ref{alg:ParameterizationLimitManifold_line:ElementBelongingAdjacent}. Now, since the element $\Tri_{\Facet}$ is not adjacent to a sharp feature, see \autoref{fig:TriangleAdjacentCurveSubdivided}, in Line \ref{alg:ParameterizationLimitManifold_line:RecomputeBarSurface}, we compute the parameter $\PointReferenceBarycentric_{\Facet}$ and the stencil $\lbrace \PointPhysical_i^{\Facet} \rbrace_{i=1, \dotsc, N}$ needed for the evaluation of the limit surface, Line \ref{alg:ParameterizationLimitManifold_line:GetSurfaceStencil}. Finally, we map the point onto its limit position, $\PointPhysical^{\infty} = \PhiSurfaceElem \left( \PointReferenceBarycentric_{\Facet} ; \lbrace \PointPhysical_i^{\Facet} \rbrace_{i=1, \dotsc, N} \right)$, Line \ref{alg:ParameterizationLimitManifold_line:SendToLimitSurface}.
\end{itemize}

\subsection{Smoothness of the Limit Model}
\label{sec:SmoothnessSurrogate}

In this section, we analyze the smoothness of the limit curves and surfaces that compose the limit model.

On the one hand, the curve subdivision scheme, see \autoref{sec:curve_subdivision}, ensures that in a feature curve the union of the inner edges, the ones not adjacent to a feature point, determines a cubic and $\mathcal{C}^2$-continuous parameterization of the limit curve. On the contrary, curved edges of the control mesh that are incident to a feature point do not feature that cubic parameterization. The limit curve is only of class $\mathcal{C}^0$ on the feature point.

\begin{figure}[t]
	\centering
	\renewcommand{\figsize}{0.3}
	\includegraphics[width=\figsize\textwidth]{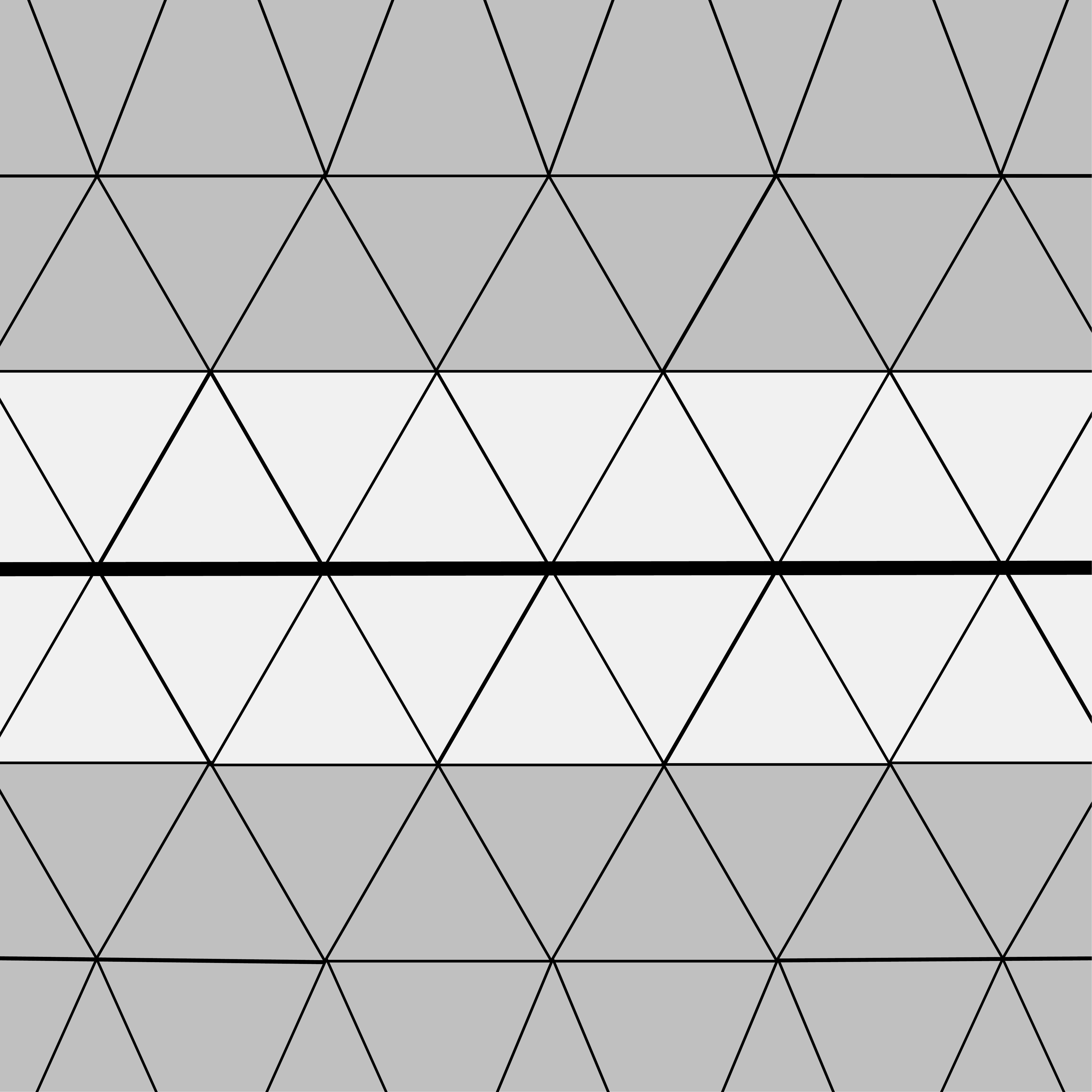}
	\caption{Regular mesh around a feature curve (bold). Elements adjacent to the feature curve are colored with light gray, while the elements not adjacent to the curve are colored in dark gray.}
	\label{fig:Interficie}
\end{figure}

On the other hand, Loop's subdivision scheme, see \autoref{sec:Loop_subdivision}, ensures that in a feature surface the union of the inner triangle, the ones not adjacent to a feature point or curve, determines a $\mathcal{C}^1$-continuous parameterization of the limit surface. Around regular vertices of the mesh, the limit surface features a quartic and $\mathcal{C}^2$-continuous parameterization. The surface triangles that are adjacent to a feature point or a curve determine the region where the limit surface is $\mathcal{C}^0$-continuous. We remark that this discontinuity in the derivatives is confined. \autoref{fig:Interficie} shows a regular mesh featuring inner triangles in dark gray and triangles adjacent to a feature curve (bold) in light gray. In this configuration, the limit surface is of class $\mathcal{C}^2$ only on the dark gray region. The limit curve determined by the bold edges is also of class $\mathcal{C}^2$.

This smooth limit model serves as surrogate geometry for the mesh curving problem. In \autoref{sec:InterpolationArbitrary}, we approximate it by means of piece-wise polynomial surface meshes that, in some cases, inherit its continuity.

\section{Approximation of the Limit Model}
\label{sec:InterpolationArbitrary}

This section is devoted to the generation and analysis of piece-wise polynomials approximations of the limit model. In \autoref{sec:HOMeshArbitrary}, we exploit the parameterization of the limit model to generate nodal high-order surface meshes interpolating the limit model. Next, in \autoref{sec:SmoothnessHOSurfaceMesh}, we analyze the smoothness of the high-order surface mesh. In \autoref{sec:DistanceToLimitManifold}, we use the parameterization of the limit model to compute the distance between the mesh and the limit model and, in \autoref{sec:AutomaticFeatureDetection}, we detail the measure used to automatically suggest to the practitioners the geometry features to smooth.

\subsection{Generation of High-order Surface Meshes}
\label{sec:HOMeshArbitrary}

\begin{algorithm}[t]
	\begin{algorithmic}[1]
			\renewcommand{\algorithmicrequire}{\textbf{Input:}}
		\renewcommand{\algorithmicensure}{\textbf{Output:}}
		\Require{Linear surface mesh $\Mesh$, Linear model $\Model^1$, Polynomial degree $\Degree$, Interpolation nodes $\lbrace \PointReferenceBarycentric_j \rbrace_{j=1, \dotsc, \NNodes}$}
		\Ensure{High-order surface mesh $\Mesh^{\Degree}$, High-order model $\Model^{\Degree}$}
		\Function{\NameAlgorithmInterpolation}{}
		\State $\Mesh_C$ $\leftarrow$ \Call{GenerateNewControlMesh}{$\Mesh$, $\Model^1$} \label{alg:InterpolateLimitManifold_line:ControlMesh}
		\State{$\lbrace \Tri_{i}^{\Degree} \rbrace$ $\leftarrow$ \Call{BuildHighOrderTopology}{$\Mesh$, $\Degree$}} \label{alg:InterpolateLimitManifold_line:BuildHighOrderTopology}
		\For{each triangle $\Tri^1_i$ of $\Mesh$}
			\State $\MeshNodePhysical_{\Tri_i^{\Degree}}^{\infty}$ $\leftarrow$ \Call{\MapOntoLimit}{$\Mesh_C$, $\Model^1$, $\Tri^1_i$, $\lbrace \PointReferenceBarycentric_j \rbrace_{j=1, \dotsc, \NNodes}$} \label{alg:InterpolateLimitManifold_line:MapNodesOntoLimitManifold}
		\EndFor
		\State{$\Mesh^{\Degree}$ $\leftarrow$ \Call{\CreateMesh}{$\lbrace \PointPhysical_i^{\infty} \rbrace$, $\lbrace \Tri_{i}^{\Degree} \rbrace$}} \label{alg:InterpolateLimitManifold_line:createMesh}
		\State{$\Model^{\Degree}$ $\leftarrow$ \Call{GenerateHOModel}{$\Mesh$, $\Model^1$, $\Mesh^{\Degree}$}} \label{alg:InterpolateLimitManifold_line:GenerateHOModel}
		\State \Return{$\Mesh^{\Degree}$, $\Model^{\Degree}$}
		\EndFunction
	\end{algorithmic}
	\caption{Generate high-order surface mesh interpolating the limit model.}
	\label{alg:InterpolateLimitManifold}
\end{algorithm}

In this section, we present one of the main contributions of this work. Given a linear surface mesh, we generate a surface mesh of polynomial degree $\Degree$ such that the nodes of the high-order element interpolate the limit model. Since there exists an explicit parameterization of the limit model, see \autoref{sec:ParameterizationLimitManifold}, it is possible to map any point in the domain onto the limit model. In \cite{jimenez_ramos_albert_2020_3653357}, we proposed a methodology that enabled us to use subdivision schemes to determine high-order meshes of $\Degree = 2^k$, $k \geq 1$, with equispaced distribution of nodes. These meshes inherited the continuity properties of subdivision schemes but were limited by the structure of those schemes. By using the parameterization of the limit model, we can generalize the usage of subdivision schemes to generate high-order meshes of arbitrary polynomial degree and nodal distribution. Thus, nodal sets with better interpolation properties can be used, as it is the case of the quasi-Lebesgue distributions presented in \cite{warburton2006explicit}. 

The main algorithm is described in \autoref{alg:InterpolateLimitManifold}. Given a linear triangular surface mesh $\Mesh$, a linear model $\Model^1$, a polynomial degree $\Degree$, and a nodal distribution of degree $\Degree$ in the master triangle in barycentric coordinates, $\lbrace \PointReferenceBarycentric_j \rbrace_{j=1, \dotsc, \NNodes}$, the function \texttt{\NameAlgorithmInterpolation} returns a surface mesh of polynomial degree $\Degree$ interpolating the limit model, and the associated high-order model $\ModelDegree$. First, in Line \ref{alg:InterpolateLimitManifold_line:ControlMesh}, we cast the triangular surface mesh to a new control mesh to ensure that the location of the initial vertices is preserved in the high-order mesh, as detailed in \ref{sec_app:interpolative}. Following, in Line \ref{alg:InterpolateLimitManifold_line:BuildHighOrderTopology}, we generate the topology $\lbrace \Tri_{i}^{\Degree} \rbrace$ of the high-order mesh. Then, in Line \ref{alg:InterpolateLimitManifold_line:MapNodesOntoLimitManifold}, for each triangle $\Tri^1_i$ of the linear mesh, we map the interpolation nodes onto the limit model by means of the evaluation of the element-wise parameterization described in \autoref{sec:ParameterizationLimitManifold}. The position of the interpolation nodes on the limit model determines the position of the nodes of the high-order element. Finally, in Line \ref{alg:InterpolateLimitManifold_line:createMesh}, we set the mesh of polynomial degree $\Degree$, and compute the associated high-order model $\ModelDegree$, Line \ref{alg:InterpolateLimitManifold_line:GenerateHOModel}.

We highlight that the surrogate geometry for the generation of the high-order mesh is determined by the initial linear mesh and the given geometry features. However, as we detail in \autoref{sec:SmoothnessHOSurfaceMesh}, prior subdivision steps improve the smoothness of the curved high-order mesh that approximates the surrogate. Therefore, if desired, after computing the new control mesh in Line \ref{alg:InterpolateLimitManifold_line:ControlMesh}, a new finer straight-sided mesh could be generated applying several subdivision steps. From this point, the curving procedure would continue as detailed in \autoref{alg:InterpolateLimitManifold}. 

We note that if the interpolation nodes follow an equispaced distribution and the polynomial degree is $\Degree = 2^k$, $k \geq 1$, this method provides the same result as the algorithm presented in \cite{jimenez_ramos_albert_2020_3653357}. In that method, the high-order nodes are generated explicitly, and therefore, it is more efficient in terms of computational time.

\subsection{Smoothness of the High-order Surface Mesh}
\label{sec:SmoothnessHOSurfaceMesh}

\begin{figure}[t]
\centering
\renewcommand{\figsize}{0.2}
\begin{tabular}{cc}
\begin{subfigure}{\figsize\textwidth}
\includegraphics[width=\textwidth]{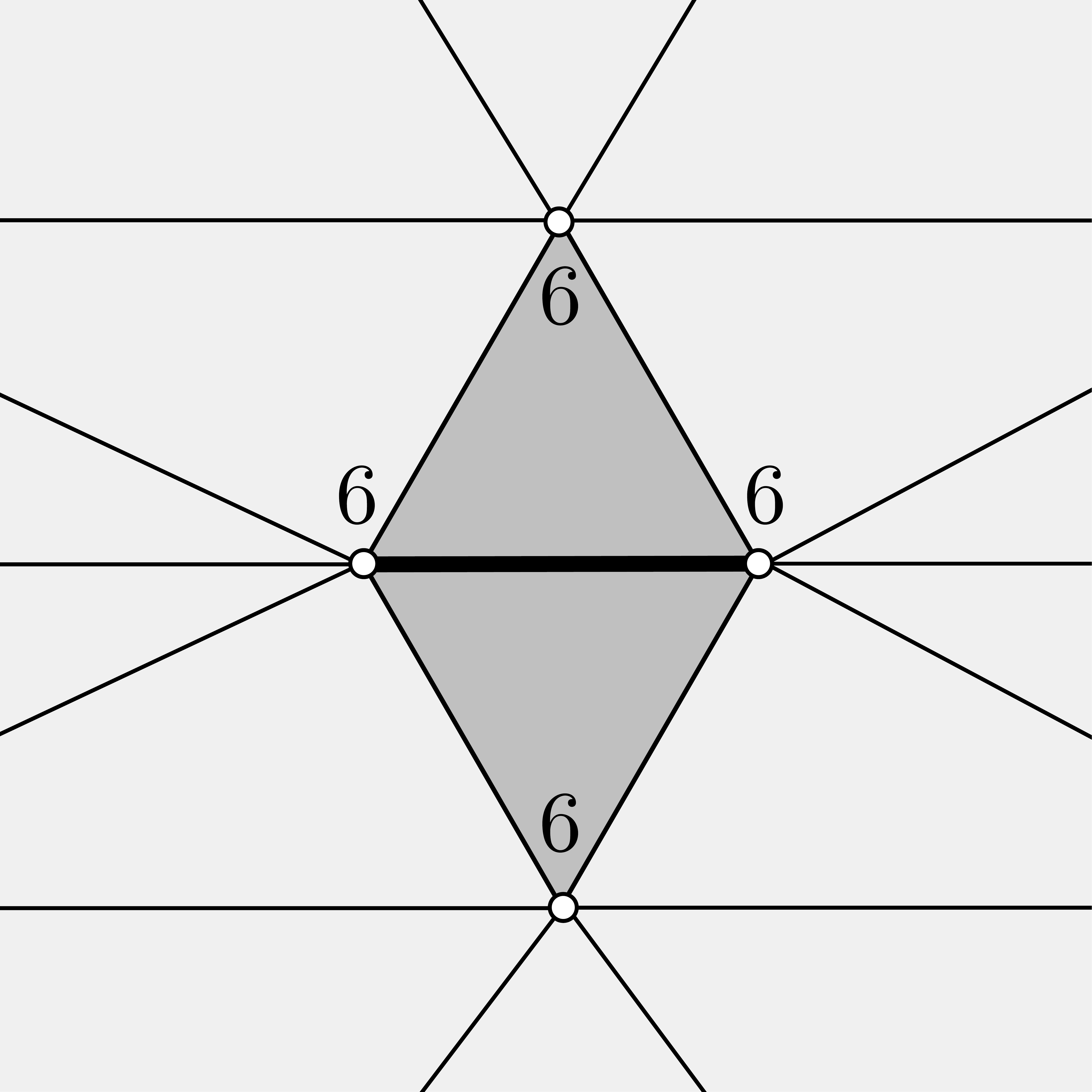}
\caption{}
\label{fig:RegularEdge}
\end{subfigure}
&
\begin{subfigure}{\figsize\textwidth}
\includegraphics[width=\textwidth]{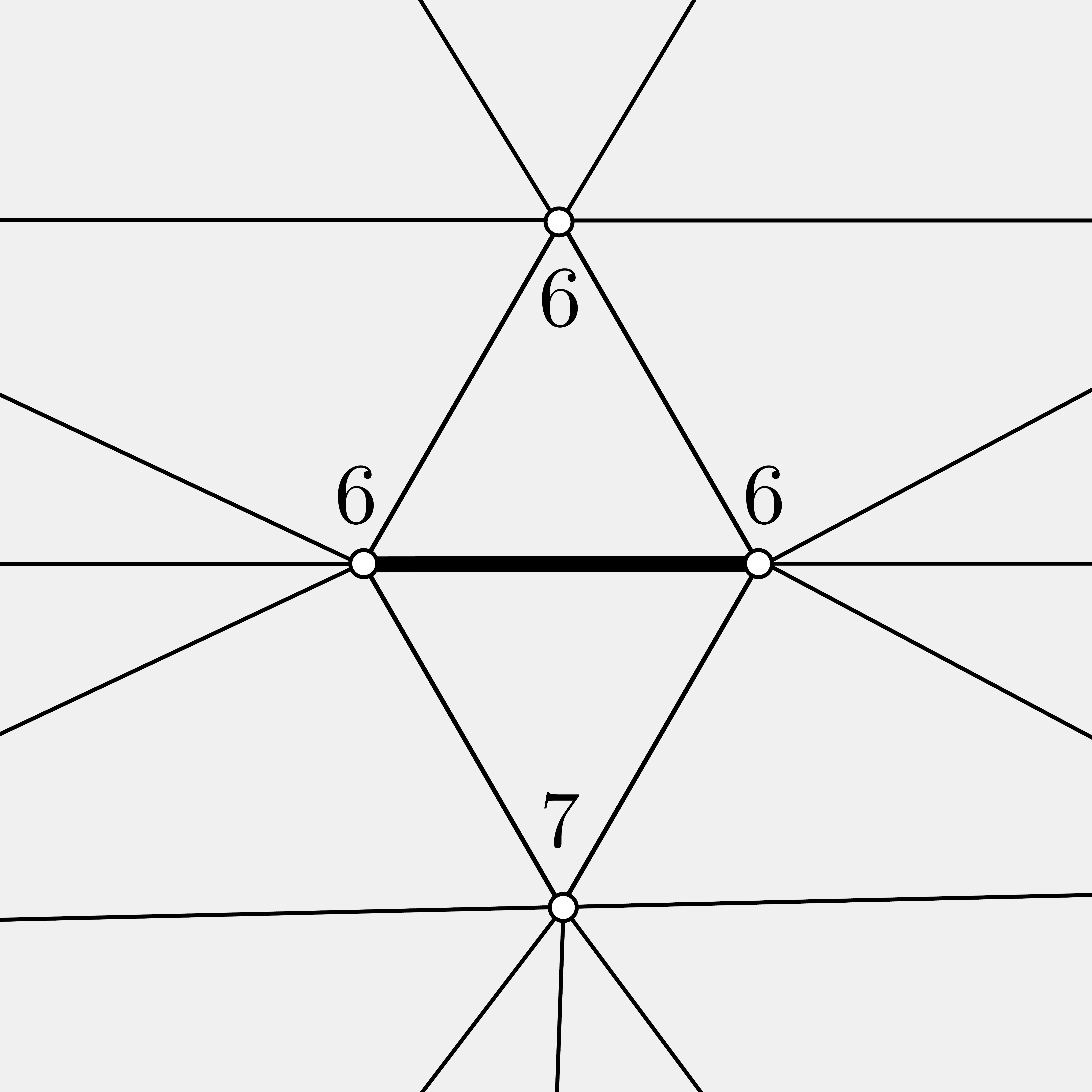}
\caption{}
\label{fig:IrregularEdge}
\end{subfigure}
\end{tabular}
\begin{tabular}{ccc}
\begin{subfigure}{\figsize\textwidth}
\includegraphics[width=\textwidth]{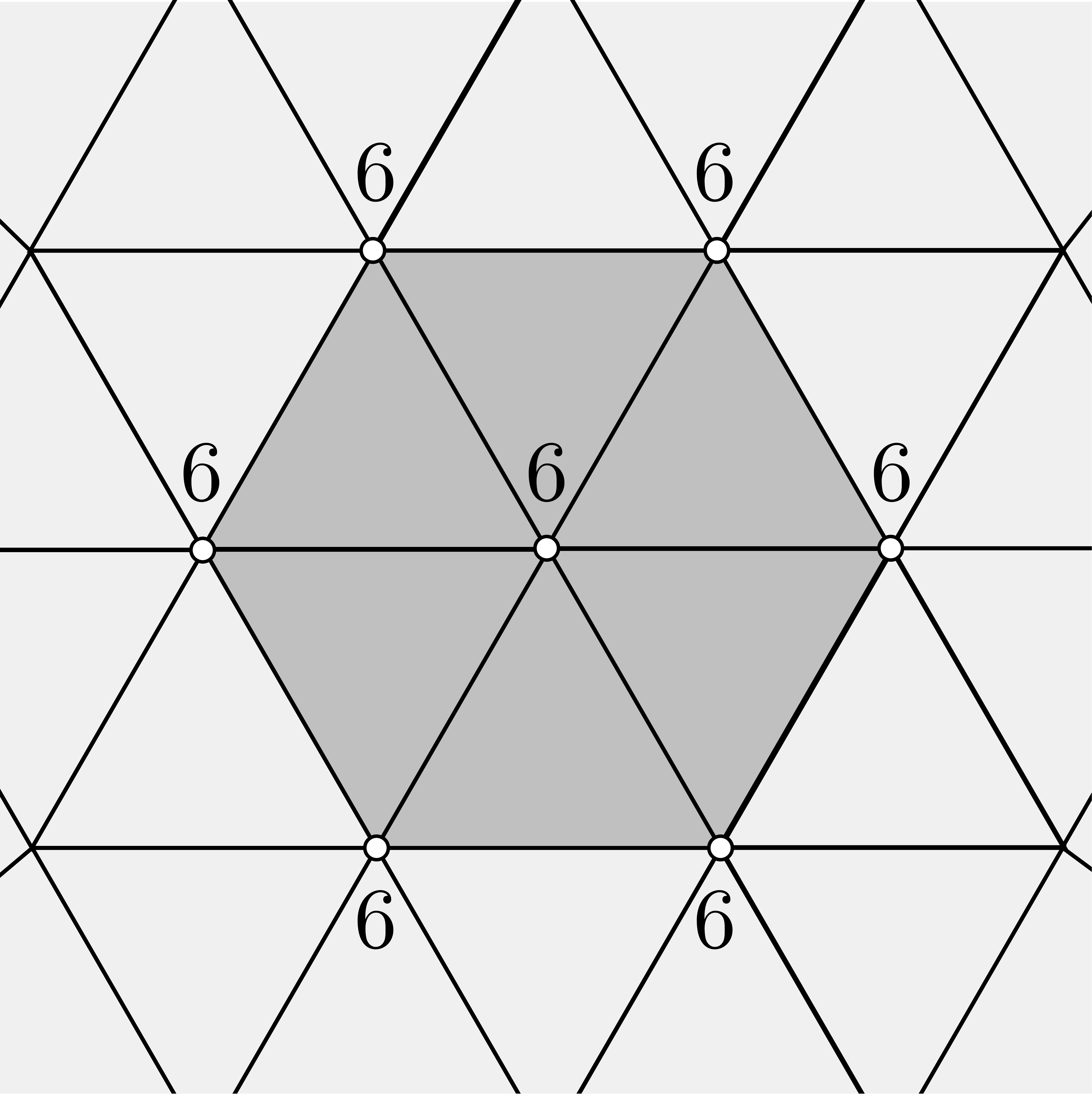}
\caption{}
\label{fig:RegularPatchNode}
\end{subfigure}
&
\begin{subfigure}{\figsize\textwidth}
\includegraphics[width=\textwidth]{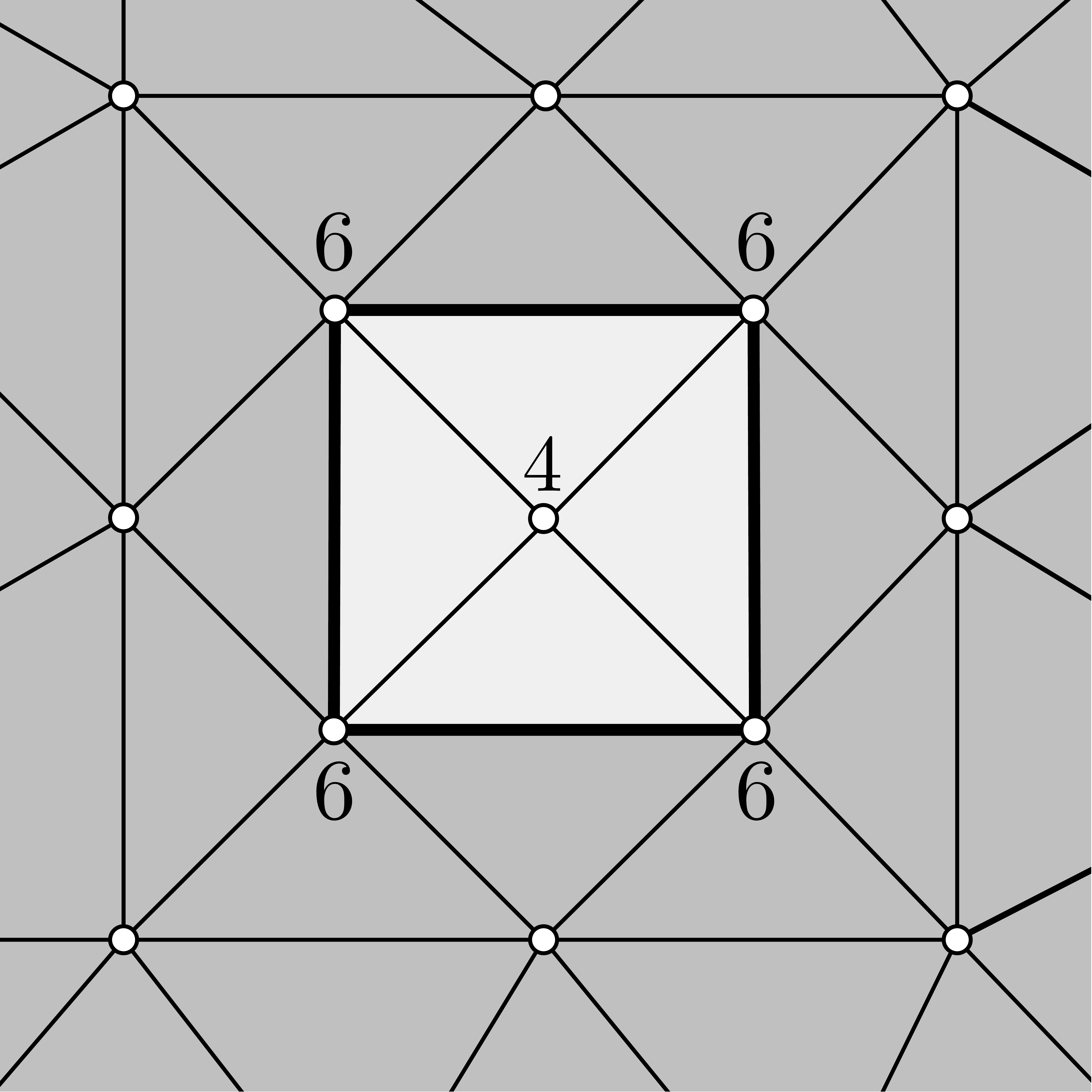}
\caption{}
\label{fig:IrregularPatchNode}
\end{subfigure}
&
\begin{subfigure}{\figsize\textwidth}
\includegraphics[width=\textwidth]{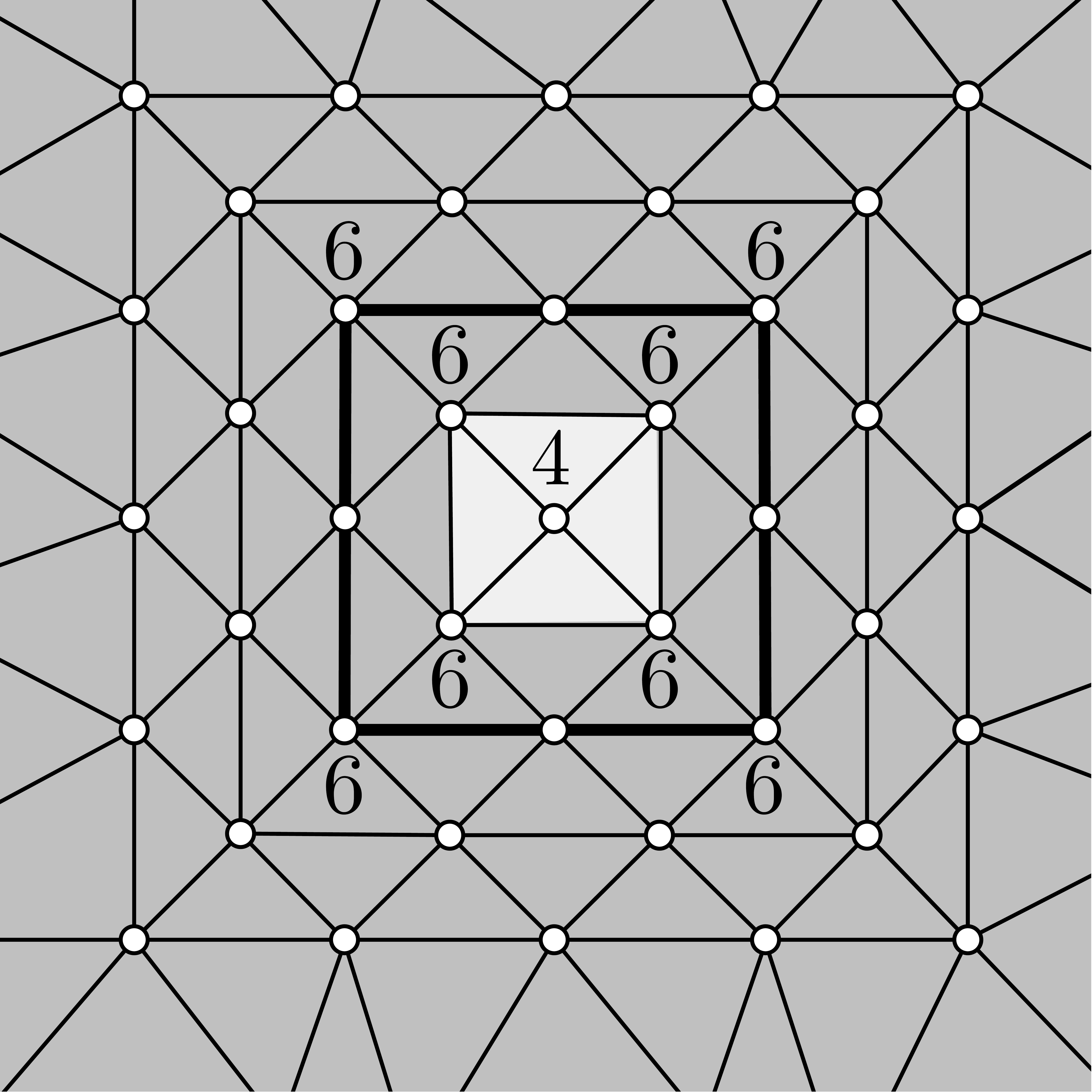}
\caption{}
\label{fig:IrregularPatchNodeSubdivided}
\end{subfigure}
\end{tabular}
\caption{
Mesh configurations to ensure $\mathcal{C}^2$ smoothness (dark gray).
\subref{fig:RegularEdge} Regular edge (bold).
\subref{fig:IrregularEdge} Irregular edge (bold).
\subref{fig:RegularPatchNode} Regular patch around a regular vertex. 
\subref{fig:IrregularPatchNode} Irregular patch (light gray) around an irregular vertex.
\subref{fig:IrregularPatchNodeSubdivided} Mitigation of the irregular patch after subdividing the mesh.
}
\end{figure}

In this section, we analyze the smoothness of the generated high-order meshes. We highlight that the smoothness of the mesh depends on the chosen polynomial degree. Three cases can be distinguished: the quadratic, the cubic, and any higher polynomial degree.

The generated meshes of polynomial degree $\Degree=2$ approximate the limit model by interpolating the limit curve (surface) with third-order accuracy. Since the limit curve (surface) features a cubic (quartic) parameterization around inner regular configurations, meshes of polynomial degree 2 are strictly $\mathcal{C}^0$-continuous, and no guarantees of the $\mathcal{C}^2$-continuity are given by the proposed subdivision-based curving method. 

Next, we analyze the smoothness of the meshes of polynomial degree three. First, the inner edges of the feature curves of the high-order mesh exactly capture the $\mathcal{C}^2$-continuous limit curve. This is so because the limit curve is parameterized by a third degree polynomial, and the elements are described by shape functions of degree three too. For edges incident to feature points, we can only guarantee $\mathcal{C}^0$-continuity. Regarding the limit surface, it is interpolated at the nodes and approximated with fourth-order accuracy.

Following, we study the smoothness of the meshes of polynomial degree $\Degree \geq 4$. As for meshes of polynomial degree three, the high-order mesh exactly captures the $\mathcal{C}^2$-continuous limit curve in the inner edges. Similarly to the curve case, the nodes of the surface mesh also interpolate the limit surface. However, the surface mesh does not inherit the smoothness of the limit surface straight-forwardly as in the curve case. This is so since the limit surface is parameterized element-wise, but the parameterization is of degree four only in a regular element, that is, in an element where its three vertices have six neighbors.

In the interior of an element, the mesh is of class $\mathcal{C}^{\infty}$. Therefore, the smoothness of the surface mesh is to be analyzed along the edges (interfaces between two inner surface elements) and vertices (interfaces between more than two inner surface elements). We can only guarantee $\mathcal{C}^0$-continuity on the edges and vertices of  triangles adjacent to sharp features. For the case of elements not adjacent to sharp features, we first analyze the case between two elements that share an edge. We say an edge is a \emph{regular edge} if all the vertices of the two triangles that share such edge are regular, \textit{i.e.} if the vertices have six neighbors, \autoref{fig:RegularEdge}. In this case, the two elements of degree $\Degree \geq 4$ that share the edge interpolate exactly the quartic limit surface. Hence both elements are exactly equal to the limit surface and, since the limit surface is $\mathcal{C}^{2}$-continuous, the interface (edge) between the two elements also is. In general, no guarantee of the continuity of the derivatives can be deduced along an edge that is not regular, see \autoref{fig:IrregularEdge}.

Now, we analyze the smoothness of the mesh around the vertices of the inner surface elements. Given a regular vertex (with six neighbors), if all the edges incident to it are regular then the surface mesh is of class $\mathcal{C}^2$ around such vertex. In particular, if all the edges incident to a regular vertex are regular, then all its neighbor vertices are also regular, as observed in \autoref{fig:RegularPatchNode}. In such regions, colored in \autoref{fig:RegularPatchNode} in dark gray, the surface mesh captures exactly the limit surface and inherits all its features. The presence of an irregular vertex, as illustrated in \autoref{fig:IrregularPatchNode}, implies the surface mesh approximates the limit surface, rather than exactly capturing it. Therefore, on the one hand, around regular patches, we are able to interpolate and exactly capture the limit surface and obtain a $\mathcal{C}^2$-continuous surface mesh. On the other hand, around irregular patches, the limit surface is interpolated and approximated with accuracy of order $\Degree + 1$ but not matched exactly.

Some prior subdivisions can be applied to the initial linear mesh to improve the smoothness of the high-order mesh. The linear mesh, \autoref{fig:IrregularPatchNode}, can be subdivided before generating the mesh of degree $\Degree$. As observed in \autoref{fig:IrregularPatchNodeSubdivided} in contrast to \autoref{fig:IrregularPatchNode}, the light gray irregular region where the limit surface (and subsequently its smoothness) is not exactly captured is reduced. Exploiting prior refinements of the linear mesh, the regions where the high-order surface mesh is not $\mathcal{C}^2$-continuous can be successively reduced. This new finer high-order mesh determines a better approximation of the surrogate geometry, and consequently, its smoothness is also improved.

\subsection{Distance to the Limit Model}
\label{sec:DistanceToLimitManifold}

In this section, we use the parameterization of the limit model, see \autoref{sec:ParameterizationLimitManifold}, to measure the accuracy of the generated high-order mesh with respect to the limit model.

For the model $\ModelDegree$, the $\IndexSurface$-th surface $\SurfaceDegree$ is a triangulation of polynomial degree $\Degree$ composed of triangular facets, $\SurfaceDegree = \bigcup_{\Facet = 1}^{n_{\Facet}} \Tri^{\Degree}_{\Facet}$. Then, we define the parameterization of the surface $\SurfaceDegree$ element-wise for each $\Tri^{\Degree}_{\Facet}$ as
\begin{align*}
\begin{split}
 \PhiDeg_{\Facet} \colon \MasterTwoD &\to \SurfaceDegree \subset \ModelDegree\\
 \PointReferenceBarycentric &\mapsto \PhiDeg_{\Facet} \left( \PointReferenceBarycentric \right),
\end{split}
\label{eq:ParamHighOrderMesh}
\end{align*}
where $\PhiDeg_{\Facet}$ is the isoparametric mapping between the master triangle and the high-order element $\Tri^{\Degree}_{\Facet}$ of $\SurfaceDegree$.

Denote by $\LimitSurface$ the $\IndexSurface$-th surface of the limit model, see \autoref{eq:ParamLimitManifold}. We define the distance between the limit surface $\LimitSurface$ and the approximation given by the high-order surface mesh $\SurfaceDegree$ as
\begin{equation}
d \left( \LimitSurface, \SurfaceDegree \right) = \max_{f = 1, \dotsc, n_{\Facet}} \sup_{\PointReferenceBarycentric \in \MasterTwoD} \norm{\PhiLim_{\Facet} \left( \PointReferenceBarycentric \right) - \PhiDeg_{\Facet} \left( \PointReferenceBarycentric \right)}_2,
\label{eq:DistanceSurfaces}
\end{equation}
where the \emph{supremum}, the least upper bound, is computed for the points $\PointReferenceBarycentric \in \MasterTwoD$, $\norm{ \cdot }_2$ is the Euclidean norm, and the maximum is taken for all the triangles of the triangulation. In practice, this value is approximated by the maximum distance over a fine grid of points $\lbrace \PointReferenceBarycentric_j \rbrace_{j = 1, \dotsc, N_d}$ in the reference domain,
\begin{equation*}
\sup_{\PointReferenceBarycentric \in \MasterTwoD} \norm{\PhiLim_{\Facet} \left( \PointReferenceBarycentric \right) - \PhiDeg_{\Facet} \left( \PointReferenceBarycentric \right)}_2 \approx \max_{j = 1, \dotsc, N_d} \norm{ \PhiLim_{\Facet} \left( \PointReferenceBarycentric_j \right) - \PhiDeg_{\Facet} \left( \PointReferenceBarycentric_j \right) }_2.
\label{eq:DistanceSurfacesDiscrete}
\end{equation*}
Then, the distance between the limit model $\ModelLimit$ and the model $\ModelDegree$ is defined as
\begin{equation}
\label{eq:distanceMeshModel}
d \left( \ModelLimit, \ModelDegree \right) = 
\frac{1}{L(\ModelLimit)}
\max_{\IndexSurface = 1, \dotsc, n_{\IndexSurface}} d \left( \LimitSurface_{\IndexSurface}, \SurfaceDegree_{\IndexSurface} \right),
\end{equation}
the maximum of the distances between the $\IndexSurface$-th surface $\LimitSurface_{\IndexSurface}$ and $\SurfaceDegree_{\IndexSurface}$, \autoref{eq:DistanceSurfaces}, $\IndexSurface = 1, \dotsc, n_{\IndexSurface}$, adimensionalized with the characteristic length of the model $L(\ModelLimit)$.

We highlight that, as detailed in \autoref{sec:SmoothnessHOSurfaceMesh}, the generated high-order mesh interpolates the limit model on the nodes and therefore, if the polynomial degree is greater than three, $\Degree \geq 3$, the polylines exactly coincide with the limit curves. If the polynomial degree is greater than four, $\Degree \geq 4$, the mesh exactly coincides with the limit surface around the regular vertices. Thus, the distance between a mesh of polynomial degree $\Degree \geq 4$ and the limit model is non-zero only around irregular vertices and along the interfaces between sharp features.

\subsection{Automatic Feature Detection}
\label{sec:AutomaticFeatureDetection}

As previously introduced in \autoref{sec:problemStatement}, the input linear model is composed of the union of features points, curves and surfaces describing the geometry to represent. However, these features may not reproduce the simulation intent, and it may be necessary to smooth them. This process has to be performed by the user and may require significant human labor, depending on the complexity of the input model. We propose to study the continuity of the normal (tangent) vector along (at) a feature curve (point) to decide whether it is suggested to be smoothed. 

Given a linear mesh which interpolates a surface or curve, we want to reconstruct the implicit high-order information of the discretization not fully available in the linear model. To this aim, we use the limit model from subdivision since it takes into account the stencil, and therefore, it reconstructs part of the initial high-order information. Consequently, to automatically detect the smooth features we propose to use the high-order model $\ModelDegree$ which approximates the limit model and provides direct and explicit access to the approximation of the high-order information.

Specifically, consider the $\IndexCurve$-th curve $\CurveDegree$ of the model $\ModelDegree$ composed of the union of poly-segments of degree $\Degree$, $\CurveDegree = \bigcup_{\Edge = 1}^{n_{\Edge}} \Seg^{\Degree}_{\Edge}$. For each segment $\Seg^{\Degree}_{\Edge}$, there are two triangles $\Tri^{\Degree}_i$ and $\Tri^{\Degree}_j$ such that $\Seg^{\Degree}_{\Edge} = \Tri^{\Degree}_i \cap \Tri^{\Degree}_j$. Then, for a point $\PointPhysical \in \Seg^{\Degree}_{\Edge}$, we consider the inner angle function
\begin{equation}
\NormalFunction \left( \PointPhysical \right) = \arccos \left( \NormalVector_i \left( \PointPhysical \right) \cdot \NormalVector_j \left( \PointPhysical \right) \right),
\label{eq:AngleNormalVectors}
\end{equation}
where $\NormalVector_{k} \left( \PointPhysical \right)$ denotes the unitary normal vector defined from triangle $\Tri^{\Degree}_k$. This function accounts for the angle between the normal vectors at point $\PointPhysical$. If the normal vectors at $\PointPhysical$ are continuous, then $\NormalVector_{i} \left( \PointPhysical \right) = \NormalVector_{j} \left( \PointPhysical \right)$, and thus, $\NormalFunction \left( \PointPhysical \right) = 0$. On the contrary, if $\NormalVector_{i} \left( \PointPhysical \right) \neq \NormalVector_{j} \left( \PointPhysical \right)$, then $\NormalFunction \left( \PointPhysical \right) > 0$. In particular, the image of $\NormalFunction$ is $\left[ 0, \pi \right]$, and it attains the minimum in a flat configuration, when $\NormalVector_{i} \left( \PointPhysical \right) = \NormalVector_{j} \left( \PointPhysical \right)$, and the maximum in a reversal configuration, when $\NormalVector_{i} \left( \PointPhysical \right) = - \NormalVector_{j} \left( \PointPhysical \right)$. 

Thus, for each curve $\FeatCurve$ of the model $\ModelDegree$, we compute
\[
\alpha_{\FeatCurve} = \frac{ \int_{\FeatCurve} \NormalFunction \left( \PointPhysical \right) \d{\PointPhysical} }{ \int_{\FeatCurve} \d{\PointPhysical} }.
\]
Integrating along the curve provides an average and therefore, we avoid undesired detection due to spurious values that may arise when performing the computation edge-wise.
The set $R_\delta = \lbrace \FeatCurve \colon \alpha_{\FeatCurve} < \delta \rbrace$ is composed of the feature curves with an angle below a desired threshold $\delta$. This set contains the potentially side features that are suggested to the practitioners to smooth. In particular, the two different surfaces adjacent to a curve feature $\FeatCurve \in R_{\delta}$ are incident with an angle less than $\delta$ and, therefore, it may indicate that the curve feature has to be smoothed. 

Once this set of curves has been identified, we smooth them in the linear model $\Model^1$ and generate a surface mesh of polynomial degree $\Degree$ and a new model $\ModelDegree$ to study the feature points to be potentially smoothed. We distinguish several cases depending on the number of curves incident to the feature point. If no curves are incident, we suggest smoothing this point. If two curves are incident to a feature point, we proceed analogously as for curves but with the tangent vector instead. We compute the tangent vector at the feature point from the two incident edges and, if the angle between the two tangent vectors is below a desired threshold, we suggest smoothing this feature point. If one or more than two curves are incident to a feature point, the smoothing operation has to be performed manually. 

We highlight that each of the feature points (curves) is characterized by a global point (curve) identifier, and therefore, the smoothing operation consists in providing a list of the identifiers of the points (curves) to smooth.

\section{Curved Volume Mesh Approximating the Limit Model}
\label{sec:volume_ho}

In this section, we detail how a linear tetrahedral mesh with marked boundary entities is curved while preserving the sharp features of the model. In \autoref{sec:curved_volume}, we detail the sharp-to-smooth modeling of the geometry features and the replacement of the straight-edged boundary of the linear mesh by the curved boundary mesh. In \autoref{sec:blending}, the curvature on the boundary is accommodated to the interior using a blending technique. This procedure leads to a high-order tetrahedral mesh where its boundary approximates a surrogate geometry composed of feature surfaces with an interior that is $\mathcal{C}^1$-continuous and $\mathcal{C}^2$-continuous almost everywhere. In addition, the vertices of the high-order mesh are kept in the same position as in the initial linear mesh.

\subsection{Substitute the Boundary of the Volume Mesh}
\label{sec:curved_volume}

\begin{algorithm}[t]
	\begin{algorithmic}[1]
		\renewcommand{\algorithmicrequire}{\textbf{Input:}}
		\renewcommand{\algorithmicensure}{\textbf{Output:}}
		\Require{Linear volume mesh $\Mesh$, Linear model $\Model^1$, Polynomial degree $\Degree$, Features to smooth R, Interpolation nodes $\lbrace \PointReferenceBarycentric_j \rbrace_{j=1, \dotsc, \NNodes}$}
		\Ensure{High-order volume mesh $\Mesh^{\Degree}$, High-order model $\Model^{\Degree}$}
		\Function{CurveVolumeMesh}{}
		\State $\Model^1$ $\leftarrow$\Call{\NameRecastFeatures}{$\Model^1$, R} \label{alg:RecastGenerateHOVolumeMesh_line:recast}
		\State $\Mesh^{\Degree}$, $\Model^{\Degree}$ $\leftarrow$ \Call{GenerateHOVolumeMesh}{$\Mesh$, $\Model^1$, $\Degree$, $\lbrace \PointReferenceBarycentric_j \rbrace_{j=1, \dotsc, \NNodes}$} \label{alg:RecastGenerateHOVolumeMesh_line:generateho}
		\State \Return $\Mesh^{\Degree}$, $\Model^{\Degree}$
		\EndFunction
	\end{algorithmic}
	\caption{Curve volume mesh smoothing features.}
	\label{alg:RecastGenerateHOVolumeMesh}
\end{algorithm}

\begin{algorithm}[t]
	\begin{algorithmic}[1]
				\renewcommand{\algorithmicrequire}{\textbf{Input:}}
		\renewcommand{\algorithmicensure}{\textbf{Output:}}
		\Require{Model $\Model^{\Degree}$, Features to smooth R}
		\Ensure{Model $\Model^{\Degree}$}
	\Function{\NameRecastFeatures}{}
	\For{each feature $f$ in R}
	\State $\Model^{\Degree}$ $\leftarrow$ \Call{RemoveFeatureFromList}{$\Model^{\Degree}$, $f$} \label{alg:RecastMarks_line:remove}
	\State $\Model^{\Degree}$ $\leftarrow$ \Call{MergeIncidentFeatures}{$\Model^{\Degree}$, $f$} \label{alg:RecastMarks_line:merge}
	\EndFor
	\State \Return{$\Model^{\Degree}$}
	\EndFunction
	\end{algorithmic}
	\caption{Smooth geometry features.}
	\label{alg:RecastMarks}
\end{algorithm}

\begin{algorithm}[t]
	\begin{algorithmic}[1]
					\renewcommand{\algorithmicrequire}{\textbf{Input:}}
		\renewcommand{\algorithmicensure}{\textbf{Output:}}
		\Require{Linear volume mesh $\Mesh$, Linear model $\Model^1$, Polynomial degree $\Degree$, Interpolation nodes $\lbrace \PointReferenceBarycentric_j \rbrace_{j=1, \dotsc, \NNodes}$}
		\Ensure{High-order volume mesh $\Mesh^{\Degree}$, High-order model $\Model^{\Degree}$}
	\Function{GenerateHOVolumeMesh}{}
	\State $\partial \Mesh$ $\leftarrow$ \Call{ExtractBoundary}{$\Mesh$} \label{alg:GenerateHOVolumeMesh_line:extractboundary}
	\State $\partial \Mesh^{\Degree}$, $\Model^{\Degree}$ $\leftarrow$ \Call{\NameAlgorithmInterpolation}{$\partial \Mesh$, $\Model^1$, $\Degree$, $\lbrace \PointReferenceBarycentric_j \rbrace$} \label{alg:GenerateHOVolumeMesh_line:generatehosurfacemesh}
	\State $\Mesh^{\Degree}$ $\leftarrow$ \Call{IncreasePolynomialDegree}{$\Mesh$, $\Degree$} \label{alg:GenerateHOVolumeMesh_line:increasepoldegree}
	\State $\Mesh^{\Degree}$ $\leftarrow$ \Call{ReplaceBoundary}{$\Mesh^{\Degree}$, $\partial \Mesh^{\Degree}$} \label{alg:GenerateHOVolumeMesh_line:replaceboundary}
	\State $\Mesh^{\Degree}$ $\leftarrow$ \Call{AccommodateCurvature}{$\Mesh^{\Degree}$, $\Mesh$} \label{alg:GenerateHOVolumeMesh_line:accommodatecurvature}
	\If{$\Mesh^{\Degree}$ is low quality} 
	\State $\Mesh^{\Degree}$ $\leftarrow$ \Call{OptimizeMesh}{$\Mesh^{\Degree}$} \label{alg:GenerateHOVolumeMesh_line:optimize}
	\EndIf
	\State \Return $\Mesh^{\Degree}$, $\Model^{\Degree}$
		\EndFunction
	\end{algorithmic}
	\caption{Generate high-order volume mesh.}
	\label{alg:GenerateHOVolumeMesh}
\end{algorithm}

\renewcommand{\figsize}{0.18}
\begin{figure}[t]
\centering
\begin{tabular}{cc}
\begin{subfigure}{\figsize\textwidth}
\centering
\includegraphics[angle=-90,width=\textwidth]{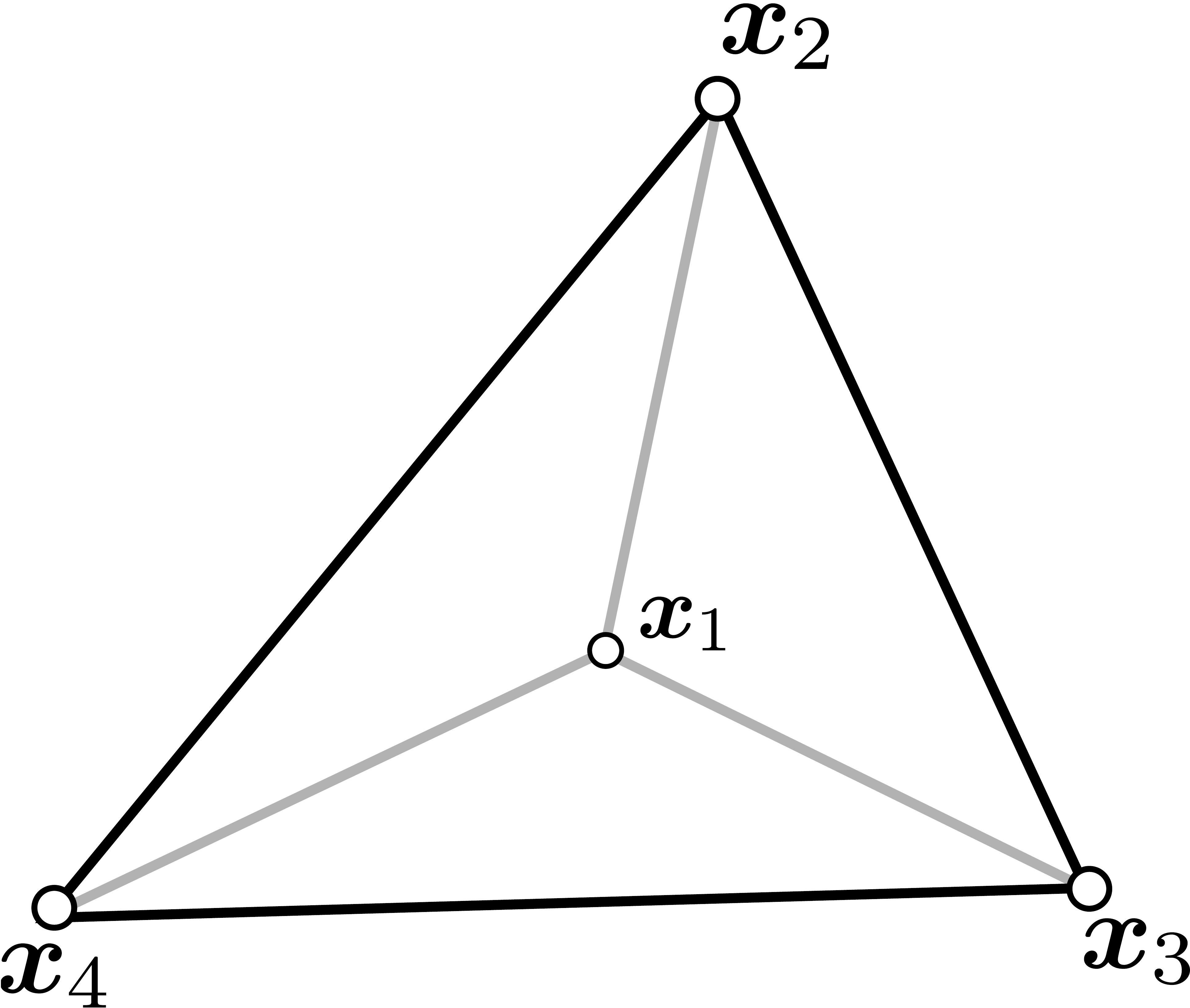}
\caption{}
\label{fig:ho_volume_boundary_surface_before_curved}
\end{subfigure}
&
\begin{subfigure}{\figsize\textwidth}
\centering
\includegraphics[angle=-90,width=\textwidth]{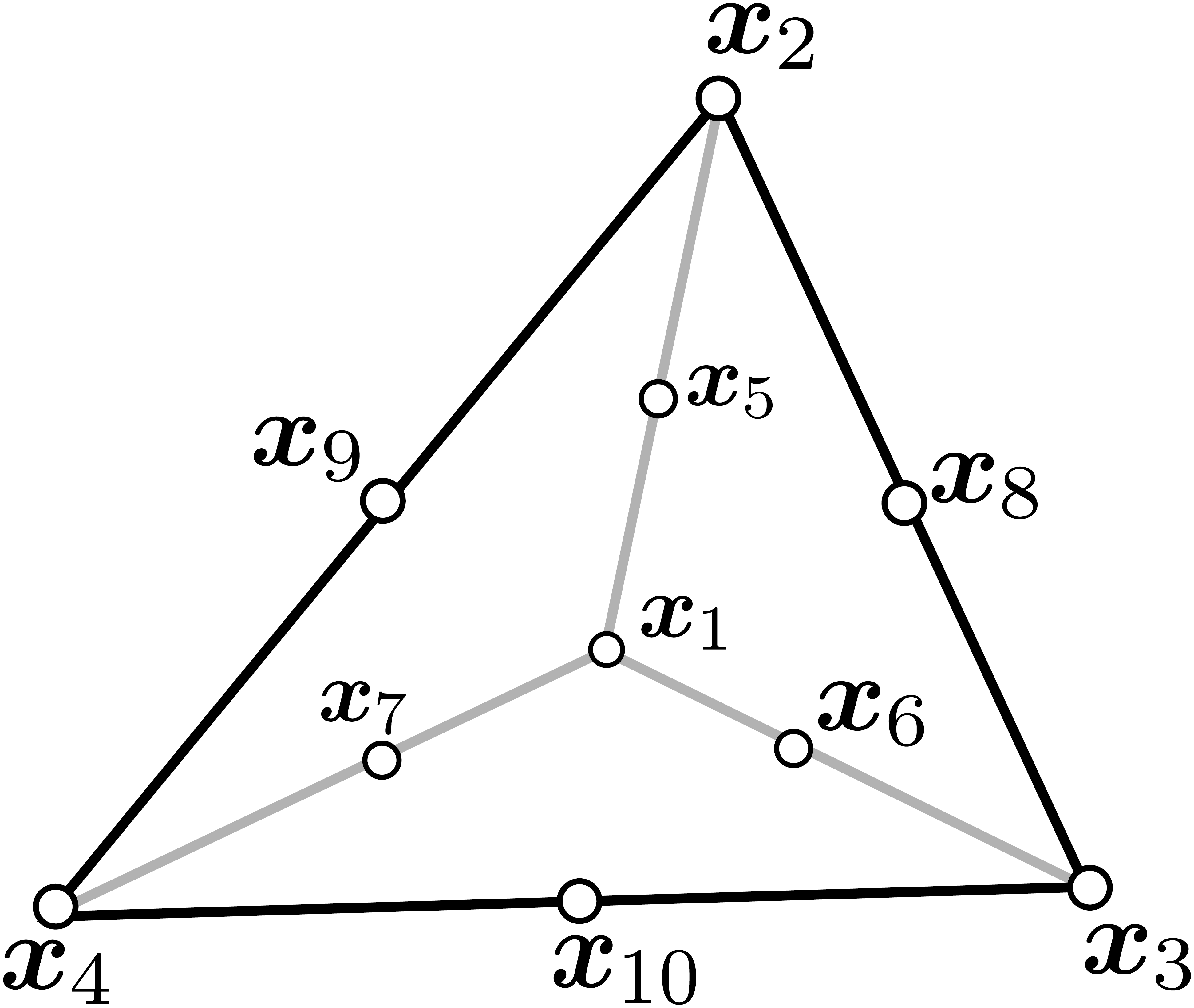}
\caption{}
\label{fig:ho_volume_reference_to_physical_nodes_curved}
\end{subfigure}
\end{tabular}
\begin{subfigure}{\figsize\textwidth}
\centering
\includegraphics[angle=-90,width=\textwidth]{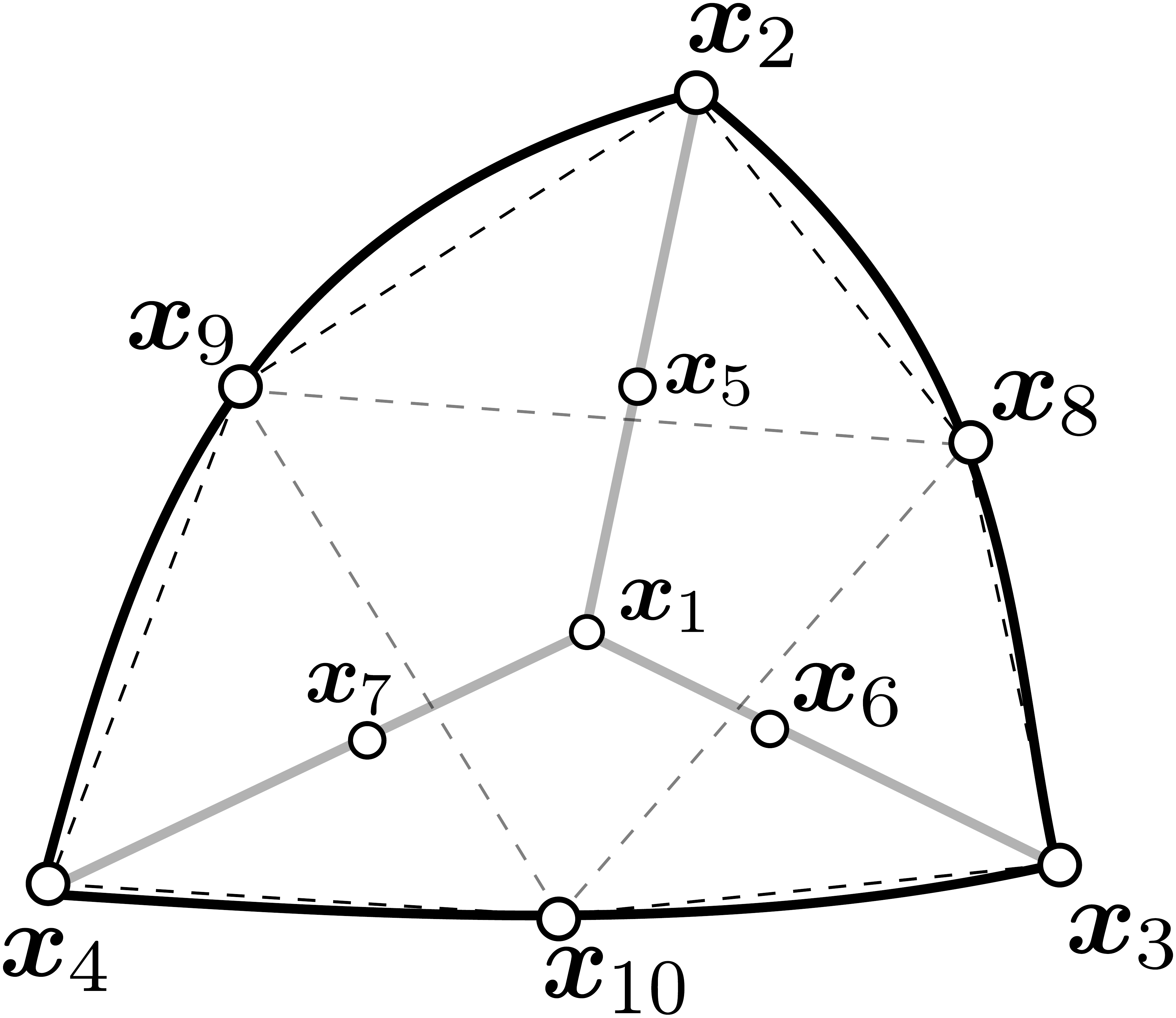}
\caption{}
\label{fig:ho_volume_reference_to_physical_curved_boundary}
\end{subfigure}
\caption{Curving of the boundary for a boundary element of polynomial degree $\Degree=2$. \subref{fig:ho_volume_boundary_surface_before_curved}{ Linear physical element, where the face $\left( 2 \; 3 \; 4 \right)$ belongs to the boundary.} \subref{fig:ho_volume_reference_to_physical_nodes_curved}{ Straight-edged physical element of polynomial degree two.} \subref{fig:ho_volume_reference_to_physical_curved_boundary}{ Curved boundary element of polynomial degree two, displaying with dashed lines the four elements from the subdivision scheme applied to the boundary.}}
\end{figure}

In this section, we detail the subdivision-based curving of a high-order tetrahedral mesh. This process is presented in \autoref{alg:RecastGenerateHOVolumeMesh}. First, in Line \ref{alg:RecastGenerateHOVolumeMesh_line:recast}, we smooth the desired feature entities. Then, in Line \ref{alg:RecastGenerateHOVolumeMesh_line:generateho}, we generate a high-order volume mesh preserving the sharp features provided by the new model once the original entities have been smoothed. These processes, denoted as \texttt{\NameRecastFeatures} and \texttt{GenerateHOVolumeMesh}, are next detailed in Algorithms \ref{alg:RecastMarks} and \ref{alg:GenerateHOVolumeMesh}.

The first step to curve the volume mesh is to smooth, if necessary, the geometry features present in the original model. Recall that points, curves, and surfaces are characterized by a unique identifier. Thus, in order to smooth a sharp feature, it is enough to provide its identifier. That is, in \autoref{alg:RecastMarks}, the variable \texttt{R} contains a list of the identifiers of the feature points and curves to be smoothed. Specifically, the smoothing of a feature is composed of two steps: removing the feature from the list of features to preserve, Line \ref{alg:RecastMarks_line:remove}; and merging the features incident to such feature, Line \ref{alg:RecastMarks_line:merge}. Since each feature is described by a unique identifier, the process of merging the incident features reduces to assigning the same identifier to these features. The features to smooth are manually provided by the user, however, in order to reduce the human labor of manually selecting the features to smooth, the indicator proposed in \autoref{sec:AutomaticFeatureDetection} can be used to determine the features that have to be potentially smoothed, as illustrated in \autoref{sec:HighLiftExample}.

Next, the curving method based on hierarchical subdivision and blending is performed. The generation of a high-order volume mesh is described in \autoref{alg:GenerateHOVolumeMesh}. First, given a linear tetrahedral mesh, the linear model $\Model^1$, the polynomial degree $\Degree$, and the interpolation nodes of the master tetrahedron in barycentric coordinates $\lbrace \PointReferenceBarycentric_j \rbrace_{j=1, \dotsc, \NNodes}$, \autoref{fig:ho_volume_boundary_surface_before_curved}, we extract its boundary, Line \ref{alg:GenerateHOVolumeMesh_line:extractboundary}. The boundary is a surface mesh that inherits the geometry features of the volume mesh. Next, in Line \ref{alg:GenerateHOVolumeMesh_line:generatehosurfacemesh}, we call the function \texttt{\NameAlgorithmInterpolation} to generate a surface mesh of polynomial degree $\Degree$ preserving the sharp features and interpolating the limit model at the boundary nodes of the interpolation set $\lbrace \PointReferenceBarycentric_j \rbrace_{j=1, \dotsc, \NNodes}$, see \autoref{sec:HOMeshArbitrary}. Third, we generate a straight-edged high-order volume mesh, Line \ref{alg:GenerateHOVolumeMesh_line:increasepoldegree}, illustrated in \autoref{fig:ho_volume_reference_to_physical_nodes_curved}. Following, in Line \ref{alg:GenerateHOVolumeMesh_line:replaceboundary}, we replace the boundary of the straight-edged mesh by the curved surface mesh, see \autoref{fig:ho_volume_reference_to_physical_curved_boundary}. Then, in Line \ref{alg:GenerateHOVolumeMesh_line:accommodatecurvature}, the curvature of the surface is accommodated to the elements adjacent to boundary, using a blending technique to be described in \autoref{sec:blending}. Finally, if the mesh contains low-quality elements, it is optimized using \cite{Gargallo-Peiro2015,gargallo2015optimization}, see Line \ref{alg:GenerateHOVolumeMesh_line:optimize}.

The methodology proposed in this work can also be used to generate, given an initial linear mesh, finer linear meshes that successively improve the approximation of the surrogate geometry. To do so, the generated high-order mesh can be reinterpreted as a linear mesh by the decomposition of each high-order element into linear elements. Specifically, the reference high-order element is decomposed into several structured linear elements determined by the high-order nodes. If the linear mesh contains low-quality elements, the optimization procedure described in \cite{Gargallo-Peiro2015} is applied to ensure a valid mesh.

\subsection{Accommodation of the Curvature of the Boundary}
\label{sec:blending}

\begin{figure}[t]
\centering
\renewcommand{\figsize}{0.2}
\begin{tabular}{cccc}
\begin{subfigure}{\figsize\textwidth}
	\centering
	\includegraphics[angle=-90,width=\textwidth]{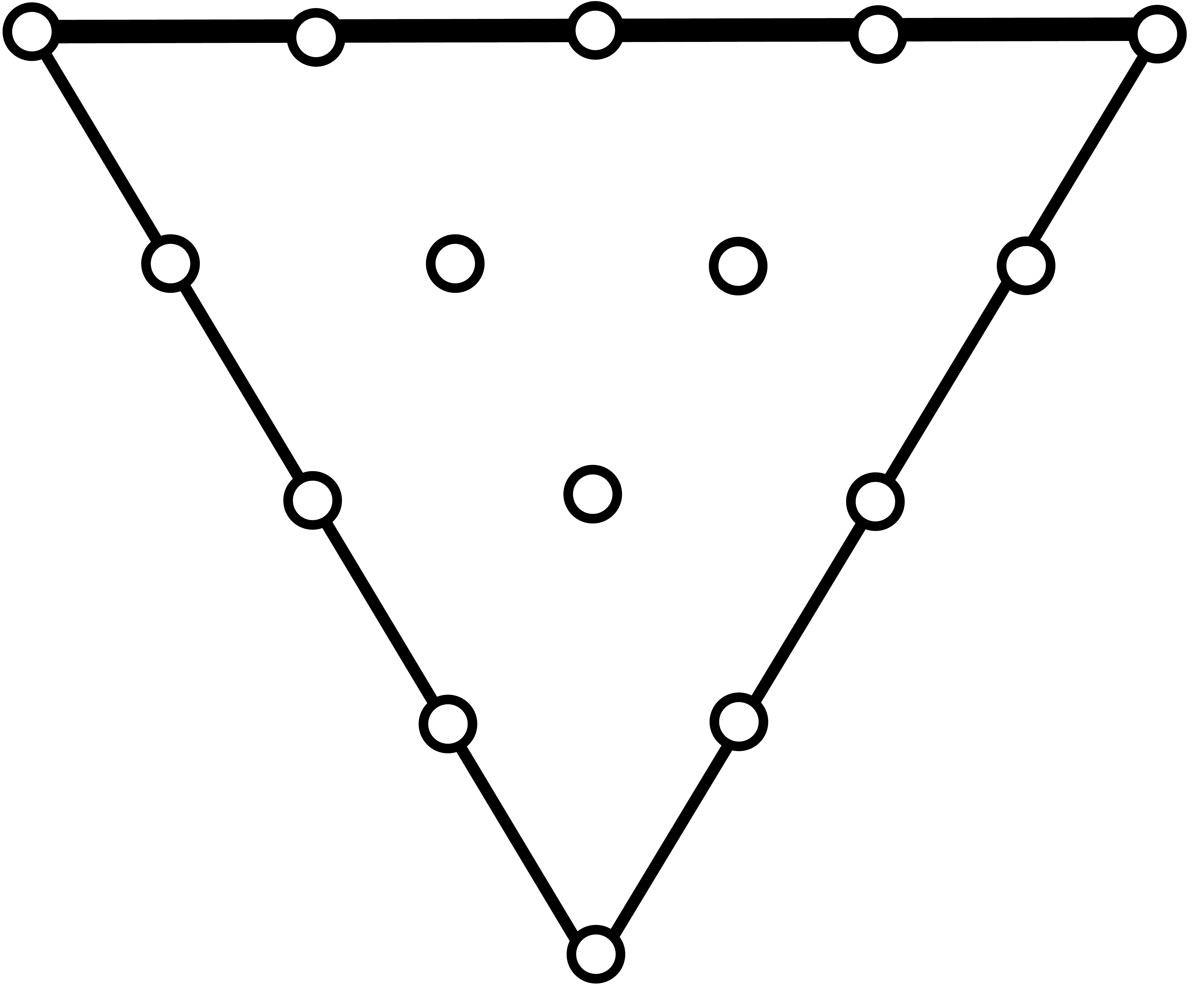}
	\caption{}
	\label{fig:TFI_straight}
\end{subfigure}
&
\begin{subfigure}{\figsize\textwidth}
\centering
\includegraphics[angle=-90,width=\textwidth]{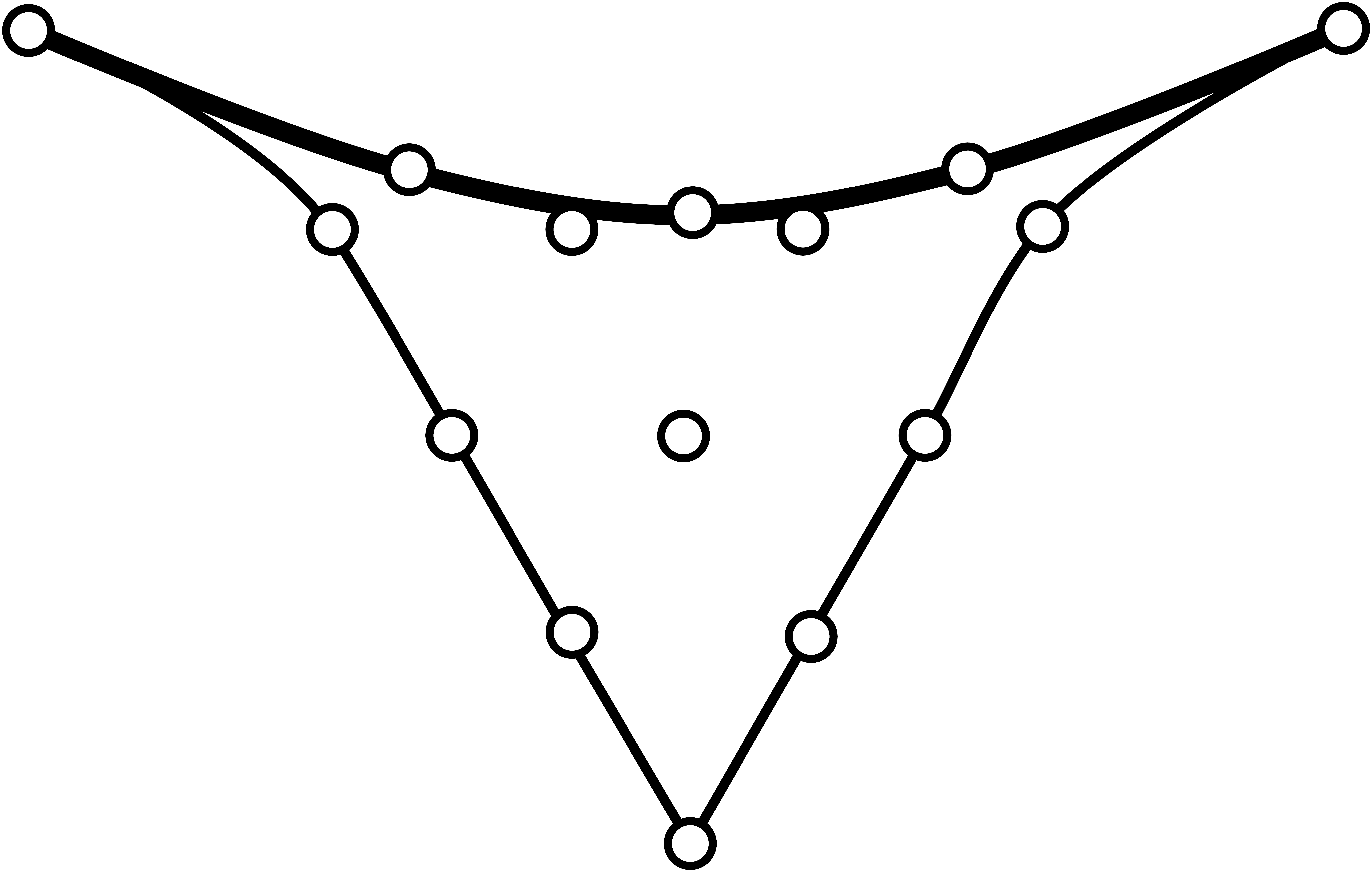}
\caption{}
\label{fig:TFI_0}
\end{subfigure}
&
\begin{subfigure}{\figsize\textwidth}
\centering
\includegraphics[angle=-90,width=\textwidth]{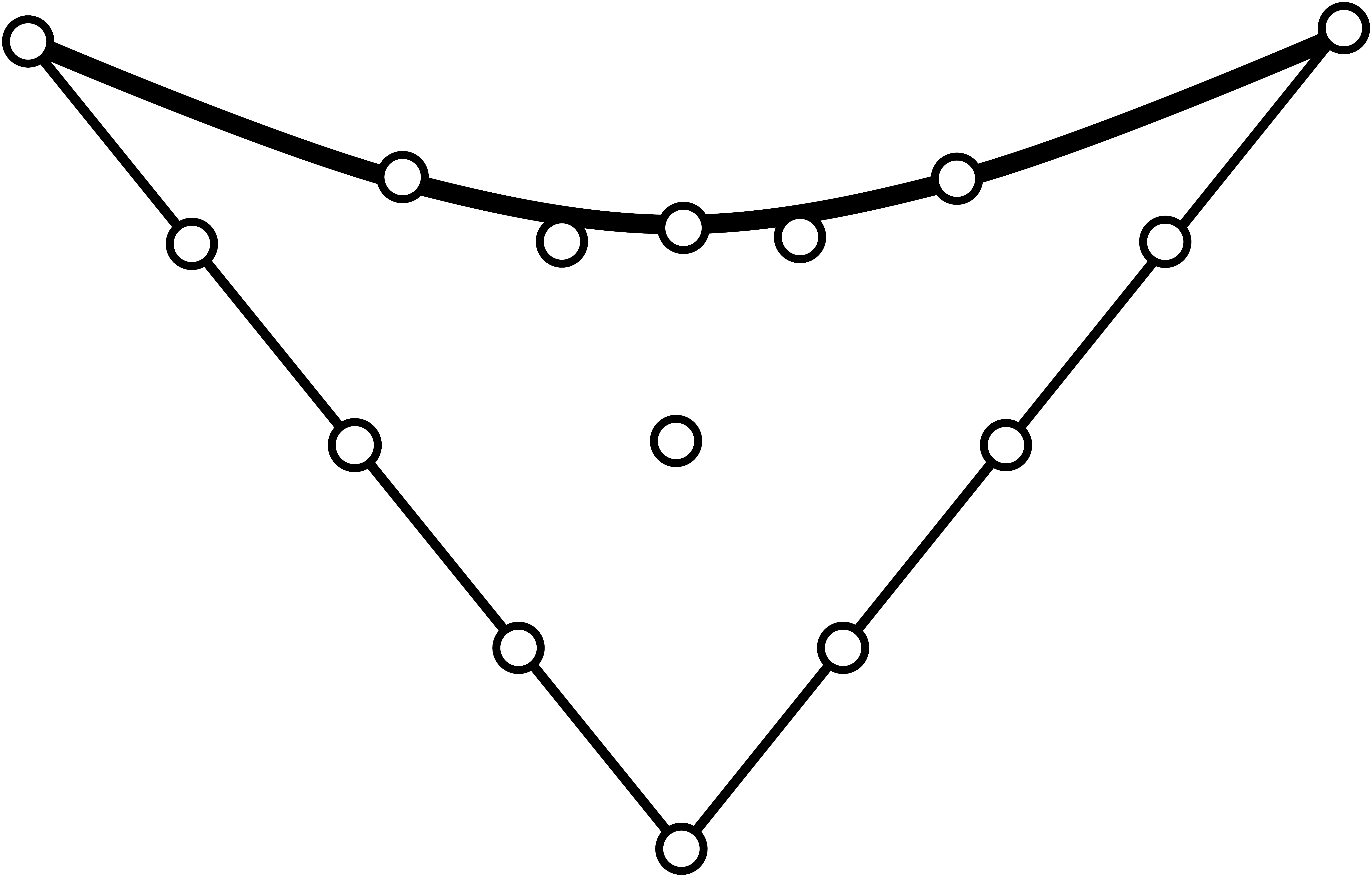}
\caption{}
\label{fig:TFI_1}
\end{subfigure}
&
\begin{subfigure}{\figsize\textwidth}
\centering
\includegraphics[angle=-90,width=\textwidth]{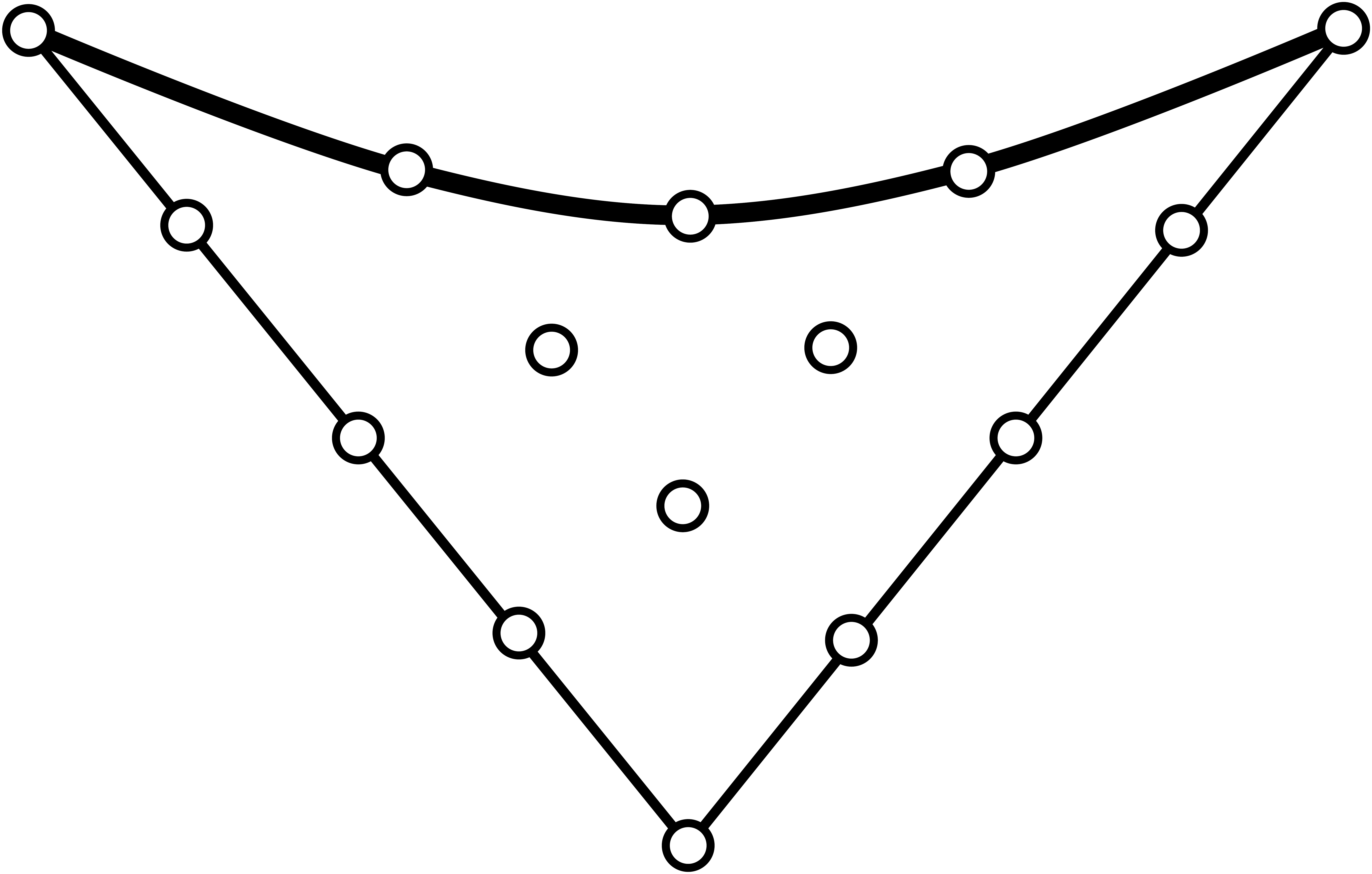}
\caption{}
\label{fig:TFI_2}
\end{subfigure}
\end{tabular}
\caption{Accommodating the curvature to a triangular element of polynomial degree $\Degree =4$. \subref{fig:TFI_straight} Straight-edged triangle with a boundary edge (in bold).
	\subref{fig:TFI_0}{ Triangle with curved boundary.} 
	\subref{fig:TFI_1}{ Transfinite interpolation applied to the edges,} 
	\subref{fig:TFI_2}{ and to the face.}}
\label{fig:TFI}
\end{figure}

In \autoref{alg:GenerateHOVolumeMesh}, after replacing the curved boundary in the straight-sided high-order mesh, Line \ref{alg:GenerateHOVolumeMesh_line:replaceboundary}, the obtained high-order mesh may contain low-quality or tangled elements, see \autoref{fig:TFI_straight} and \ref{fig:TFI_0}. In isotropic mesh generation, only boundary tetrahedra are affected and therefore, the number of invalid elements is small compared with the scale of the generated meshes. Thus, similarly to \cite{johnennew}, as an attempt to improve the mesh quality in a fast and explicit manner, in Line \ref{alg:GenerateHOVolumeMesh_line:accommodatecurvature} in function \texttt{AccommodateCurvature}, we use Transfinite Interpolation (TFI) \cite{perronnet1998interpolation} to accommodate the curved surface to those entities of the boundary elements not present in the surface mesh. Specifically, given a boundary element, we use transfinite interpolation hierarchically on its entities to accommodate the curving of the boundary. That is, first we relocate the nodes on edges, then nodes on faces, and finally, nodes in the interior of tetrahedra. We explicitly detail the position of all the interior nodes of the boundary elements in \ref{sec_app:blending}.

The analogous procedure for triangular meshes is illustrated in \autoref{fig:TFI}. In \autoref{fig:TFI_straight}, we show a straight-sided boundary triangle of degree $\Degree = 4$, with the boundary edge highlighted in bold. After the curving of the boundary, the boundary nodes have been relocated, see \autoref{fig:TFI_0}. In \autoref{fig:TFI_1}, we illustrate the relocation of the nodes on the interior edges of the triangle, and the interior nodes in \autoref{fig:TFI_2} account for the curvature propagated through the edges.

We remark that this method does not guarantee to repair the invalid elements, neither ensures an increase of the element quality. However, it is an explicit and fast formulation, which in practice represents a good initial condition for mesh curving methods when no geometry is available \cite{gargallo2015optimization,ruiz2019automatically,MOXEY2016130,toulorge2013robust,PerssonPeraire}. In all the tested applications, see \autoref{sec:results}, the procedure improves significantly the quality of the meshes. Once the TFI-based relocation process is finalized, if low quality or inverted elements are present, we perform the non-linear quality optimization procedure presented in \cite{gargallo2015optimization}.

\section{Results}
\label{sec:results}

In this section, we present several examples to illustrate the main features of the methods presented in this work. As a proof of concept, the proposed algorithms have been developed in Anaconda Python \cite{python}. The prototyping code is sequential (one execution thread) and non-vectorized. All the examples have been run on a MacBook Pro (with one dual-core Intel Core i5 CPU, a clock frequency of 2.3 GHz, and a total memory of 16 GBytes).

In all the examples, we validate both the high-order boundary and volume meshes using the Jacobian-based distortion measure proposed in \cite{gargallo2016distortion,Gargallo-Peiro2015}. In particular, the quality of a high-order element is computed with respect to the corresponding original straight-edged element in the linear mesh.

\subsection{Convergence of the High-order Surface Mesh to the Limit Model}
\label{sec:ExampleSphere}

\newcommand{\LebCt}{\Lambda}
\newcommand{\LebEq}{\LebCt_{\text{Eq}}}
\newcommand{\LebWB}{\LebCt_{\text{WB}}}
\newcommand{\LowBound}{L}
\newcommand{\LowBoundEq}{\LowBound_{\text{Eq}}}
\newcommand{\LowBoundWB}{\LowBound_{\text{WB}}}
\newcommand{\DistEqui}{d \left( \ModelLimit, \ModelDegree_{\text{Eq}} \right)}
\newcommand{\DistWB}{d \left( \ModelLimit, \ModelDegree_{\text{WB}} \right)}
\begin{table}[t]
\centering
\begin{tabular}{ccccccc}
\hline
$\Degree$ & $\LebEq$ & $\DistEqui$ & $\LowBoundEq$ & $\LebWB$ & $\DistWB$ & $\LowBoundWB$ \\
\hline
1  & $1.00$     & $2.10 \cdot 10^{-2}$     & $1.05 \cdot 10^{-2}$ & $1.00$    & $2.10 \cdot 10^{-2}$       & $1.05 \cdot 10^{-2}$      \\
2  & $1.66$  & $5.64  \cdot 10^{-3}$ & $2.12 \cdot 10^{-3}$ & $1.66$ & $5.64 \cdot 10^{-3}$ & $2.12 \cdot 10^{-3}$ \\
3  & $2.27$  & $2.67 \cdot 10^{-3}$ & $8.15 \cdot 10^{-4}$ & $2.11$ & $2.52 \cdot 10^{-3}$ & $8.10 \cdot 10^{-4}$ \\
4  & $3.47$  & $1.88 \cdot 10^{-3}$ & $4.20 \cdot 10^{-4}$ & $2.66$ & $1.67 \cdot 10^{-3}$ & $4.57 \cdot 10^{-4}$ \\
5  & $5.45$  & $1.31 \cdot 10^{-3}$ & $2.04 \cdot 10^{-4}$ & $3.12$ & $1.14 \cdot 10^{-3}$ & $2.78 \cdot 10^{-4}$ \\
6  & $8.75$  & $1.09 \cdot 10^{-3}$ & $1.11 \cdot 10^{-4}$ & $3.70$ & $8.79 \cdot 10^{-4}$ & $1.87 \cdot 10^{-4}$ \\
7  & $14.35$ & $9.92 \cdot 10^{-4}$ & $6.46 \cdot 10^{-5}$ & $4.27$ & $6.36 \cdot 10^{-4}$ & $1.21 \cdot 10^{-4}$ \\
8  & $24.01$ & $8.63 \cdot 10^{-4}$ & $3.45 \cdot 10^{-5}$ & $4.96$ & $4.50 \cdot 10^{-4}$ & $7.55 \cdot 10^{-5}$ \\
9  & $40.92$ & $5.26 \cdot 10^{-4}$ & $1.25 \cdot 10^{-5}$ & $5.74$ & $2.98 \cdot 10^{-4}$ & $4.43 \cdot 10^{-5}$  \\
10 & $70.89$ & $7.02 \cdot 10^{-4}$ & $9.76 \cdot 10^{-6}$ & $6.67$ & $1.91 \cdot 10^{-4}$ & $2.49 \cdot 10^{-5}$ \\
\hline
\end{tabular}
\caption{Interpolation with equispaced and non-equispaced nodes of a sphere limit model for polynomial degree $\Degree$, $\Degree = 1, \dotsc, 10$: Lebesgue constant, distance to the limit model, and lower bound of the distance.}
\label{tab:ConvergenceSphere}
\end{table}

\begin{figure}[t]
\centering
\renewcommand{\figsize}{0.45}	
\begin{tabular}{cc}
\begin{subfigure}{\figsize\textwidth}
	\centering
	\includegraphics[width=\textwidth]{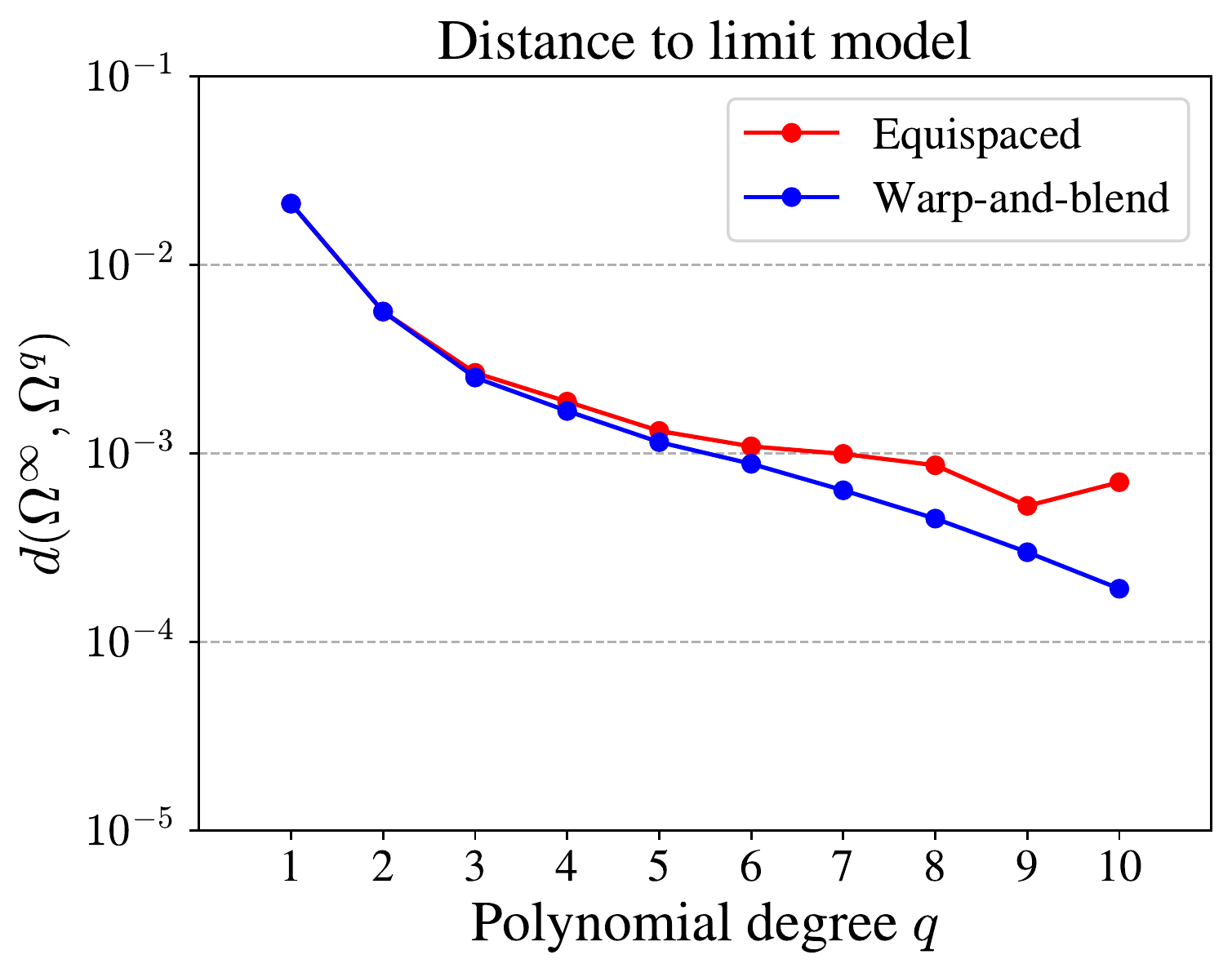}
	\caption{}
	\label{fig:ConvergenceSphere}
\end{subfigure}
&
\begin{subfigure}{\figsize\textwidth}
\centering
\includegraphics[width=\textwidth]{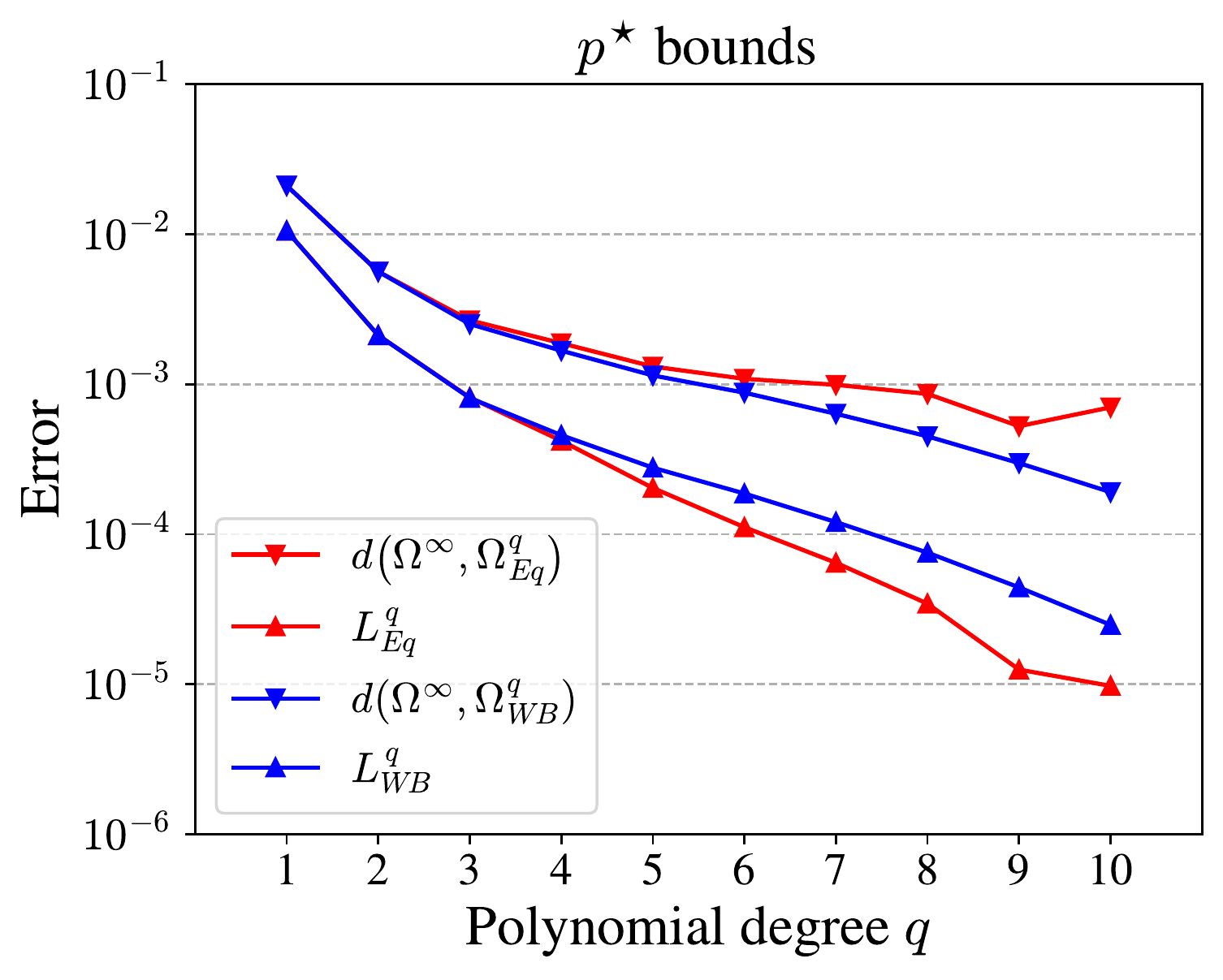}
\caption{}
\label{fig:SphereBounds}
\end{subfigure}
\end{tabular}
\caption{Convergence of the bounds and the distance to the limit model for the sphere meshes of polynomial degree  $\Degree$, $\Degree = 1, \dotsc, 10$, with equispaced and non-equispaced distribution sets: \subref{fig:ConvergenceSphere} distance, and \subref{fig:SphereBounds} lower and upper bounds of the distance of the best approximating polynomial.}
\label{fig:SphereGraphs}
\end{figure}

\begin{figure}[t]
\centering
\renewcommand{\figsize}{0.4}
\begin{tabular}{cc}
	\begin{subfigure}{\figsize\textwidth}
		\centering
		\includegraphics[width=\textwidth]{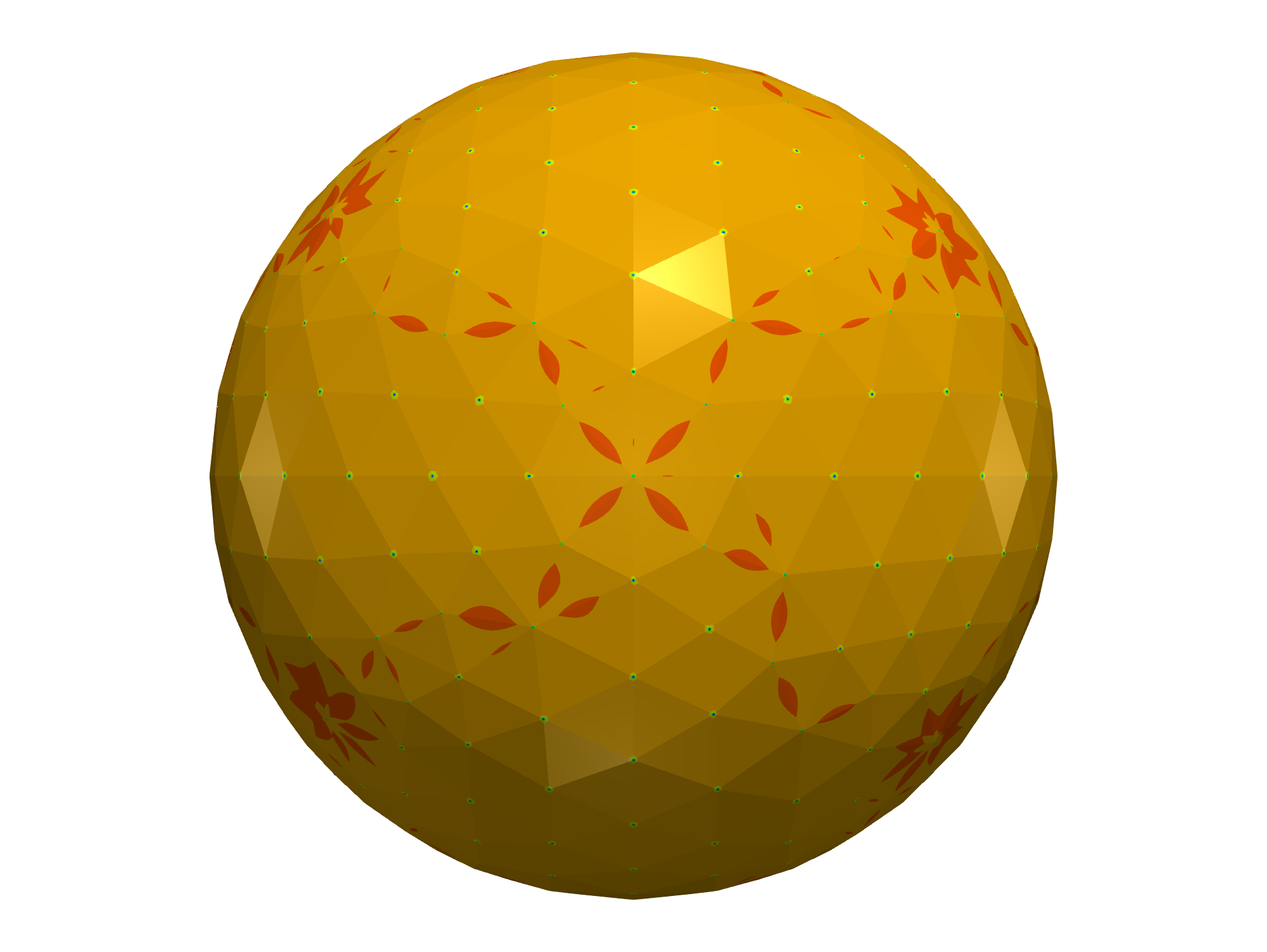}
		\caption{}
		\label{fig:SphereLinear}
	\end{subfigure}
	&
	\begin{subfigure}{\figsize\textwidth}
	\centering
	\includegraphics[width=\textwidth]{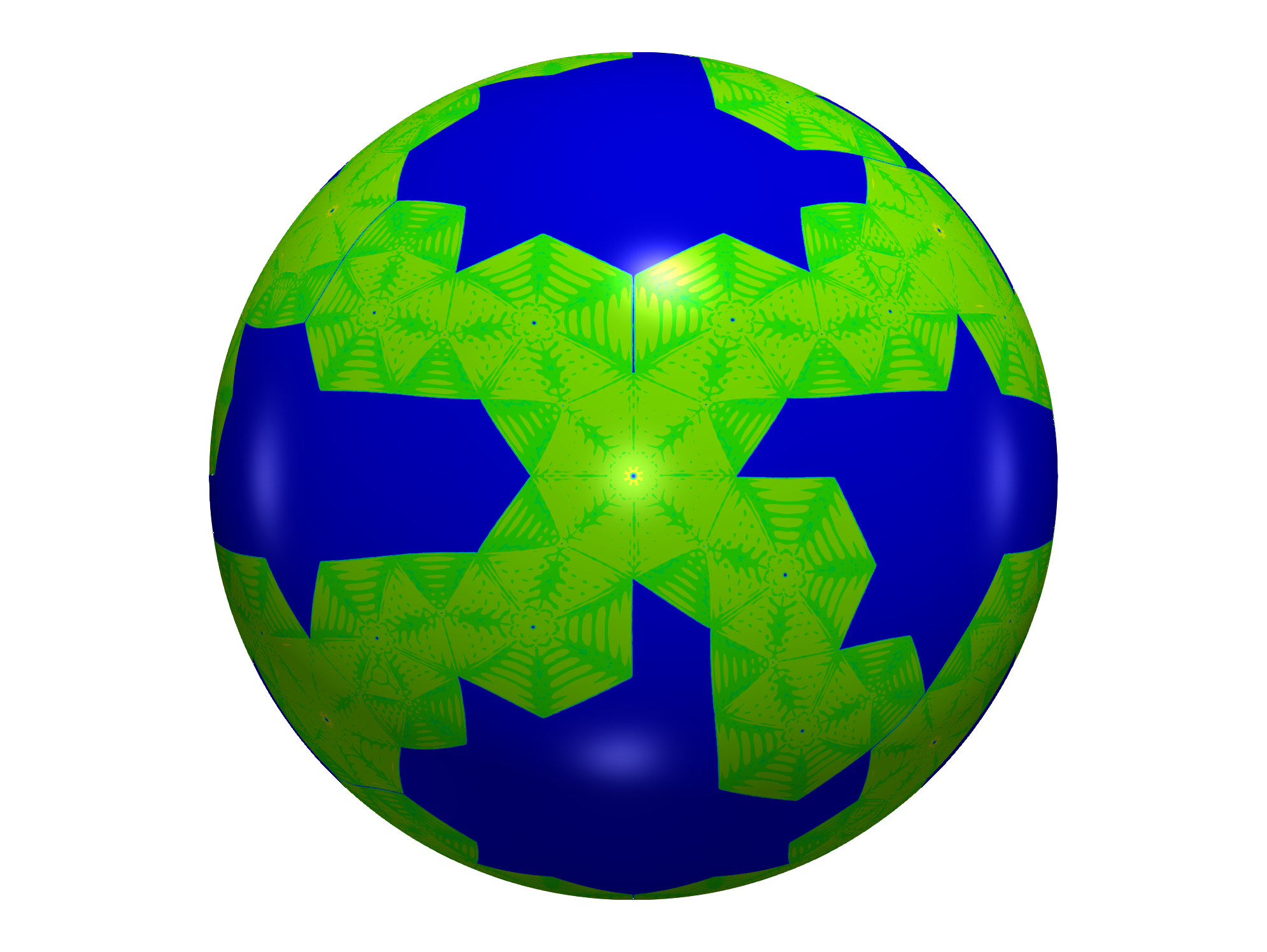}
	\caption{}
	\label{fig:SphereHighOrderDistanceToLimit}
	\end{subfigure}
\end{tabular}
	\begin{subfigure}{\textwidth}
	\centering
	\includegraphics[width=0.4\textwidth]{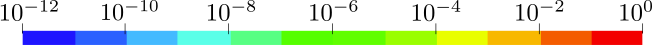}
	\end{subfigure}
\caption{Mesh of a sphere of polynomial degree \subref{fig:SphereLinear} one, $\Degree=1$, and	\subref{fig:SphereHighOrderDistanceToLimit} ten, $\Degree = 10$, both colored according to the distance to the limit model.}
\label{fig:SphereMeshExample}
\end{figure}

In this example, we analyze the convergence of the surface meshes of polynomial degree $\Degree$ to the limit model when $\Degree$ is increased, $\Degree = 1, \dotsc, 10$. This kind of convergence is useful in such applications where the initial partition of elements has to be preserved. We consider a sphere of radius 1 discretized with a linear triangular mesh composed of 450 nodes and 896 elements, see \autoref{fig:SphereLinear}. The mesh is colored according to the distance to the limit model, see \autoref{eq:distanceMeshModel}, with $L(\ModelLimit)=1$, the radius of the sphere. Note that the distance at the vertices is zero due to the interpolation property of the scheme. Then, for each polynomial degree $\Degree$, we generate two surface meshes with two different nodal distributions: one equispaced, and the other defined using a warp-and-blend technique aimed at quasi-minimizing interpolation errors presented in \cite{warburton2006explicit}.

\newcommand{\BestApp}{p^{\star}}
For each polynomial degree and interpolation set, we compute the distance between the high-order surface mesh and the limit model, see \autoref{eq:distanceMeshModel}. In \autoref{tab:ConvergenceSphere}, we report the distances between the meshes and the limit model, for the different degrees and nodal distributions. In \autoref{fig:ConvergenceSphere}, we plot the logarithm of the distance in terms of the polynomial degree using red for the data of the equispaced distribution and blue for the data of the warp-and-blend distribution. We observe that for the warp-and-blend distribution the logarithm of the distance seems to converge linearly with the polynomial degree $\Degree$. In \autoref{tab:ConvergenceSphere}, we also show the Lebesgue constant of the equispaced distribution, $\LebEq$, and of the warp-and-blend distribution, $\LebWB$. If we denote by $\BestApp$ the best approximating polynomial of the limit model $\PhiLim$, since it is optimal, we have the inequality
\[
\norm{\PhiLim - \BestApp }_{\infty} \leq \norm{\PhiLim - \PhiDeg}_{\infty}.
\]
Furthermore, for a given interpolation set, the Lebesgue constant bounds the interpolation error in terms of the error made by the best approximating polynomial,
\begin{equation}
\norm{\PhiLim - \PhiDeg}_{\infty} \leq \left( 1 + \LebCt \right) \norm{\PhiLim - \BestApp }_{\infty}.
\label{eq:LebesgueBound}
\end{equation}
Combining these inequalities, we bound the error made by $\BestApp$,
\begin{equation}
\frac { \norm{\PhiLim - \PhiDeg}_{\infty} }{1 + \LebCt} \leq \norm{\PhiLim - \BestApp }_{\infty} \leq \norm{\PhiLim - \PhiDeg}_{\infty}.
\label{eq:BoundsBestApp}
\end{equation}
The term on the left is denoted as $\LowBoundEq^{\Degree}$ and $\LowBoundWB^{\Degree}$ for the equispaced and warp-and-blend distribution, respectively, and they are also reported in \autoref{tab:ConvergenceSphere}. In \autoref{fig:SphereBounds}, we plot the lower and upper bound for the two interpolation sets. For low polynomial degrees, both distributions provide similar results. However, as the polynomial degree increases, the equispaced distribution becomes more uncertain, while the warp-and-blend distribution is more accurate and the region is more restricting. The error of the best approximating polynomial $\BestApp$ is between the two bounds and, more concretely, in the more restricting region. In this particular example, this region is given by the bounds of the warp-and-blend distribution and, in general, we would expect the same behavior since the warp-and-blend distribution has better interpolation properties. 

We might intuitively expect that for the higher polynomial degrees, equidistant point distributions lead to significant error oscillations. Nevertheless, in this example, we can see that both point distributions lead to almost comparable errors. To understand this error similarity, we can use the upper bound of the interpolation error in terms of the Lebesgue constant, see \autoref{eq:LebesgueBound}. In particular, for polynomial degree 10, if we divide the upper bound of the error for both point distributions, we see that equidistant points feature an error at most $71 / 7$ times larger than the error for warp and blend points. This error relation, around $10$, agrees with the numerical results of  \autoref{tab:ConvergenceSphere}.

Finally, in \autoref{fig:SphereHighOrderDistanceToLimit}, we illustrate the mesh of polynomial degree $\Degree = 10$ colored according to the pointwise distance to the limit model. It can be observed that, in the neighborhood of the regular vertices, the mesh exactly captures the limit model and, consequently, the distance is zero. On the contrary, around irregular vertices, the mesh approximates the limit model and the distance is non-zero. 

\subsection{Volume Curving of a Structured Topography Mesh}

\begin{figure*}[t]
\centering
\begin{tabular}{cc}
\begin{subfigure}[t]{0.49\textwidth}
\centering
\includegraphics[width=\textwidth]{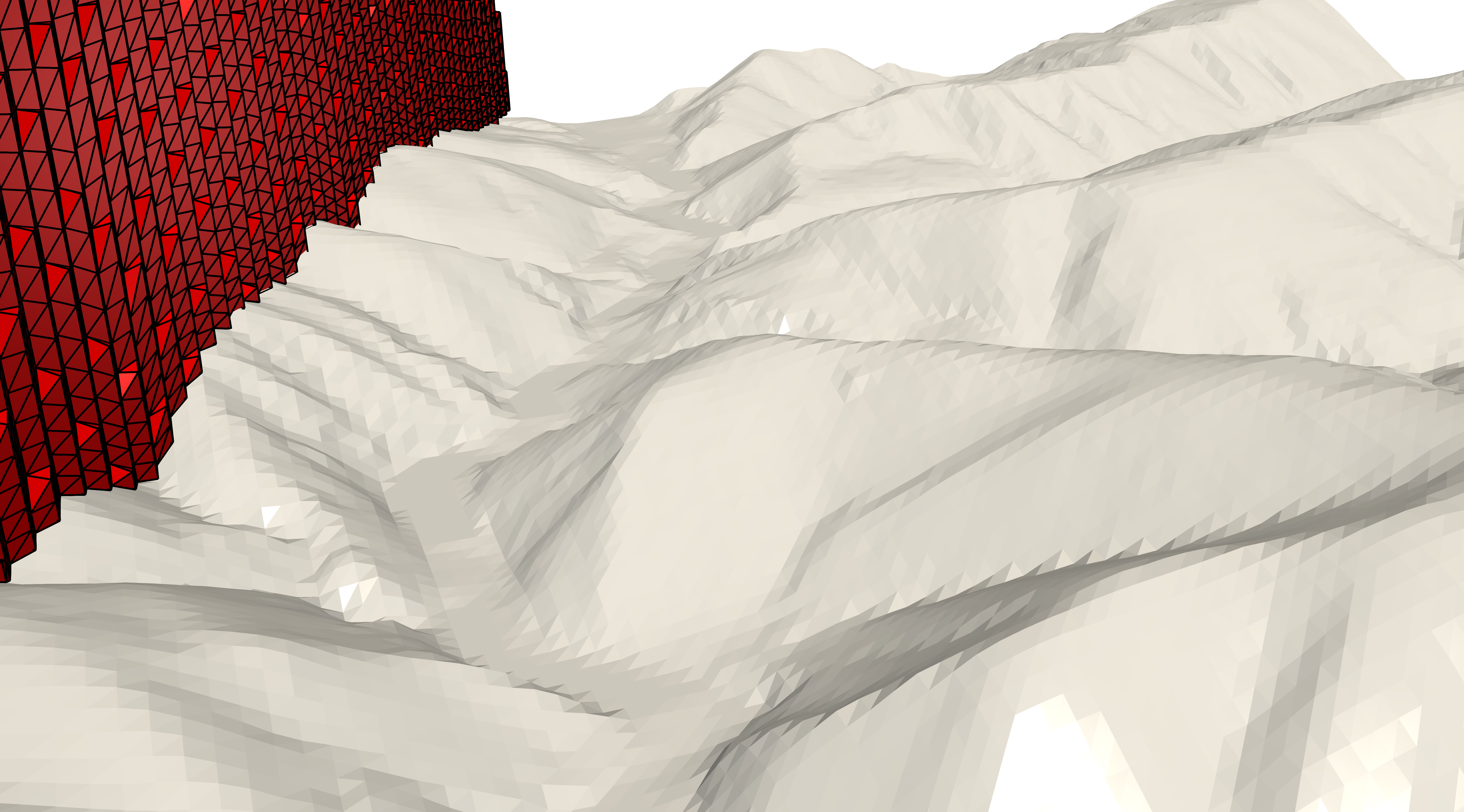}
\caption{}
\label{fig:Escudo_Linear}
\end{subfigure}
&
\begin{subfigure}[t]{0.49\textwidth}
\centering
\includegraphics[width=\textwidth]{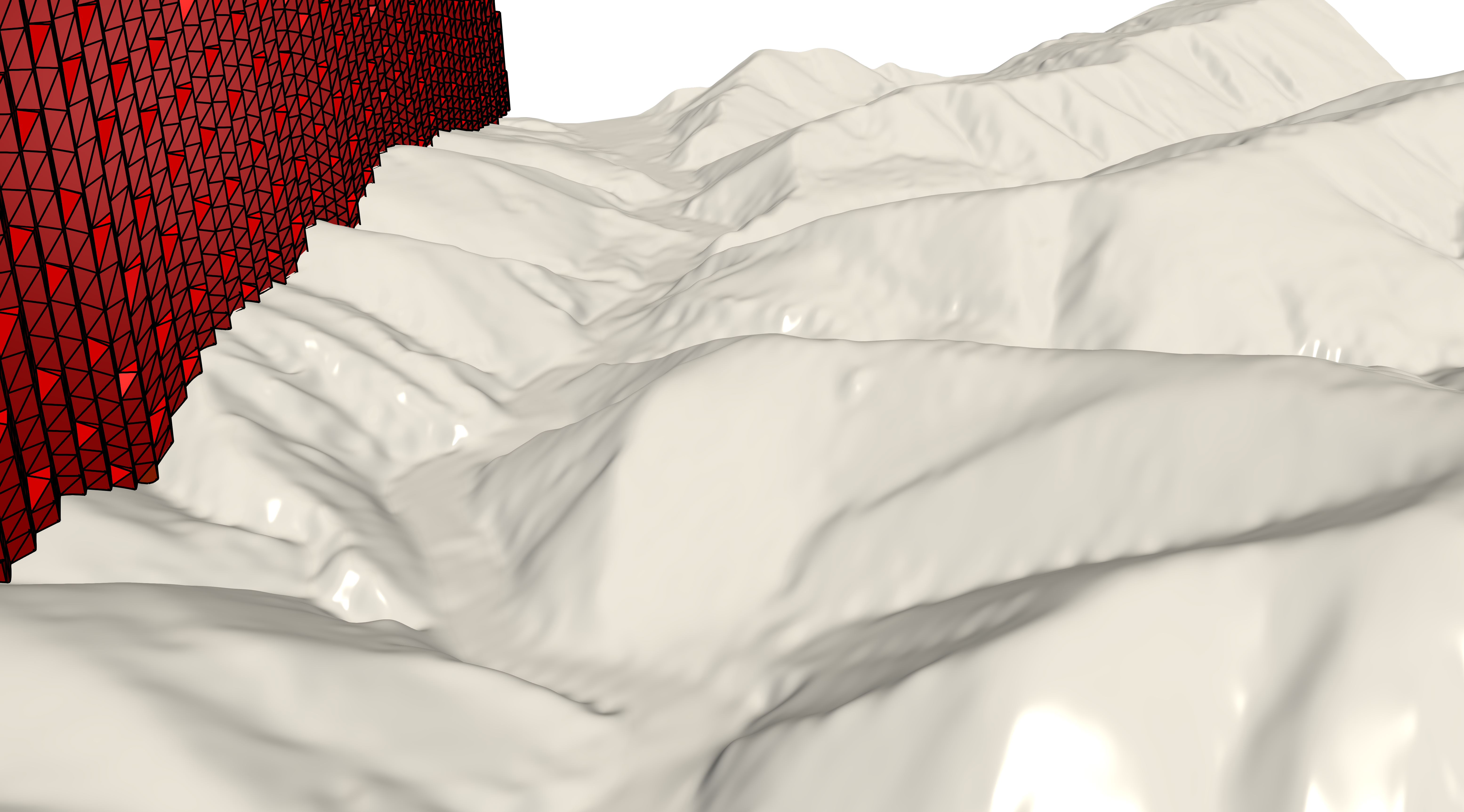}
\caption{}
\label{fig:Escudo_HO}
\end{subfigure}
\end{tabular}
\centering
\begin{subfigure}{\textwidth}
\centering
\includegraphics[width=0.25\textwidth]{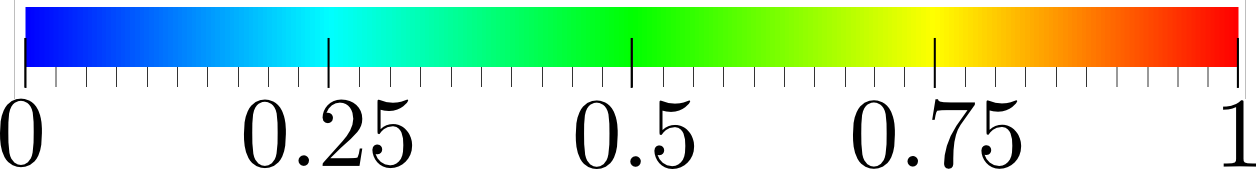}
\end{subfigure}
\caption{Curving of a tetrahedral mesh of the Escudo mountain range (Spain).  \subref{fig:Escudo_Linear}{ Linear mesh.} \subref{fig:Escudo_HO}{ Curved mesh of polynomial degree $\Degree=4$.}
Elements of the volume meshes are colored with their elemental quality.
} 
\label{fig:Escudo}
\end{figure*}

\begin{table}[t]
\centering
\begin{tabular}{lcc}
\hline
\multicolumn{1}{c}{} & Min Q & \# inv\\
\hline
Boundary & 0.93 & 0 \\
Volume (no TFI) & 0 & 169 \\
Volume (TFI)& 0.80& 0 \\
\hline
\end{tabular}
\caption{Quality statistics of a mesh of polynomial degree $\Degree=4$ for the Sierra del Escudo (Spain) topography.}
\label{tab:TopoStats}
\end{table}

In this example, we illustrate the features of our method with a structured linear mesh that discretizes the topography on the Escudo mountain range (Spain). The original data is provided as a level curve map that is cast to a structured mesh with the level values on the nodes.

We consider a tetrahedral mesh that discretizes the region enclosed by the topography and a planar ceiling, located at the desired height. The tetrahedral mesh is regular and is generated using the mesher presented in \cite{gargallo:meshForABLandWindFarms,gargallo2018mesh}. From the linear mesh, we generate a curved high-order mesh of polynomial degree four using the procedure detailed in \autoref{sec:volume_ho}. Since the surface nodes are regular, the high-order topography surface mesh is $\mathcal{C}^2$-continuous and exactly captures the limit surface, as detailed in \autoref{sec:SmoothnessHOSurfaceMesh}. To check the implementation, the distance 
between the mesh of polynomial degree four and the limit model, see \autoref{eq:distanceMeshModel}, has been computed, verifying that the result is zero.

In \autoref{fig:Escudo_Linear}, we show the initial linear mesh, composed of $4.8 \cdot 10^5$ nodes and $2.6 \cdot 10^6$ tetrahedra. The high-order boundary mesh is generated in 106 minutes and is composed of $1.1 \cdot 10^6$ nodes and $1.4 \cdot 10^5$ triangles, featuring a minimum quality of 0.93. The curved high-order volume mesh takes 18 minutes to be generated and contains 169 inverted elements. The process of accommodating the curvature of the boundary detailed in \autoref{sec:blending} is performed to $4.1 \cdot 10^5$ elements abutting the boundary. This blending untangles all the invalid elements in 7 minutes, attaining a minimum quality of 0.8. Finally, in \autoref{fig:Escudo_HO}, we show the mesh of polynomial degree four composed of $2.9 \cdot 10^7$ nodes and $2.6 \cdot 10^6$ elements. The statistics regarding the mesh quality on the different steps of the mesh generation procedure are presented in \autoref{tab:TopoStats}.

\subsection{Smoothness of the Boundary Mesh}
\label{sec:NormalsSphere}

\begin{table}[t]
\centering
\begin{tabular}{cccccc}
	\hline
	$\Degree$ & 1 & 2 & 3 & 4 & 5\\
	\hline
	$\max_{\Edge} \NormalFunction_{\infty, \Edge}$ & 9.00& 5.49 & 1.72  & 1.54 & 1.02\\
	\hline
\end{tabular}
\caption{Maximum angle between the normal vectors of adjacent elements for the surface mesh of the sphere of polynomial degree $\Degree$, $\Degree = 1, \dotsc, 5$.}
\label{tab:NormalsSphere}
\end{table}

In this example, we analyze the smoothness of the interpolative model $\ModelDegree$ for the mesh of the sphere in \autoref{sec:ExampleSphere}. When the polynomial degree is greater than four, $\Degree \geq 4$, the surface mesh features $\mathcal{C}^2$-continuity around regular patches, see \autoref{sec:SmoothnessHOSurfaceMesh}. Thus, normal vectors are continuous along regular edges. On the contrary, we may observe discontinuous normal vectors along edges around irregular points which indicate that the mesh is not $\mathcal{C}^1$-continuous.

To quantify the smoothness of the surface mesh, we compute the distortion of the normal vector. Specifically, consider the $\IndexSurface$-th surface $\SurfaceDegree$ composed of triangular facets, $\SurfaceDegree = \bigcup_{\Facet = 1}^{n_{\Facet}} \Tri^{\Degree}_{\Facet}$. Then, for each internal edge $\Seg^{\Degree}_{\Edge}$ such that $\Seg^{\Degree}_{\Edge} = \Tri^{\Degree}_i \cap \Tri^{\Degree}_j$, we compute 
\[
\NormalFunction_{\infty, \Edge} = \max_{k=1, \dotsc, N} \NormalFunction \left( \PointPhysical_k \right),
\]
where $\PointPhysical_k \in \Seg^{\Degree}_{\Edge}$ is a sampling point, and $ \NormalFunction \left( \PointPhysical \right)$ is defined in  \autoref{eq:AngleNormalVectors}. The value $\NormalFunction_{\infty, \Edge}$ approximates the maximum angle formed by the normal vectors along the edge $\Seg^{\Degree}_{\Edge}$.

In \autoref{tab:NormalsSphere}, we show the maximum angle between normal vectors for the mesh of the sphere in \autoref{sec:ExampleSphere} for different polynomial degrees. We observe a significant improvement with cubic meshes with respect to the original linear mesh. For quartic and quintic polynomial meshes, the maximum angle formed by two normal vectors from adjacent elements is less than two degrees. 

\subsection{Sharp-to-smooth Modeling and Mesh Volume Curving}
\label{sec:ExampleFalcon}

\begin{figure*}[t]
\renewcommand{\figsize}{0.40}
\centering
\begin{tabular}{cc}
\begin{subfigure}[t]{\figsize\textwidth}
\centering
\includegraphics[width=\textwidth]{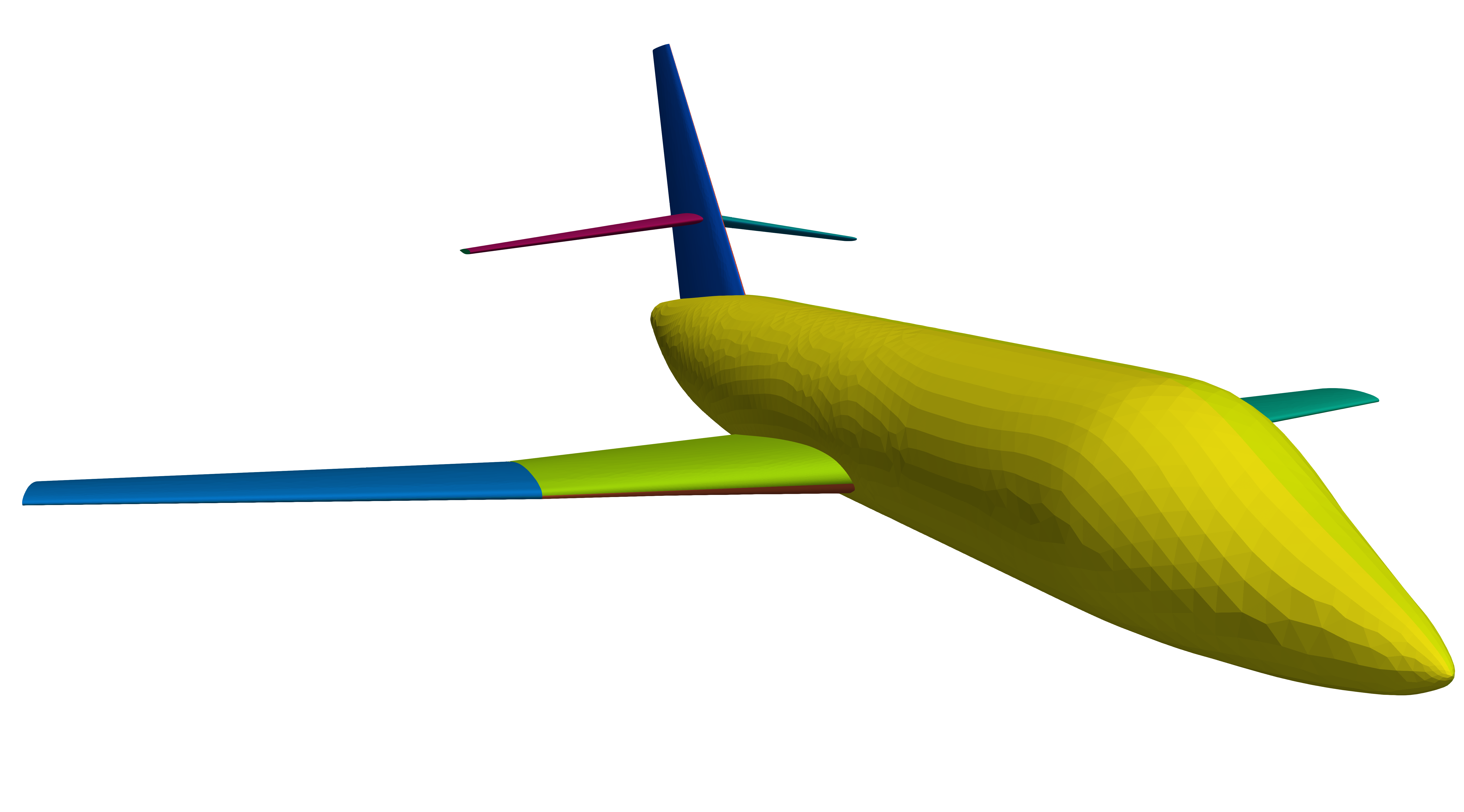}
\caption{}
\label{fig:Falcon.Surface.Colors}
\end{subfigure}
&
\begin{subfigure}[t]{\figsize\textwidth}
\centering
\includegraphics[width=\textwidth]{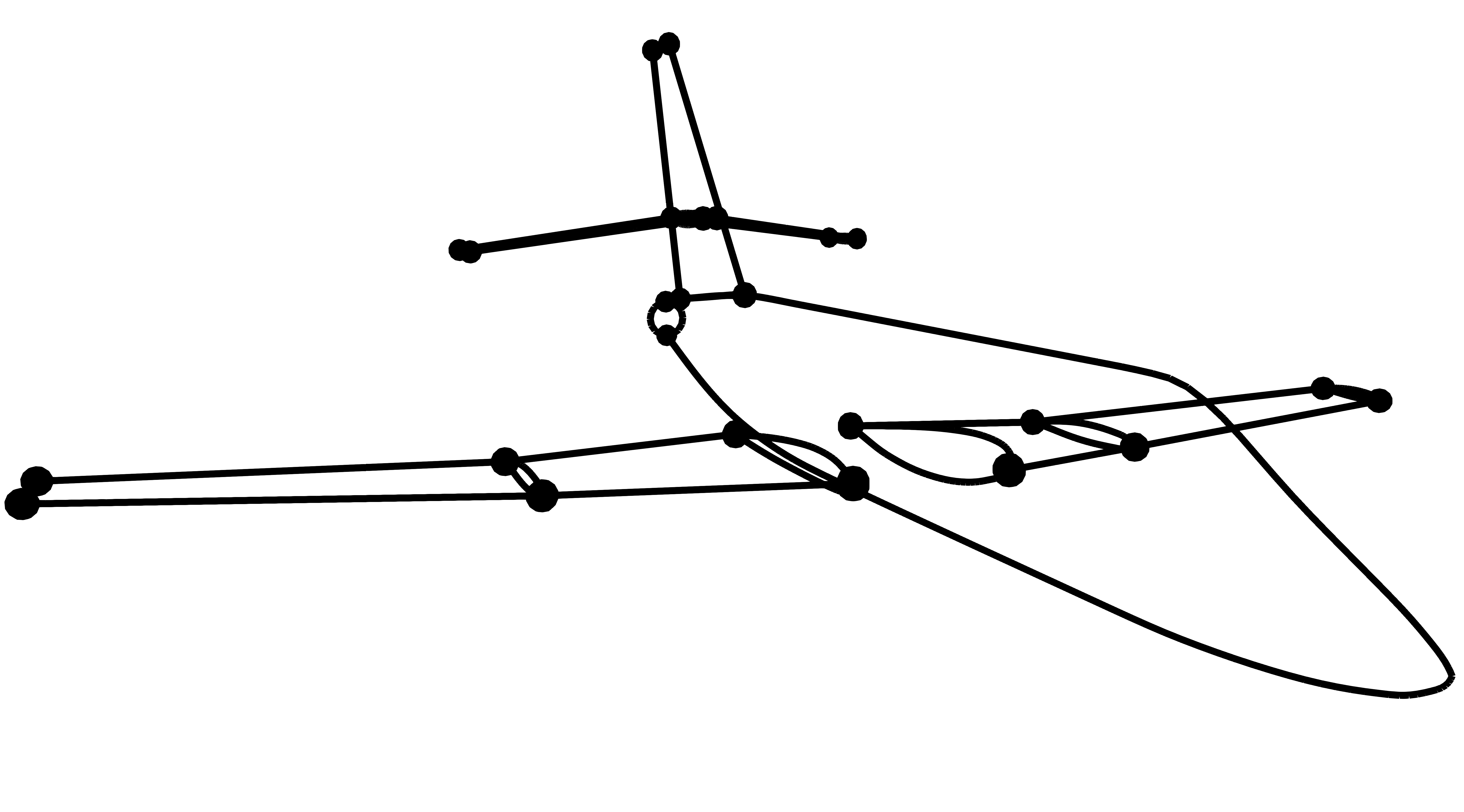}
\caption{}
\label{fig:Falcon.Curves.Colors}
\end{subfigure}
\\
\begin{subfigure}[t]{\figsize\textwidth}
\centering
\includegraphics[width=\textwidth]{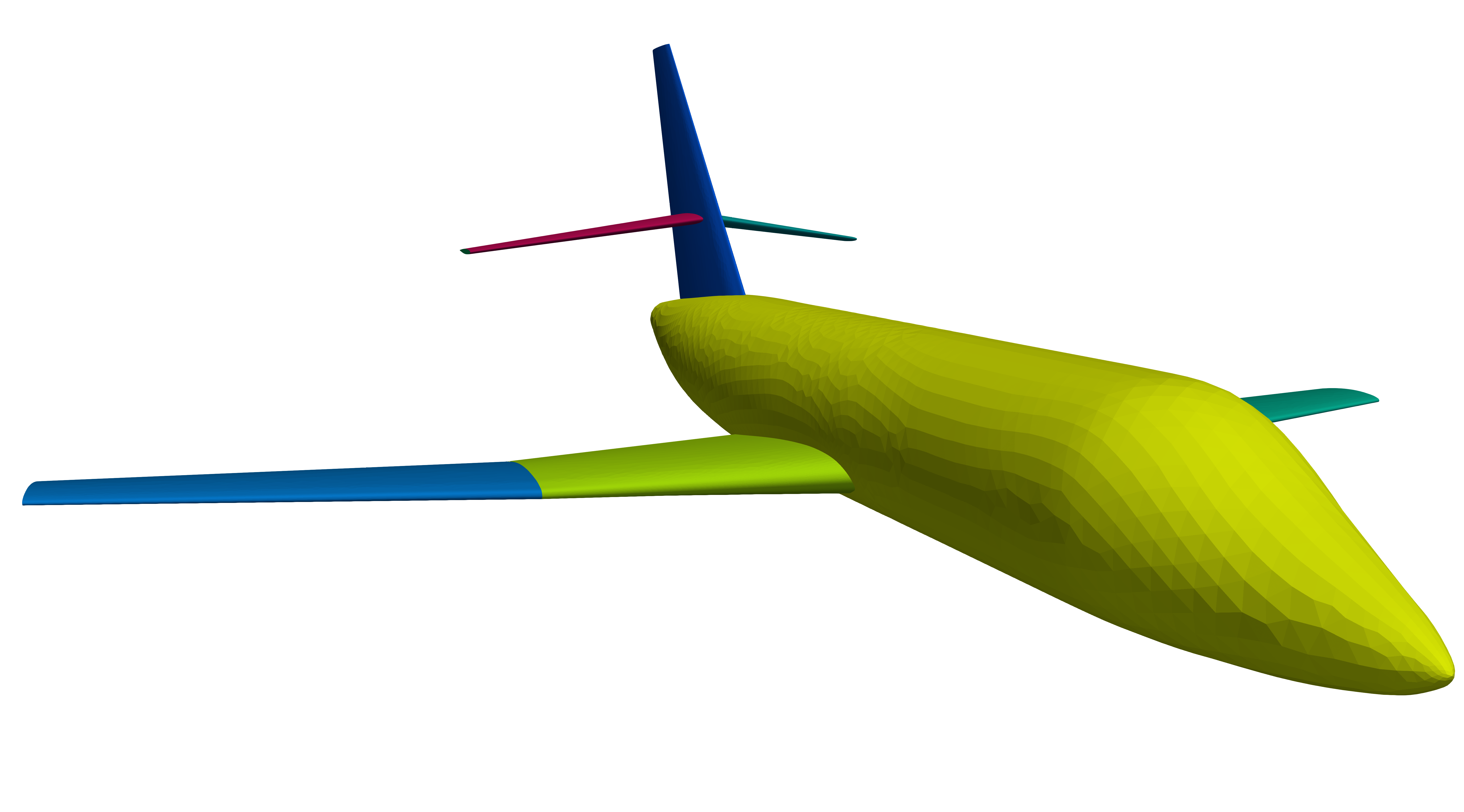}
\caption{}
\label{fig:Falcon.Surface.Model}
\end{subfigure}
&
\begin{subfigure}[t]{\figsize\textwidth}
\centering
\includegraphics[width=\textwidth]{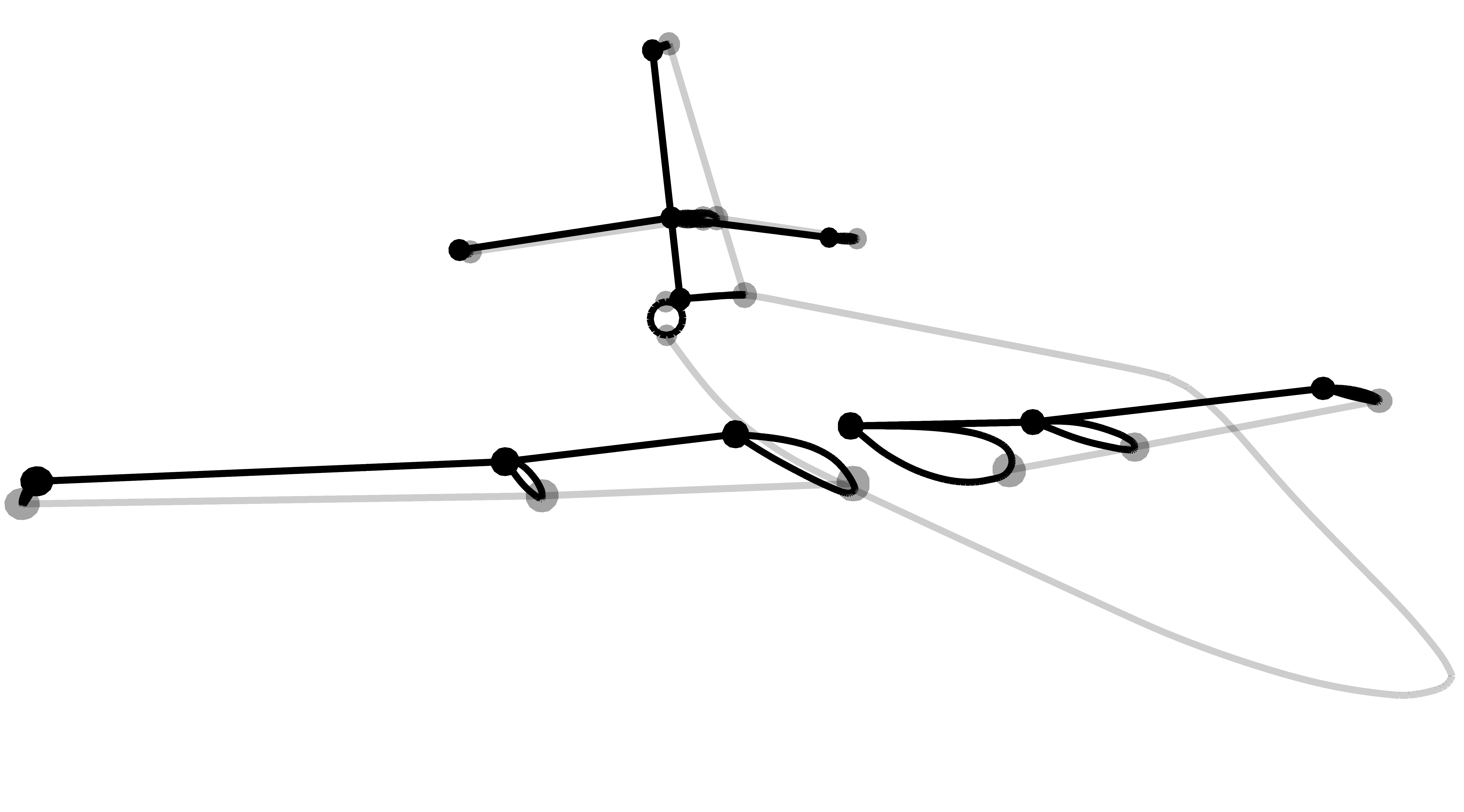}
\caption{}
\label{fig:Falcon.Curves.Model}
\end{subfigure}
\end{tabular}
\caption{Initial and final linear mesh model of a Falcon aircraft. Initial model: \subref{fig:Falcon.Surface.Colors} surface features colored with their surface identifier, and \subref{fig:Falcon.Curves.Colors} curve and point features. Final model: \subref{fig:Falcon.Surface.Model} virtual surface features colored with their surface identifier, \subref{fig:Falcon.Curves.Model} curve and point features smoothed (gray) and preserved (black).} 
\label{fig:Falcon}
\end{figure*}

\begin{figure*}[t]
\centering
\begin{tabular}{cc}
\begin{subfigure}[t]{0.49\textwidth}
\centering
\includegraphics[width=\textwidth]{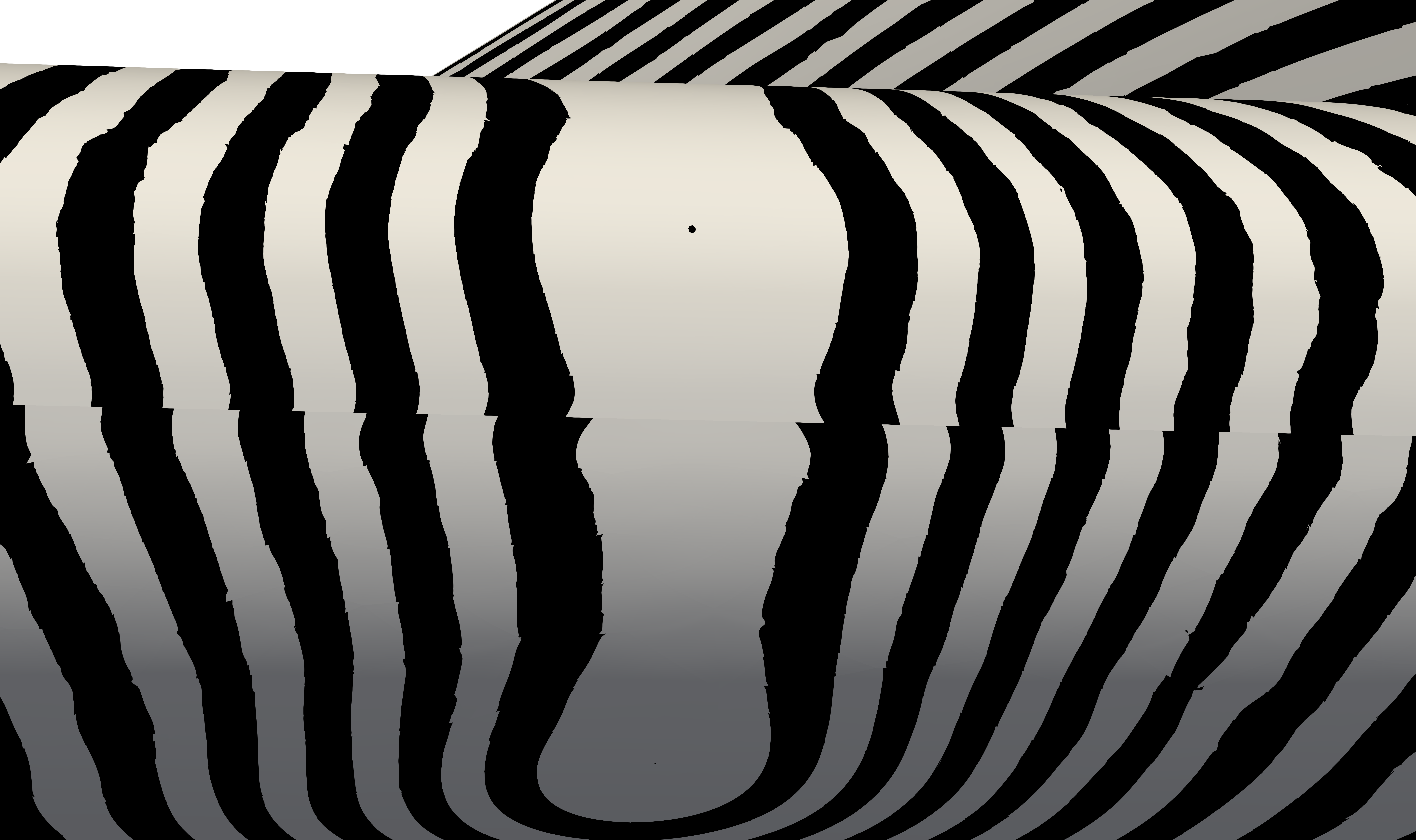}
\caption{}
\label{fig:Falcon.Wing}
\end{subfigure}
&
\begin{subfigure}[t]{0.49\textwidth}
\centering
\includegraphics[width=\textwidth]{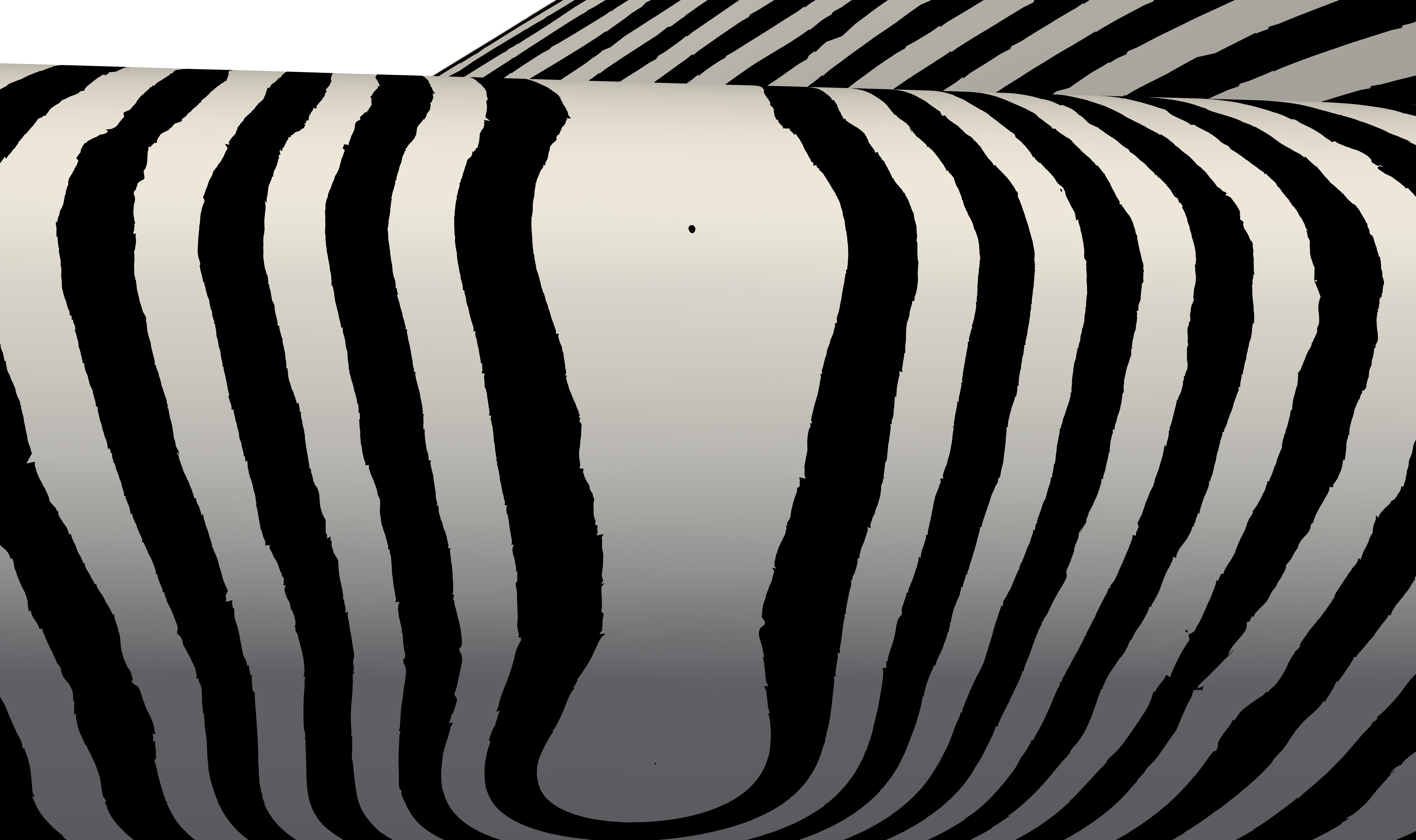}
\caption{}
\label{fig:Falcon.Wing.ErasedEntities}
\end{subfigure}
\end{tabular}
\caption{Close-up view of the leading edge of a Falcon aircraft wing. Zebra mapping on the mesh of polynomial degree four with \subref{fig:Falcon.Wing} the initial model, and \subref{fig:Falcon.Wing.ErasedEntities} the final model with the leading edge smoothed.}
\label{fig:Falcon.Wing.Example}
\end{figure*}

\begin{figure}[t]
\centering
\renewcommand{\figsize}{0.5}
\includegraphics[width=\figsize\textwidth]{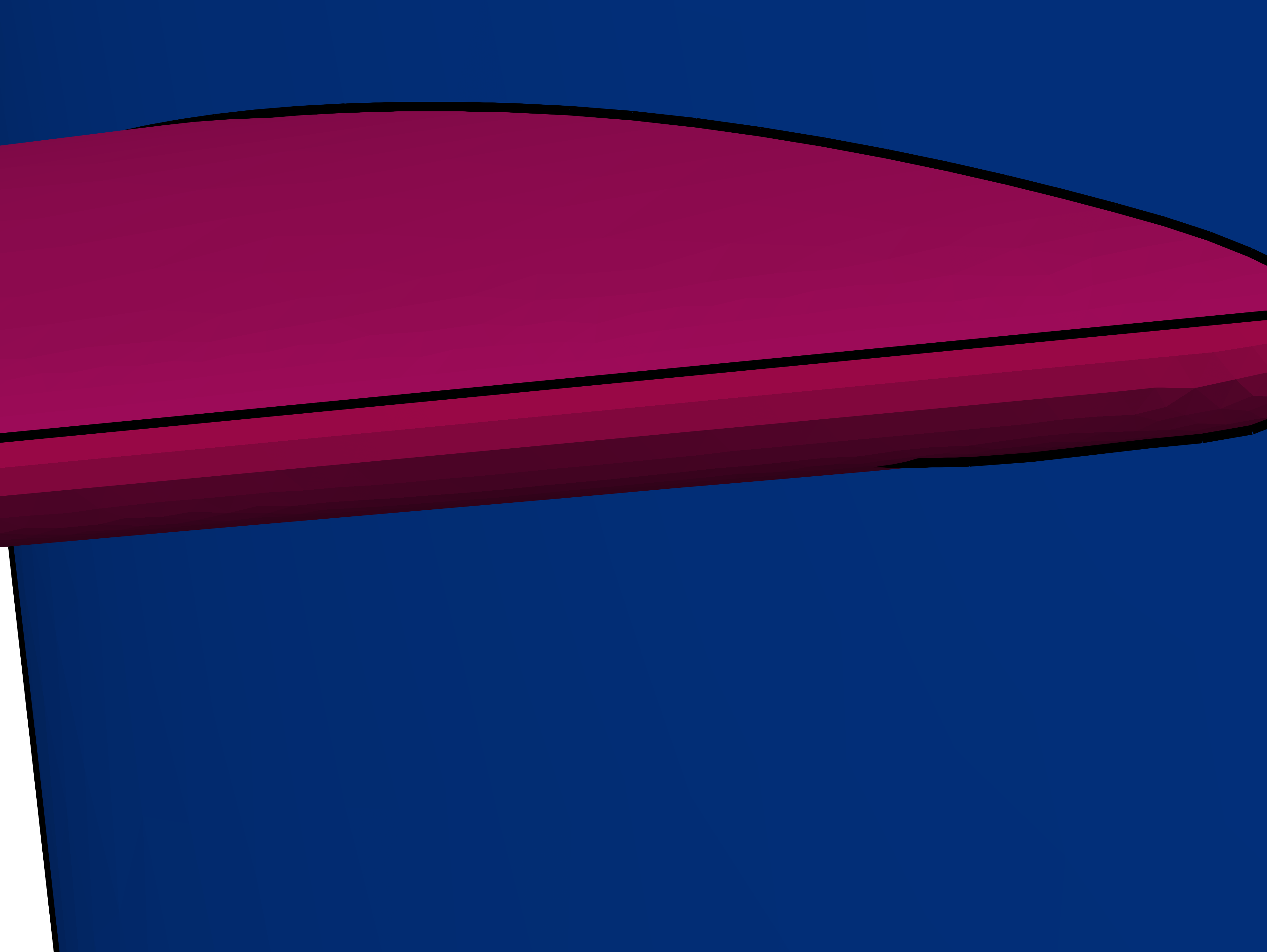}
\caption{Close-up view of the leading edge of a Falcon aircraft rear wing. Surfaces are colored according to their surface identifier.}
\label{fig:Falcon.RepairCAD}
\end{figure}

\begin{table}[t]
\centering
\begin{tabular}{cccc}
\hline
$\Degree$ & $\LebWB$ & $\DistWB$ & $\LowBoundWB$ \\
\hline
1  & $1.00$    & $1.48 \cdot 10^{-2}$       & $7.42 \cdot 10^{-4}$      \\
2  & $1.66$ & $4.09 \cdot 10^{-4}$ & $1.54 \cdot 10^{-4}$ \\
3  & $2.11$ & $1.98 \cdot 10^{-4}$ & $6.38 \cdot 10^{-5}$ \\
4  & $2.66$ & $9.93 \cdot 10^{-5}$ & $2.71 \cdot 10^{-5}$ \\
5  & $3.12$ & $5.14 \cdot 10^{-5}$ & $1.25 \cdot 10^{-5}$ \\
\hline
\end{tabular}
\caption{Interpolation with non-equispaced nodes of a Falcon aircraft limit model for polynomial degree $\Degree$, $\Degree = 1, \dotsc, 5$: Lebesgue constant, distance to the limit model, and lower bound of the distance.}
\label{tab:FalconDistance}
\end{table}

\begin{figure}[t]
\centering
\renewcommand{\figsize}{0.5}
\includegraphics[width=\figsize\textwidth]{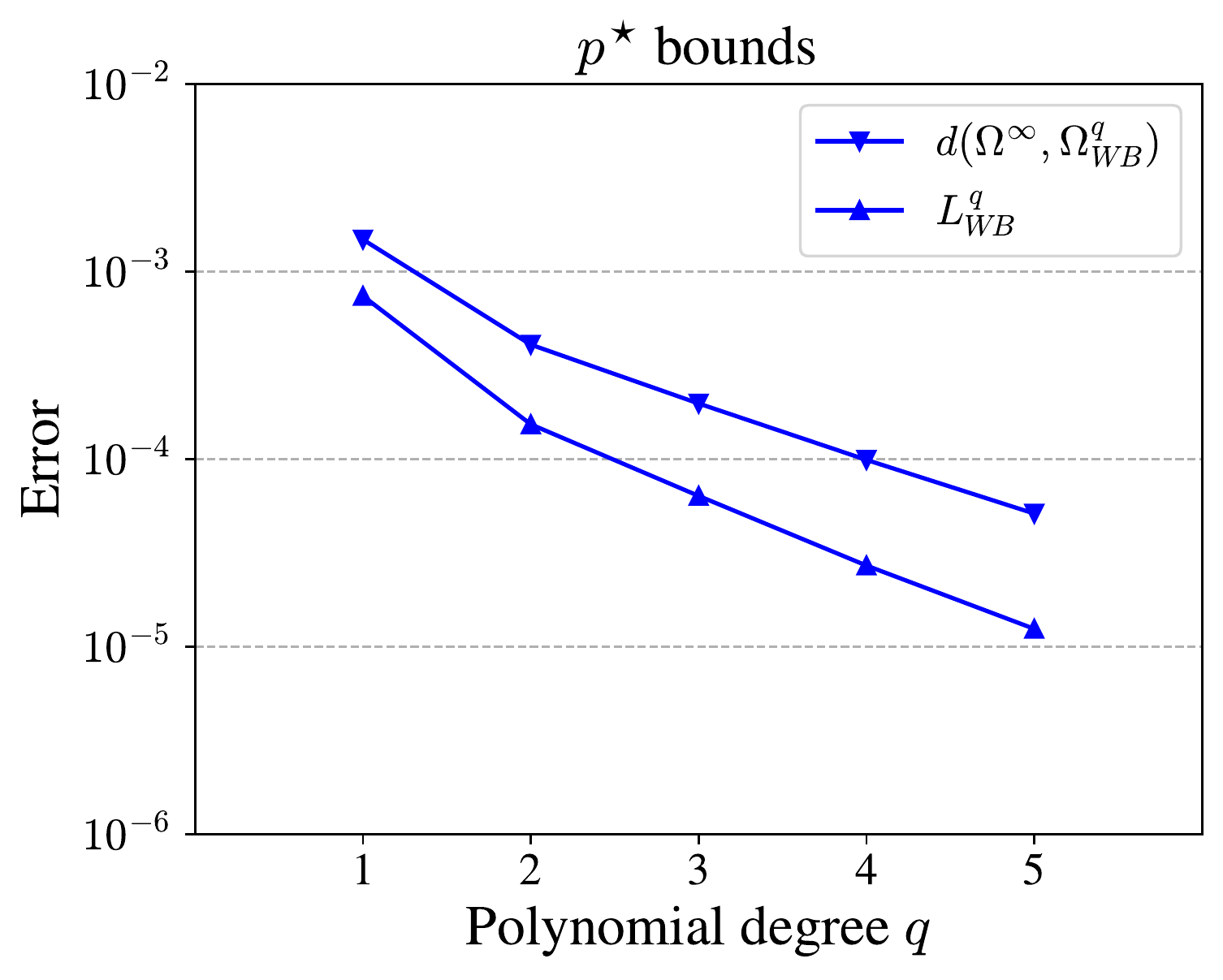}
\caption{Lower and upper bounds of the distance of the best approximating polynomial in terms of the polynomial degree $\Degree$ for the Falcon aircraft.}
\label{fig:Falcon.Convergence}
\end{figure}

\begin{figure}[t]
\centering
\renewcommand{\figsize}{0.5}
\includegraphics[width=\figsize\textwidth]{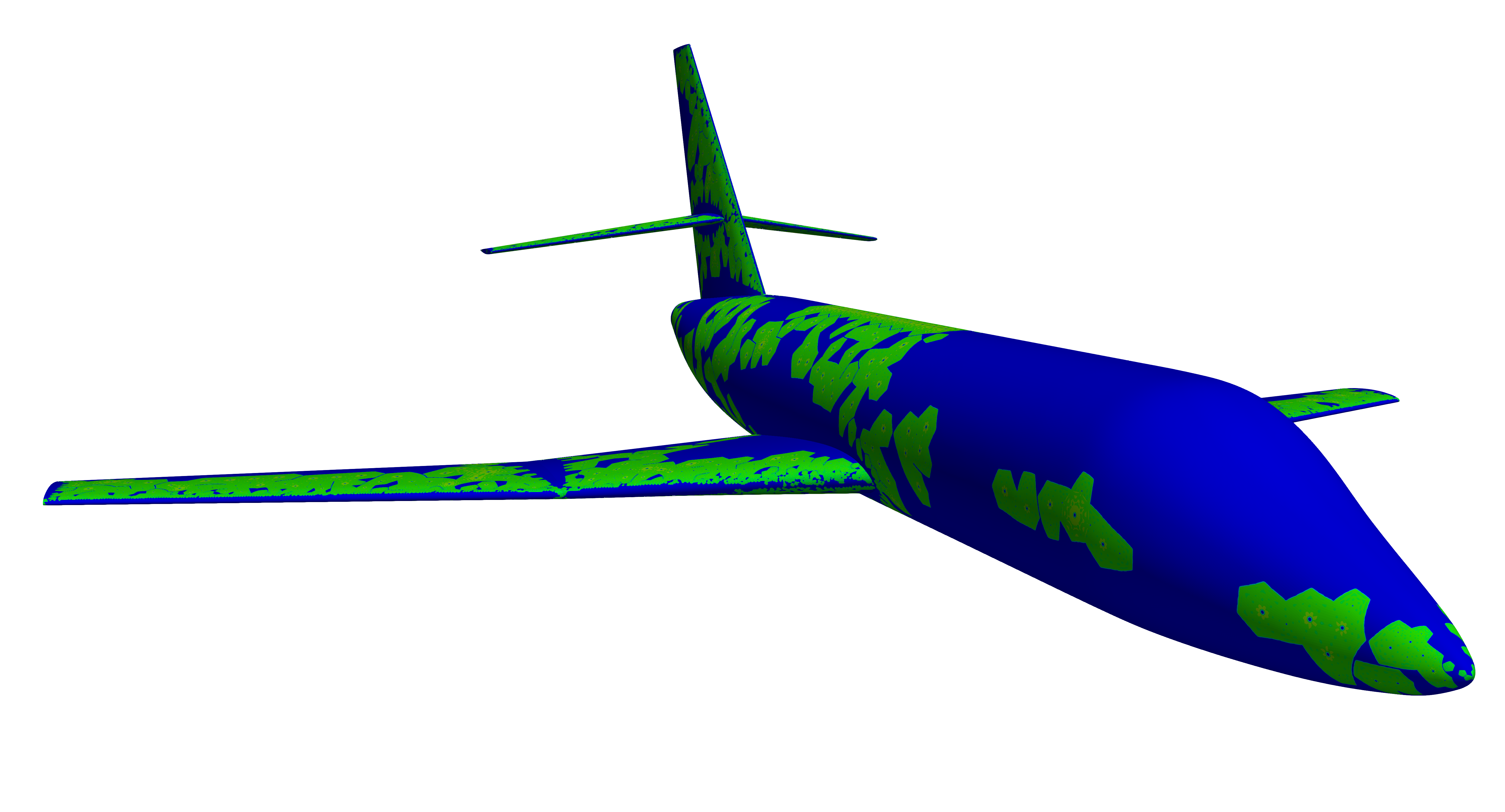}
\begin{subfigure}{\textwidth}
\centering
\includegraphics[width=0.4\textwidth]{ColorBarDiscretized_Horizontal.png}
\end{subfigure}
\caption{Mesh of polynomial degree $\Degree = 5$ of a Falcon aircraft colored according to the distance to the limit model.}
\label{fig:Falcon.DistanceToLimit}
\end{figure}

In this example, we illustrate the capability of our method to perform a sharp-to-smooth modeling in different features of the geometry. To preserve the simulation intent, we smooth some of the feature entities present in the original model and thus provide a new model improving the smoothness of the surrogate geometry. Each feature point (node of the mesh), curve (set of edges of the mesh), and surface (set of triangles of the mesh) is associated with a unique identifier. Therefore, to smooth a feature, it is enough to know its identifier.

We consider a linear tetrahedral mesh from a CAD legacy model of a simplified Falcon aircraft. The boundary triangles are marked identifying the feature surfaces. The complete linear model is composed of 28 surfaces, 54 curves, and 34 points.
As shown in \autoref{fig:Falcon.Surface.Colors} and \autoref{fig:Falcon.Curves.Colors}, the main part of the fuselage is composed of two surfaces and a curve. However, such curve is not desirable since, ideally, we would desire a smooth model along each section of the fuselage. To address this issue, we smooth the curve indicating its unique identifier. The first step consists in removing the curve from the list of feature curves. Following, the two surfaces initially incident to this curve, see \autoref{fig:Falcon.Surface.Colors}, are merged by identifying the identifiers of the two surfaces with a new but equal identifier, see \autoref{fig:Falcon.Surface.Model}. As a result, the whole fuselage is modeled as a smoother virtual surface.

Similarly, we observe that each section of the wing is described by a top and bottom surface, and a leading and trailing edge curve. To preserve the simulation intent, we smooth the feature curve describing the leading edge. This way, the surface at the top and the bottom are merged and join smoothly in the front part of the wing. We highlight that the curve describing the trailing edge is maintained as a sharp feature.

Note that the lateral wing joins the fuselage in a profile described by two curves (top and bottom) and two vertices (front and back). We smooth the feature point in the front. Thus, this point is removed from the list of feature points, and the two curves are merged by identifying their identifiers as a unique one. As a result, we obtain a single closed curve with a sharp endpoint on the trailing edge. 

Similar changes are made in similar features of the mesh to generate the proper model for flow simulation, see \autoref{fig:Falcon.Surface.Model} and \autoref{fig:Falcon.Curves.Model}. As highlighted in \autoref{sec:curved_volume}, once the identifiers of all the features to smooth are located, the smoothing process is straight-forward. Given the list of identifiers, the smoothing process consists in removing these features from the list of features to preserve and automatically identify the features adjacent to the smoothed feature as a single one. We remark that the smoothing of some of the geometry features does not modify the mesh, only the number of feature points, curves, and surfaces changes. Specifically, the original model of the presented Falcon aircraft contains 34 feature points, 54 curves, and 28 surfaces; while the model with the smoothed features contains 20 points to preserve, 32 curves, and 20 surfaces.

To illustrate the difference between these two models, we take a close look at the leading edge of the wing. In \autoref{fig:Falcon.Wing}, we show a zebra mapping on the mesh of polynomial degree $\Degree=4$ generated with the initial marks. Specifically, we see the isophote bands for a spotlight in front of the leading edge. We observe a discontinuity in the normal vector of the wing along the leading edge. In \autoref{fig:Falcon.Wing.ErasedEntities}, we show a mesh generated with a model in which the leading edge has been smoothed. The nodes originally present in the leading edge are still on the leading edge, but the new points are generated to interpolate the almost everywhere $\mathcal{C}^2$-continuous surrogate geometry. The second mesh model is smoother than the original one along those regions where the features have been smoothed. The leading edge now belongs to the interior of the surface, and therefore, all the nodes interpolate a $\mathcal{C}^1$-continuous surface. 

The modeling capability not only allows for the smoothing of geometric features that prevented the model from describing the simulation intent, but it also removes the sharp features that were artificially created by CAD engines. That is, it is possible to repair a CAD model with artificial features. In this example, we observe that the intersection of the two surfaces (top and bottom) describing the rear wing does not occur exactly at the leading edge of the wing, but at the top of the wing, see \autoref{fig:Falcon.RepairCAD}. Indeed, this curve does not appropriately describe the simulation intent and should be smoothed. To this end, we proceed as described above by locating the curve identifier, removing the curve from the list of feature curves to preserve, and automatically assigning the same identifier to the adjacent surfaces.

In \autoref{tab:FalconDistance}, for each polynomial degree $\Degree$, $\Degree = 2, \dotsc 5$, we report the distances between the limit model and the high-order surface meshes, see \autoref{eq:distanceMeshModel}. In this example, $L\left(\ModelLimit\right)$ is set to the aircraft length. Note that as the polynomial degree $\Degree$ increases, the distance to the limit model is reduced. In \autoref{fig:Falcon.Convergence}, we plot the lower and upper bound of the best approximation polynomial, see \autoref{eq:BoundsBestApp}. The upper curve coincides with the logarithm of the distance which seems to converge linearly with the polynomial degree. In \autoref{fig:Falcon.DistanceToLimit}, we show the point-wise distance between the mesh of polynomial degree five and the limit model. Note that around the regular vertices, the distance between the limit model and the surface mesh of polynomial five is zero. Around the irregular vertices, the distance is smaller than $10^{-4}$.

\begin{table}[t]
	\centering
	\begin{tabular}{cccccc}
		\hline
		$\Degree$ & 1 & 2 & 3 & 4 & 5\\
		\hline
		$\max_{\Edge} \NormalFunction_{\infty, \Edge}$ & 63.92 & 24.08 & 8.19  & 5.04 & 5.60\\
		\hline
	\end{tabular}
	\caption{Maximum angle between the normal vectors of adjacent elements for the surface mesh of the Falcon aircraft of polynomial degree $\Degree$, $\Degree = 1, \dotsc, 5$.}
	\label{tab:NormalsFalcon}
\end{table}

\begin{figure}[t]
	\centering
	\begin{tabular}{cc}
		\begin{subfigure}[t]{0.49\textwidth}
			\centering
			\includegraphics[width=\textwidth]{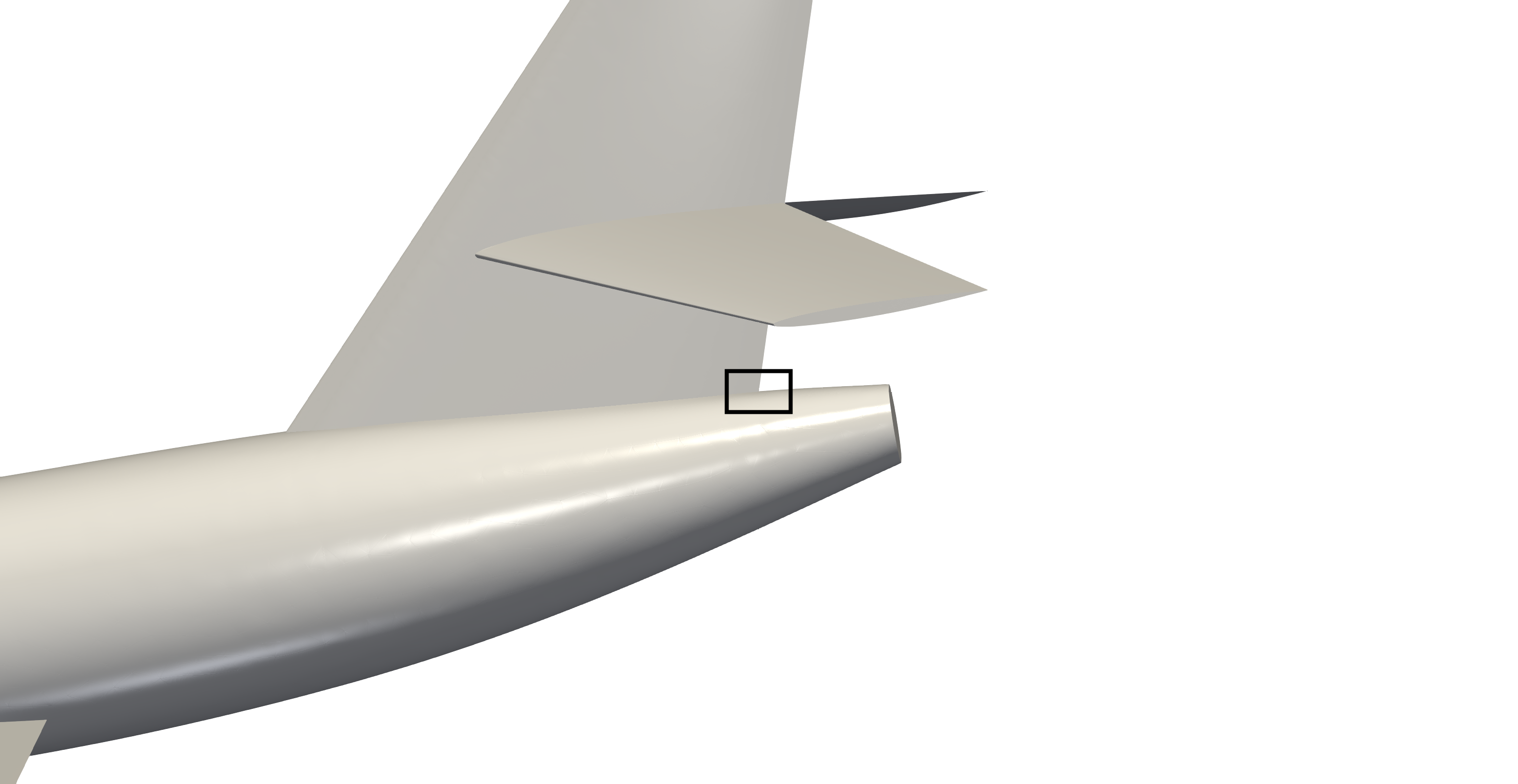}
			\caption{}
			\label{fig:Falcon.Fin.Big}
		\end{subfigure}
		&
		\begin{subfigure}[t]{0.49\textwidth}
			\centering
			\includegraphics[width=\textwidth]{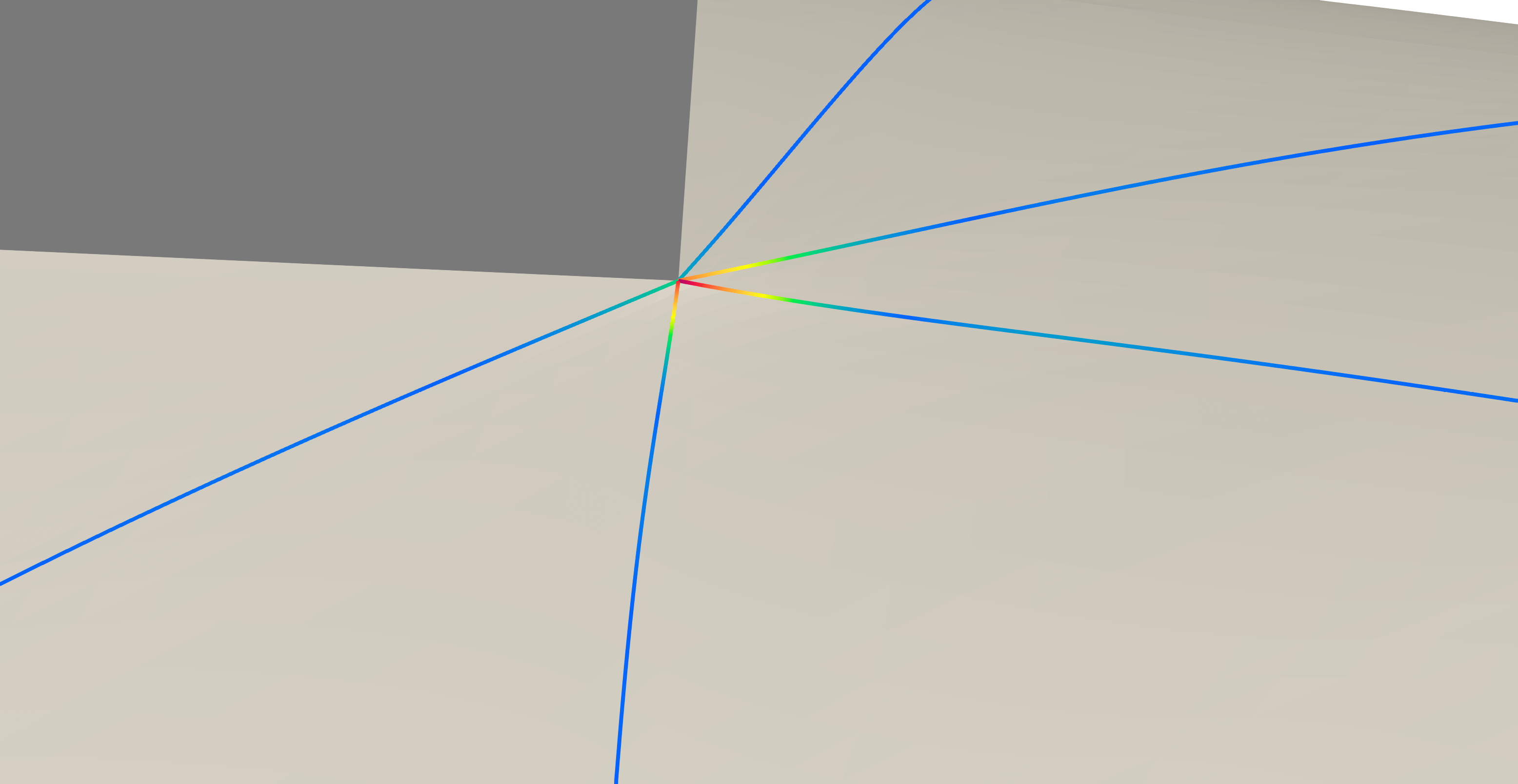}
			\includegraphics[width=0.5\textwidth]{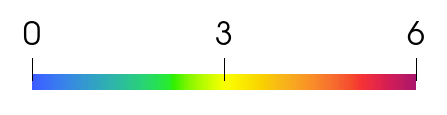}
			\caption{}
			\label{fig:Falcon.Fin.ZoomAngle}
		\end{subfigure}
	\end{tabular}
	\caption{Close look of the rear point of the vertical stabilizer of the mesh of polynomial degree $\Degree = 5$ of a Falcon aircraft. \subref{fig:Falcon.Fin.Big} Rear part of the aircraft. \subref{fig:Falcon.Fin.ZoomAngle} Angle between the normal vectors along the interior edges incident to the sharp point.}
	\label{fig:Falcon.Fin}
\end{figure}

The maximum angle between normal vectors along interior edges is reduced significantly as we increase the polynomial degree, see \autoref{tab:NormalsFalcon}. The linear mesh features an angle of 60 degrees along the leading edge of the rear wing. In contrast, the mesh of polynomial degree five attains a maximum angle smaller than 2 degrees in the same region. 
However, the interpolative property of the method affects the smoothness of the high-order mesh.  For instance, consider the rear point where the vertical stabilizer joins the fuselage, see \autoref{fig:Falcon.Fin}. All the high-order nodes of the triangles around the sharp point interpolate the $\mathcal{C}^1$-continuous limit surface determined by the surface subdivision scheme, except for the vertices that coincide with the sharp point, which remain fixed. In that region, we can only guarantee $\mathcal{C}^0$-continuity and the angle between the normal vectors is greater than 5 degrees.

\begin{table}[t]
	\centering
	\begin{tabular}{lcc}
		\hline
		\multicolumn{1}{c}{} & Min Q & \# inv\\
		\hline
		Boundary & 0.55 & 0  \\
		Volume (no TFI) & 0 & 2179  \\
		Volume (TFI)& 0 & 24  \\
		\multirow{2}{*}{\shortstack[l]{Volume \\ (TFI + Optimization)}} & \multirow{2}{*}{0.69} & \multirow{2}{*}{0} \\ & & \\
		\hline
	\end{tabular}
	\caption{Quality statistics of a mesh of polynomial degree $\Degree=5$ for a Falcon aircraft.}
	\label{tab:FalconStats}
\end{table}

\begin{figure*}[t]
	\centering
	\includegraphics[width=0.8\textwidth]{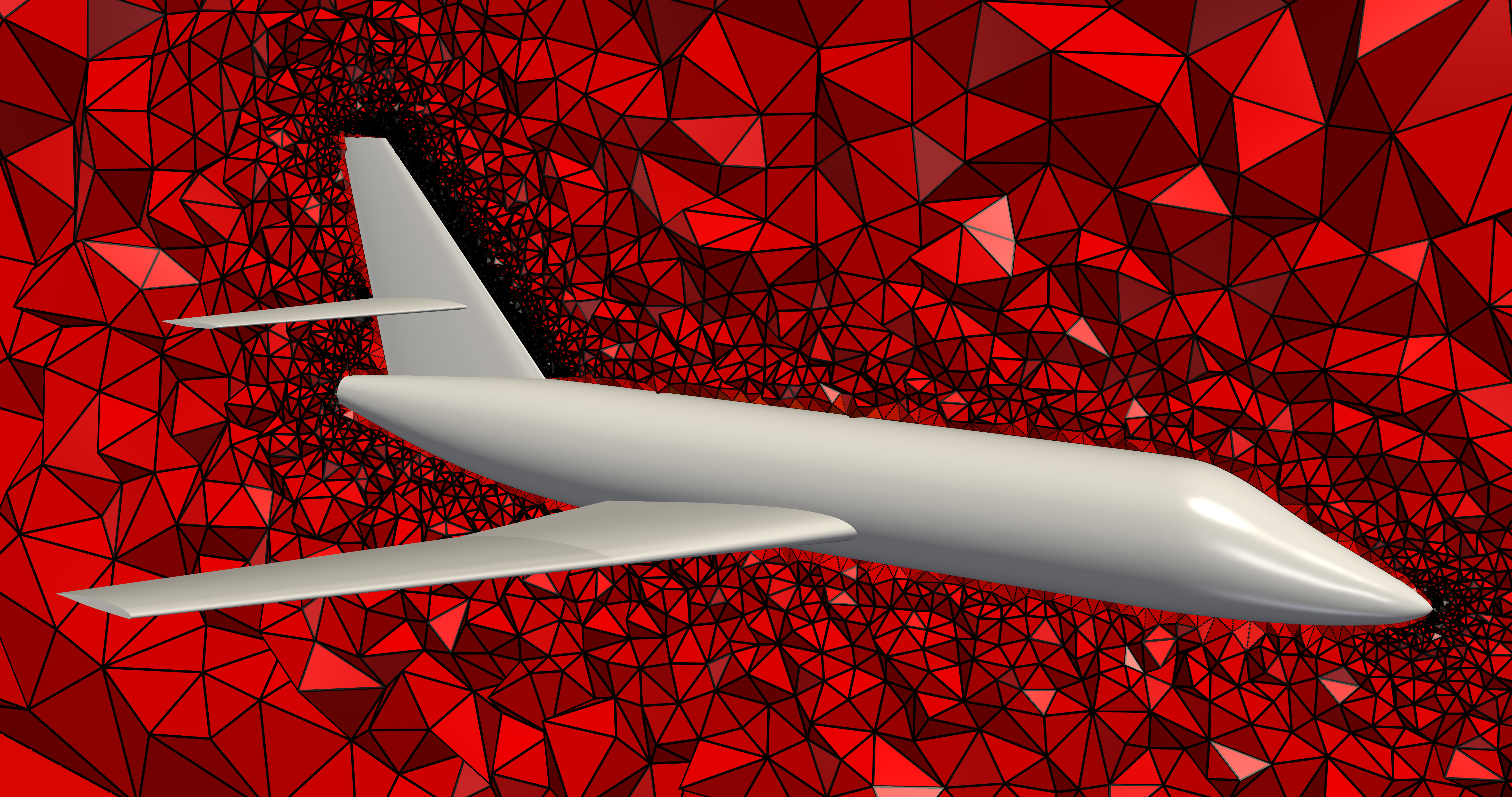}
	\centering
	\begin{subfigure}{\textwidth}
		\centering
		\includegraphics[width=0.25\textwidth]{qualBarParaview_color.png}
	\end{subfigure}
	\caption{Curved tetrahedral mesh of polynomial degree $\Degree=5$ of a Falcon aircraft with no invalid elements.}
	\label{fig:Falcon.HO}
\end{figure*}

A summary of the mesh quality can be found in \autoref{tab:FalconStats}. The linear mesh is composed of $3.1 \cdot 10^5$ nodes and $1.7 \cdot 10^6$ tetrahedra, and the volume mesh of polynomial degree $\Degree=5$ is generated in 6.5 hours and is composed of $3.6 \cdot 10^7$ nodes and $1.7 \cdot 10^6$ elements. This mesh, prior to the blending technique, contains 2179 tangled elements. In this example, there are $3.0 \cdot 10^5$ boundary elements and in 8 minutes the TFI reduces to 24 the number of invalid elements, that is, 99\% of the invalid elements have been untangled. Now, we apply the optimization technique presented in \cite{Gargallo-Peiro2015,gargallo2015optimization} to optimize locally the quality of the inverted elements. Since the mesh after the TFI is close to optimal, it is a good initial condition for the implicit optimization. This process takes 41 minutes and the mesh becomes valid achieving a minimum quality of 0.69. In \autoref{fig:Falcon.HO}, we show the valid curved tetrahedral mesh of polynomial degree $\Degree=5$ with the volume elements colored according with their quality.

\subsection{Assisted Sharp-to-smooth Boundary Modeling}
\label{sec:HighLiftExample}

\begin{figure*}[t]
\renewcommand{\figsize}{0.45}
\centering
\begin{tabular}{cc}
\begin{subfigure}[t]{\figsize\textwidth}
\centering
\includegraphics[width=\textwidth]{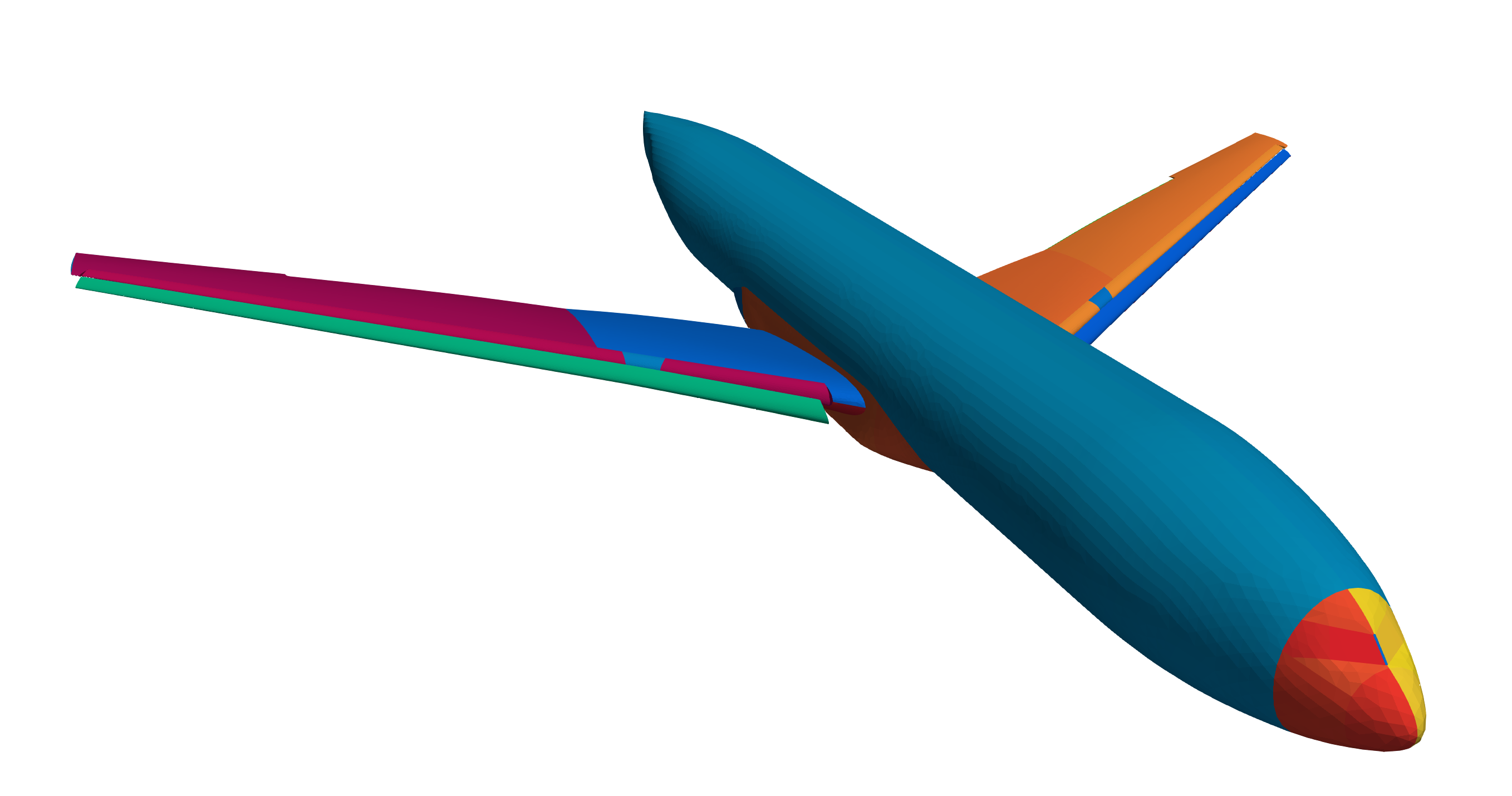}
\caption{}
\label{fig:HighLift.Surface.Colors}
\end{subfigure}
&
\begin{subfigure}[t]{\figsize\textwidth}
\centering
\includegraphics[width=\textwidth]{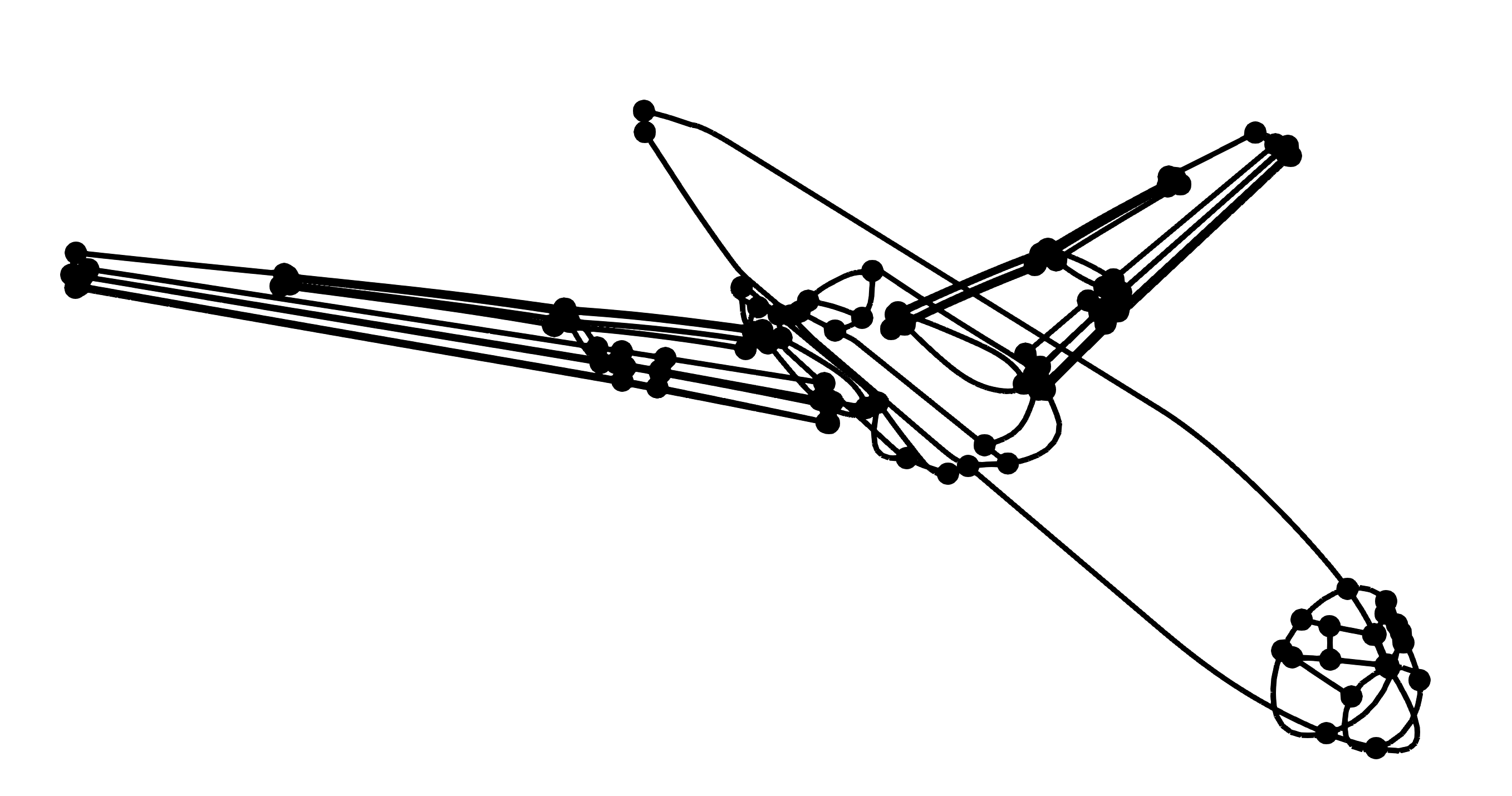}
\caption{}
\label{fig:HighLift.Curves.Colors}
\end{subfigure}
\\
\begin{subfigure}[t]{\figsize\textwidth}
\centering
\includegraphics[width=\textwidth]{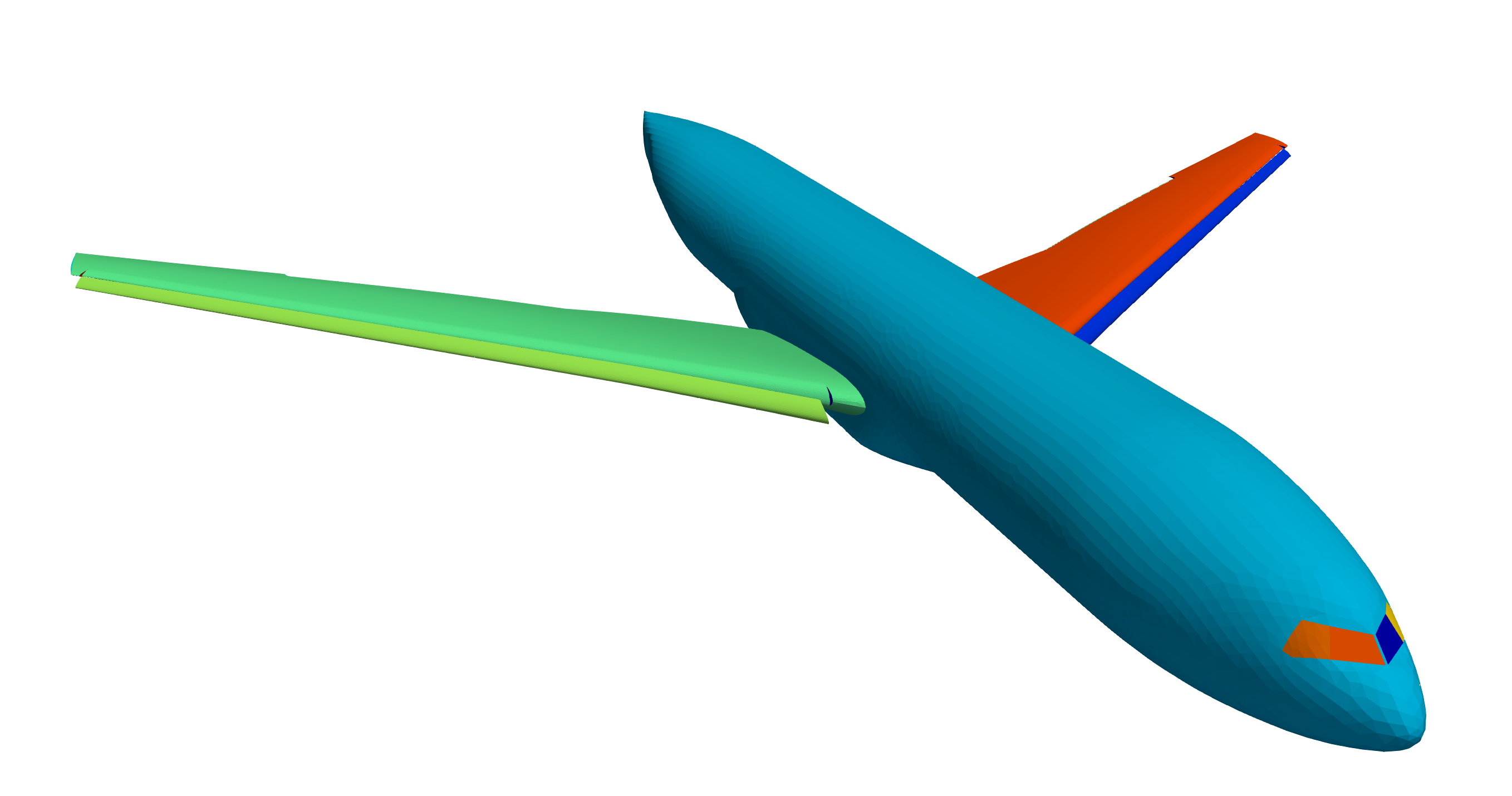}
\caption{}
\label{fig:HighLift.Surface.Model}
\end{subfigure}
&
\begin{subfigure}[t]{\figsize\textwidth}
\centering
\includegraphics[width=\textwidth]{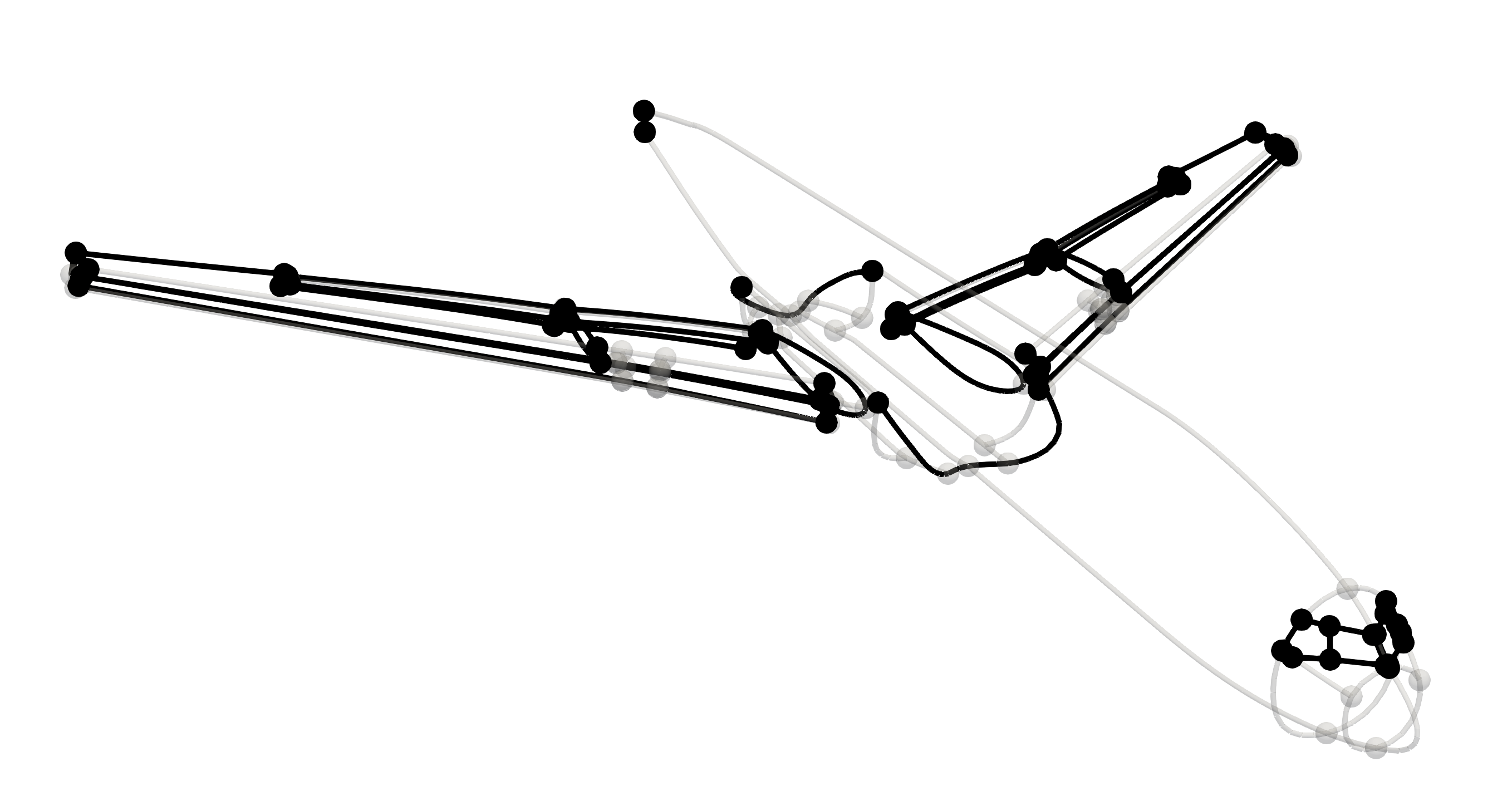}
\caption{}
\label{fig:HighLift.Curves.Model}
\end{subfigure}
\end{tabular}
\caption{Initial and final linear mesh model of an aircraft in high-lift configuration. Initial model: \subref{fig:HighLift.Surface.Colors} surface features colored with their surface identifier, and \subref{fig:HighLift.Curves.Colors} curve and point features. Final model: \subref{fig:HighLift.Surface.Model} virtual surface features colored with their surface identifier, \subref{fig:HighLift.Curves.Model} curve and point features smoothed (gray) and preserved (black).} 
\label{fig:HighLift}
\end{figure*}

\begin{figure*}[t]
\renewcommand{\figsize}{0.85}
\centering
\includegraphics[width=\figsize\textwidth]{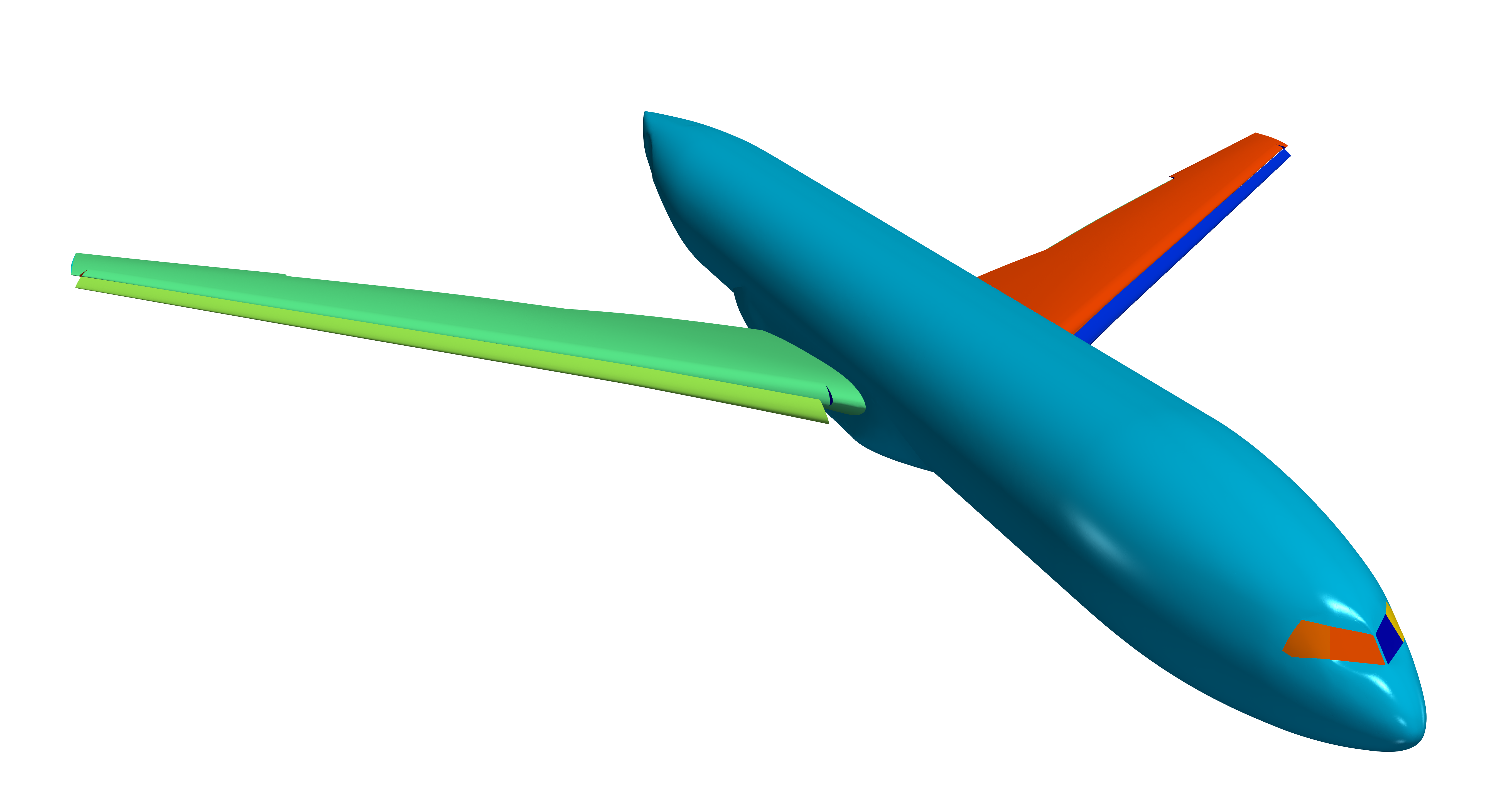}
\centering
\caption{Curved triangular mesh of polynomial degree $\Degree=4$ of an aircraft in high-lift configuration colored with the surface identifiers.}
\label{fig:HighLift.HO}
\end{figure*}

As detailed in \autoref{sec:AutomaticFeatureDetection}, it is possible to automatically suggest to the practitioners the features to smooth. In this example, we consider a mesh-based boundary representation of an aircraft in high-lift configuration, a common research model presented in the 3rd High-Lift Prediction Workshop \cite{rumsey2019overview}. The model, see \autoref{fig:HighLift.Surface.Colors} and \autoref{fig:HighLift.Curves.Colors}, is represented by 182 feature points, 282 curves, and 118 surfaces. For models with a significant complexity, we use the automatic feature detection capability detailed in \autoref{sec:AutomaticFeatureDetection} to reduce the human labor required to design a model preserving the simulation intent.

First, using the original model, we generate a surface mesh of polynomial degree $\Degree = 4$. Then, for each feature curve, we compute the normal vector along the curve and decide whether this curve has to be smoothed. The threshold used in this example is $\delta = 17$ degrees. We highlight that this is a parameter that can be easily tuned to determine the curves in the set $R_{\delta}$. Next, we manually decide if these candidates have to be indeed smoothed or not. Some of the curves present in this set are the ones representing the leading edge of the wings and flaps that indeed have to be smoothed. However, there are some curves in the set that we do not want to smooth. For example, we do not smooth the two curves that describe where the swept wing starts to angle, and the curves that represent the cabin windows. Thus, these curves are not smoothed. Second, once the model has been updated, we generate again the mesh to detect the smooth points. The final model is composed of 122 feature points, 169 curves, and 67 surfaces, see \autoref{fig:HighLift.Surface.Model} and \autoref{fig:HighLift.Curves.Model}. 

In \autoref{fig:HighLift.HO}, we show the generated mesh of polynomial degree four using the smoothed model where the surfaces have been colored with their surface identifier. Since the curves describing the leading edge have been automatically smoothed, the wing surface describes the simulation intent. On the contrary, we can appreciate the desired discontinuity on the normal vectors of the surface along the curves describing the cabin windows.

\section{Discussion}
\label{sec:discussion}

After detailing the methods and results to interpolate the subdivision features for curved geometry modeling, we present a discussion on many aspects related to: $\mathcal{C}^1$-continuous shape functions, evaluating a limit curve around a feature point, the feature detection threshold, the TFI node relocation, polynomial degrees, and exploiting successive subdivisions.

\paragraph{$\mathcal{C}^{1}$-continuous shape functions} In this work, we indirectly approximate the $\mathcal{C}^{1}$-continuity of the limit surface by using interpolative triangular elements. An alternative approach might be to directly impose $\mathcal{C}^{1}$-continuity using the Argyris \cite{argyris1968tuba} or the Bell \cite{bell1969refined} triangular elements. Both elements have polynomial degree 5 and feature one degree of freedom per edge, and six degrees of freedom per vertex. The degrees of freedom of the vertices correspond to the value, the two first derivatives, and the three second derivatives required to determine the symmetric Hessian. Since the limit surface has $\mathcal{C}^{1}$-continuity everywhere, it is possible to query at the vertices the value and the first derivatives from the limit surface, and set the corresponding vertex degrees of freedom. However, since at the irregular points the limit surface might not be $\mathcal{C}^{2}$-continuous, the Hessian at the irregular points of the limit surface is not defined. Since it is undefined, to apply this approach, one would need to decide a criterion to force a Hessian at the irregular nodes. Herein, to not force the Hessian, we favor the numerical approximation of the $\mathcal{C}^{1}$-continuous limit surface.

\paragraph{Limit curve around a sharp point} To evaluate the limit curve close to a sharp point, we can replace the successive refinement in \autoref{alg:ParameterizationLimitManifold_curve} with an explicit evaluation. To perform this evaluation, we can consider a stencil featuring three segments and four points. To this end, one straightforward approach is to add to the original control poly-line a ghost segment on the other side of each sharp point. Each ghost segment is limited by the original sharp point and an additional point having the same coordinates. Now, we can evaluate the standard limit curve expression on this new control poly-line since the original segment becomes internal. Other evaluation alternatives are detailed in \cite{de1978practical}.

\paragraph{Feature detection threshold} The automatic feature detector presented in \autoref{sec:AutomaticFeatureDetection} uses the same threshold $\delta$ for the whole mesh. Even though the current normal disparity is a helpful tool to set up a ranking of sharp features, it might be worth considering a variable weight accounting for the local mesh size. This improvement may be studied in the near future.

\paragraph{TFI node relocation} In \autoref{sec:blending}, we present a technique to accommodate the curvature of the surface mesh to the volume elements adjacent to the boundary. By discarding the TFI node relocation, we could simplify the mesh curving procedure by directly untangling and optimizing the volume mesh. However, the TFI node relocation pre-process helps, for isotropic meshes, to accelerate the whole curving process. This is so since the TFI relocation is explicit, fast, and provides a fair initial approximation. This approximation reduces the number of non-linear iterations required by the implicit and slower optimization solver. Finally, note that in all the examples, the TFI relocation reduces the number of invalid elements.

\paragraph{Polynomial degree} In this work, we present a method to interpolate the limit model with arbitrary polynomial degree. In practice, one might prefer polynomial degree four. This preference for degree four is so since the interpolation reproduces the $\mathcal{C}^{2}$-continuity of the limit model around the regular points. Furthermore, we have seen that around the irregular points with degree four, the interpolative approximation to the limit model has a sufficiently small error. If we need a smaller error, we can increase the polynomial degree of the limit model approximation. Nevertheless, this approximation cannot be guaranteed to be $\mathcal{C}^{1}$-continuous around irregular points. Thus, since the limit model is just a geometry surrogate, degree four could be adequate for practical purposes.

\paragraph{Exploiting successive subdivisions} Renouncing to interpolate the limit model, we could simplify the interpolation of subdivision features by skipping the limit model parameterization \cite{Evaluation-of-Loop-Subdivision-Surfaces}. To this end, we could interpolate the limit model with a polynomial degree equal to a power of two on equispaced nodal distribution exploiting the structure of the subdivision scheme \cite{jimenez_ramos_albert_2020_3653357}. Then, we would interpolate this surrogate with an inferior polynomial degree to obtain an alternative approximation. The resulting surface mesh might not interpolate the limit model, but it might provide similar accuracy measurements.

\section{Concluding Remarks}
\label{sec:concludingRemarks}

We have proposed a method to interpolate a subdivision model with arbitrary degree and nodal distribution. The method evaluates one time the limit model and thus, skips successive refinement on posterior geometry queries. The results show that we can generate, from an initial straight-edged mesh, a piece-wise polynomial mesh that approximates smooth curves and surfaces while preserving the required sharp features. We also show that the resulting curved high-order surface mesh is ready to prescribe the boundary in mesh curving methods. We have to highlight that for regular meshes, as it is the case for digitized topographies, the resulting curved surface mesh has class $\mathcal{C}^2$. 

We have numerical evidence that the proposed distance from the nodal interpolation to the limit model converges geometrically with the polynomial degree for nodal distributions with sub-optimal Lebesgue constant. Hence, we can expect accurate approximations of the limit model for sufficiently high polynomial degrees. We understand that this geometrical convergence rate, to a parameterization with up to continuous derivatives, seems to be originated by the natural alignment of the triangulation entities with the subdivision continuity interfaces. That is, irregular points of the initial triangulation are aligned with $\mathcal{C}^1$ irregular points of the limit model. Analogously, the interior of segments and triangles of the initial triangulation are aligned with  the interior of  the $\mathcal{C}^2$ segments and the $\mathcal{C}^\infty$ triangles of the limit model, respectively.

The results show that the proposed assisted sharp-to-smooth modeling capability reduces the human labor required to prescribe the simulation intent.  Specifically, it facilitates assigning sharp and smooth features to a linear model composed of points, polylines, and triangulations. To illustrate it, we have assigned the desired flow simulation intent to an aircraft model in a high-lift configuration.

In conclusion, we have presented a methodology to model and represent curved geometry of practical interest for flow simulation with unstructured high-order methods. In perspective, high-order methods might benefit from using curved meshes that approximate our curved boundary representation, which we devised to describe the flow simulation intent.

\section*{Acknowledgments}

This project has received funding from the European Research Council (ERC) under the European Union's Horizon 2020 research and innovation programme under grant agreement No 715546. This work has also received funding from the Generalitat de Catalunya under grant number 2017 SGR 1731. The work of the third author has been partially supported by the Spanish Ministerio de Econom\'ia y Competitividad under the personal grant agreement RYC-2015-01633. Special thanks to Eloi Ruiz-Giron\'es.

\appendix

\section{Non-interpolative to Interpolative}
\label{sec_app:interpolative}

Given a control mesh, the schemes in \autoref{sec:curve_subdivision} and \autoref{sec:Loop_subdivision} for mesh subdivision generate a hierarchy of subdivided meshes all of them tending to the same limit model (curve or surface). These schemes do not preserve the position of the initial vertices of the mesh. However, a new control mesh can be computed so that the limit model contains the nodes of the initial mesh \cite{On-the-Use-of-Loop-Subdivision-Surfaces-for-Surrogate-Geometry, jimenez_ramos_albert_2020_3653357}. That is, new points can be found such that the limit curves and surfaces interpolate the given data points.

Given a polygon, the curve subdivision scheme presented in \cite{LaneRiesenfeld1980} generates a sequence of refined polygons, all of them converging to the same cubic $\mathcal{C}^2$-continuous curve. We remark that the initial control mesh determines the limiting curve, so all the refined polygons converge to the same curve. The expression to map a node of the polygon onto the limit curve becomes linear when applied to the nodes of the polygon. Specifically, denote by $\PointPhysical_i^l$ the position of node $i$ at refinement level $l$, and by $\PointPhysical^l_{i-1}$ and $\PointPhysical^l_{i+1}$ its neighbor nodes. Then, the position of node $\PointPhysical^l_{i}$ on the limit curve, $\PointPhysical_i^{l,\infty}$, is determined by the following linear expression
\begin{equation}
\PointPhysical_i^{l,\infty} = \frac{1}{6} \left( \PointPhysical^l_{i-1} + 4 \PointPhysical_i^l + \PointPhysical^l_{i+1} \right).
\label{eq:limit_curve}
\end{equation}

In the case of surfaces, Loop's subdivision scheme \cite{smooth-subdivision-surfaces-based-on-triangles} defines a hierarchy of control meshes, all of them converging to the same limit surface. Analogously to the curve case, we can compute the limiting location for the nodes at any refinement level. In particular, given a linear mesh, the limit position for the node $\Vertex$ at any refinement level $l$, denoted by $\PointPhysical_{\Vertex}^{l,\infty}$, is given by the following linear expression
\begin{equation}
\PointPhysical_{\Vertex}^{l,\infty} = \left( 1- k \chi_k \right) \PointPhysical^l_{\Vertex} + \chi_k \sum_{i=1}^k \PointPhysical^l_i,
\label{eq:limit_position}
\end{equation}
where $\lbrace \PointPhysical^l_i \rbrace_{i=1, \dotsc, k} $ are the positions of the $k$ neighbor nodes of $\Vertex$ at level $l$, and where the weights are computed as
\[
\chi_k = \frac{1}{k+\frac{3}{8  \omega_k}},
\]
with $\omega_k = \frac{1}{k} \left( \frac{5}{8}-\left( \frac{3}{8}+ \frac{1}{4} \cos \left( \frac{2 \pi}{k} \right) \right)^2 \right)$. 

\newcommand{\LinearOperator}{\bm{L}}
To ensure that the limit model contains the nodes of the initial mesh, we compute a new control mesh. Specifically, let $\LinearOperator$ be the operator that maps the nodes of the initial mesh, $\bm{X}^{0}$, onto their limit position. Row $i$ of matrix $\bm{X}^{0}$ corresponds to the position of node $i$ of the boundary mesh $\PointPhysical^0_i$. Then, we compute a new control mesh, with nodes position denoted by $\bm{X}^{C}$, such that
\[
\LinearOperator \bm{X}^{C} = \bm{X}^{0},
\]
where the $i$th row of matrix $\bm{X}^{C}$ corresponds to the unknown position of node $i$ of the control mesh, $\PointPhysical^C_i$. In the case of the subdivision schemes considered in this work, $\LinearOperator$ is a linear application with rows given by the coefficients in \autoref{eq:limit_curve} or \autoref{eq:limit_position}, for curve and surface points, respectively. This matrix is sparse and the solution of the linear system can be computed using a sparse direct solver. In our Python implementation, we call the sparse direct solver of the SuperLU library \cite{superlu99,superlu_ug99} through the Python SciPy package \cite{scipy}.
Recall that this operation is performed on the boundary of a tetrahedral mesh. Therefore, the dimension of the linear system is of the order of the number of boundary nodes and not of the order of nodes of the volume mesh.

\section{Accommodate the Curvature of the Boundary}
\label{sec_app:blending}

In this appendix, we detail the computation TFI-based approach proposed in this work to accommodate the curving of the boundary surface mesh into the interior of the volume mesh in the mesh approximation process of the limit model, see \autoref{sec:volume_ho}.

First, consider an edge of a high-order boundary element with one of its endpoints on the curved surface. The new location of the edge nodes is given by the linear isoparametric mapping between the one-dimensional reference domain and the physical edge, denoted as  $\phi^1$. Specifically, the new position of the $k$th node of the edge, $\PointPhysical_k$, for $k=0, \dotsc, \Degree$, is determined as
\[
\PointPhysical_k = \phi^1 \left( \PointReferenceCartesian_k \right) = \sum_{l=0}^{1} \PointPhysical_{\Vertex_l} N_{\Vertex_l} \left( \PointReferenceCartesian_k \right)
\]
where $\PointReferenceCartesian_k$ is the position of the $k$th interpolation node of the reference domain, $\PointPhysical_{\Vertex_l}$ is the position of the $l$th endpoint of the edge, and $N_{\Vertex_l}$ is the linear nodal shape function of the interval associated to the $l$th endpoint. In \autoref{fig:TFI_1}, we illustrate the relocation of the nodes of the edges in a triangle when one of its edges (in bold) is on the boundary.

Now, we are interested in relocating the nodes on the interior of a face of a boundary tetrahedron. On the one hand, if such face belongs to the boundary, its nodes have been already relocated onto the surrogate geometry, see bold edge in \autoref{fig:TFI_1}. On the other hand, if such face does not belong to the boundary, its edge nodes have been modified using the transfinite interpolation for edges explained above, see non-bold edges in \autoref{fig:TFI_1}. For the latter, we apply the transfinite interpolation for faces, that is, we accommodate the deformation of the curving of the edges to the interior of the triangular faces.

We denote by $f_i$ the edge of the triangle opposite to vertex $i$, $i=1,2,3$. Let us denote as $\xIsoEdge{i}{k}$ the coordinates of the $k$th node of the edge $f_i$ in the physical element, $k=0,\ldots, \Degree$. These nodes are fixed now, since as previously detailed their location has already been computed. Therefore, each curved or relocated edge can be parametrized using the restriction to such edge of the two-dimensional isoparametric mapping of degree $\Degree$, $\PhiDeg$, as:
\begin{equation}
\PhiEdge{i} \left( \PointReferenceCartesian \right) := 
\left. \PhiDeg \right|_{f_i} \left( \PointReferenceCartesian \right) =
 \sum_{k=0}^{\Degree} \xIsoEdge{i}{k} \NSF{i}{k} \left( \PointReferenceCartesian \right)
\label{eq:isoparam_1d}
\end{equation}
where  $\NSF{i}{k} \left( \xi \right)$ is the high-order nodal shape function of the triangle associated to the $k$th node of the edge $f_i$. In particular, the boundary of the triangle is fixed and parameterized by the three mappings $\lbrace \PhiEdge{i} \rbrace_{i=1,2,3}$.

Consider a point $\PointPhysical$ in the physical triangle, to which we want to compute its displaced position in terms of the location of the boundary edge nodes. Denote by $\PointReferenceCartesian$ the position of the point in the reference triangle expressed in cartesian coordinates such that $\PhiDeg \left( \PointReferenceCartesian \right) = \PointPhysical$. Now, denote by $\PointReferenceBarycentric$ the same point in the reference domain expressed in barycentric coordinates $ \left( \PointReferenceBarycentricLetter_1, \PointReferenceBarycentricLetter_2, \PointReferenceBarycentricLetter_3 \right)$, $\sum_{i=1}^{3} \PointReferenceBarycentricLetter_i = 1$. Following \cite{perronnet1998interpolation}, we compute the projection of the point to the edges. A point on an edge can be parametrized as a function of the barycentric coordinates of the two vertices of the triangle defining the edge. Therefore, two different projections ($\PRBProj{*}{j}$ for $j \in f_{*}$) are computed for each one of the three edges of the triangle (rows, $\PRBProj{i}{*}$ for $i=1,2,3$):
\begin{equation*}
\begin{split}
f_1 &= \left(2,3 \right):
\begin{cases}
\PRBProj{1}{2} = \left( 0,1-\PointReferenceBarycentricLetter_3,\PointReferenceBarycentricLetter_3 \right),\\
\PRBProj{1}{3} = \left( 0,\PointReferenceBarycentricLetter_2, 1-\PointReferenceBarycentricLetter_2 \right),  
\end{cases}
\\
f_2 &= \left(1,3 \right): 
\begin{cases}
\PRBProj{2}{1} =\left(1-\PointReferenceBarycentricLetter_3,0,\PointReferenceBarycentricLetter_3 \right),\\
\PRBProj{2}{3} =  \left( \PointReferenceBarycentricLetter_1,0,1-\PointReferenceBarycentricLetter_1 \right),
\end{cases}
\\
f_3 &= \left(1,2 \right): 
\begin{cases}
 \PRBProj{3}{1} = \left( 1-\PointReferenceBarycentricLetter_2,\PointReferenceBarycentricLetter_2,0 \right), \\
 \PRBProj{3}{2} = \left( \PointReferenceBarycentricLetter_1,1- \PointReferenceBarycentricLetter_1 , 0\right).
 \end{cases}
\end{split}
\end{equation*}

Note that $\PRBProj{i}{j}$ belongs to edge $f_i$ and has the $j$th component expressed in terms of the other components. Then, we express these six projections of the point at the edges, computed in barycentric coordinates, back in the reference coordinates $\PointReferenceCartesian$. As previously remarked, we denote the change from barycentric coordinates of a point $\PRBProj{i}{j}$ to reference coordinates as $\PRCProj{i}{j}$, for $i=1,2,3$. Since these points are on the edges of the triangle, they can be mapped onto the physical triangle through the mappings $\PhiEdge{i}$ of the edges, $i=1,2,3$, presented in \autoref{eq:isoparam_1d} as:
\begin{equation*}
\PPProj{i}{j} :=\PhiEdge{i} \left( \PRCProj{i}{j} \right)
\end{equation*}
We highlight that given a point $\PointPhysical$, $\PPProj{i}{j}$ corresponds to the coordinates on the physical element of the projection $j$ of the point to the edge $f_i$, $i=1,2,3$.

Finally, the new position of point $\PointPhysical$ in the physical triangle, denoted as $\hat{\PointPhysical}$, is given in  \cite{perronnet1998interpolation} as: 
\begin{equation*}
\begin{split}
\hat{\PointPhysical}&= \PointReferenceBarycentricLetter_1 \left( \PPProj{2}{1} + \PPProj{3}{1} -  \PointPhysical_{\Vertex_1} \right)
+ \PointReferenceBarycentricLetter_2 \left( \PPProj{1}{2} + \PPProj{3}{2} - \PointPhysical_{\Vertex_2} \right) \\
&+  \PointReferenceBarycentricLetter_3 \left( \PPProj{1}{3} + \PPProj{2}{3} -  \PointPhysical_{\Vertex_3} \right)
\end{split}
\end{equation*}
where $\PointPhysical_{\Vertex_j}$ is the position of the $j$th vertex of the physical triangle. We highlight that the transfinite interpolation for triangles can be expressed as a function of the isoparametric mapping of the edges and the location of the vertices of the triangle.

In order to relocate the nodes in the interior of the high-order physical faces, the steps detailed above are applied to the interpolation nodes in the interior of the high-order reference triangle, see \autoref{fig:TFI_2}. This procedure is repeated for all the faces with a boundary node or edge.

Lastly, we follow an analogous approach to modify the position of the nodes in the interior of the boundary tetrahedra. The boundary of a tetrahedron is composed of four faces and six edges. These faces and edges have already been curved with the procedures detailed above. Therefore, we relocate the interior nodes according to the curved boundary already accommodated to the edges and faces. Similarly to the triangle case, the transfinite interpolation for tetrahedra can be expressed as a function of the isoparametric mapping of the edges, the isoparametric mapping of the faces,  and the location of the vertices of the tetrahedron.

Denote by $T$ the set of vertices of a tetrahedron. We define the entity $f_{i_1,\ldots,i_k}$ as the entity of dimension $d-k$, $d=3$, with vertices given by the nodes $T \setminus \lbrace i_1, \ldots, i_k \rbrace$. Note that the face opposite to node $i$ is denoted by $f_i$, and the edge shared by the faces $f_i$ and $f_k$ is $f_{ik}$. Given the three-dimensional isoparametric mapping of degree $\Degree$, $\PhiDeg$, analogously to \autoref{eq:isoparam_1d}, we denote the restriction to the face $f_i$ as $\PhiEdge{i}$, and the restriction to the edge $f_{ik}$ as $\PhiEdge{ik}$.

Similarly to the two-dimensional case, consider a point $\PointPhysical$ in the physical tetrahedron, and denote by $\PointReferenceCartesian$ and $\PointReferenceBarycentric$ its preimage in the reference tetrahedron expressed in cartesian and barycentric coordinates, respectively.  Now, denote by  $\PRBProj{i}{j}$ the projection of the point to the face $f_i$ that has the $j$th component expressed in terms of the other components. $\PRBProj{ik}{j}$ denotes the projection of the point to the edge $f_{ik}$ that has the $j$th component expressed in terms of the others. These projections in the reference domain expressed in cartesian coordinates are denoted by $\PRCProj{i}{j}$ and $\PRCProj{ik}{j}$, respectively.

Since these points are on the faces and edges of the tetrahedron, they can be mapped onto the physical element through the mappings $\PhiEdge{i}$ on the faces and $\PhiEdge{ik}$ on the edges as:
\begin{equation*}
\PPProj{i}{j} :=\PhiEdge{i} \left( \PRCProj{i}{j} \right), \;\;\; \PPProj{ik}{j} :=\PhiEdge{ik} \left( \PRCProj{ik}{j} \right)
\end{equation*}

Finally, the new position of point $\PointPhysical$ in the physical triangle, denoted as $\hat{\PointPhysical}$, is given in  \cite{perronnet1998interpolation} as: 
\begin{equation*}
\begin{split}
\hat{\PointPhysical} &=
 \PointReferenceBarycentricLetter_1 \left( \PPProj{2}{1} + \PPProj{3}{1} + \PPProj{4}{1} - \PPProj{23}{1}-\PPProj{24}{1}-\PPProj{34}{1} +  \PointPhysical_{\Vertex_1} \right)\\
&+ \PointReferenceBarycentricLetter_2 \left( \PPProj{1}{2} + \PPProj{3}{2} + \PPProj{4}{2} - \PPProj{13}{2}-\PPProj{14}{2}-\PPProj{34}{2} +  \PointPhysical_{\Vertex_2} \right) \\
&+  \PointReferenceBarycentricLetter_3 \left( \PPProj{1}{3} + \PPProj{2}{3} + \PPProj{4}{3} - \PPProj{12}{3}-\PPProj{14}{3}-\PPProj{24}{3} +  \PointPhysical_{\Vertex_3} \right)\\
&+  \PointReferenceBarycentricLetter_4 \left( \PPProj{1}{4} + \PPProj{2}{4} + \PPProj{3}{4} - \PPProj{12}{4}-\PPProj{13}{4}-\PPProj{23}{4} +  \PointPhysical_{\Vertex_4} \right)
\end{split}
\label{eq:TFI_tets}
\end{equation*}

\bibliographystyle{elsarticle-num} 
\bibliography{bibtex} 

\end{document}